\documentclass[11pt,a4paper]{article}

\usepackage{epsfig}
\usepackage{amsmath}
\usepackage{amssymb}
\usepackage{verbatim}
\usepackage{bm}
\usepackage{dsfont}
\usepackage[footnotesize]{caption}
\usepackage{subfigure}
\usepackage{cancel}
\usepackage{cite}
\usepackage{textcomp}
\usepackage{calc}
\usepackage{geometry}

\def\Mp{M_{\rm Pl}}
\def\mG{m_{3/2}}

\begin{document}


\noindent April 2014 \hfill \parbox{\widthof{SISSA  16/2014/FISI}}{DESY 14-005}

\hfill SISSA  16/2014/FISI

\hfill \parbox{\widthof{SISSA  16/2014/FISI}}{IPMU 14-0083}

\vskip 1.5cm

\begin{center}
{\LARGE\bf Hybrid Inflation in the Complex Plane}

\vskip 2cm

{\large W.~Buchm\"uller$^a$, V.~Domcke$^b$, K.~Kamada$^c$,  K.~Schmitz$^d$}\\[3mm]
{\it{
a  Deutsches Elektronen-Synchrotron DESY, 22607 Hamburg, Germany\\
b SISSA/INFN, 34100 Trieste, Italy \\
c Institut de Th\'eorie des Ph\'enom\`enes Physiques,
\'Ecole Polytechnique F\'ed\'erale de Lausanne, 1015 Lausanne, Switzerland \\
d Kavli IPMU (WPI), TODIAS, University of Tokyo, Kashiwa 277-8583, Japan}
}
\end{center}

\vskip 1cm


\begin{abstract}


\noindent Supersymmetric hybrid inflation is an exquisite framework to connect
inflationary cosmology to particle physics at the scale of grand unification.
Ending in a phase transition associated with spontaneous symmetry breaking, 
it can naturally explain the generation of entropy, matter and dark matter.
Coupling F-term hybrid inflation to soft supersymmetry breaking distorts the
rotational invariance in the complex inflaton plane---an
important fact, which has been neglected in all previous studies.
Based on the $\delta N$ formalism, we analyze the cosmological perturbations for the first
time in the full two-field model, also taking into account the fast-roll dynamics at and
after the end of inflation.
As a consequence of the two-field nature of hybrid inflation,
the predictions for the primordial fluctuations depend not only on the parameters
of the Lagrangian, but are eventually fixed by  the choice of the inflationary trajectory.
Recognizing hybrid inflation as a two-field model resolves two shortcomings often
times attributed to it: The fine-tuning problem of the initial conditions is greatly
relaxed and a spectral index in accordance with the PLANCK data can be achieved in a
large part of the parameter space without the aid of supergravity corrections.
Our analysis can be easily generalized to other (including large-field) scenarios
of inflation in which soft supersymmetry breaking transforms an initially
single-field model into a multi-field model.
\end{abstract}.


\thispagestyle{empty}

\newpage


\tableofcontents

\newpage


\section{Introduction}


Supersymmetric hybrid inflation is a promising framework for
describing the very early universe.
Not only does it account for a phase of accelerated
expansion; it also provides a detailed picture of the subsequent
transition to the radiation dominated phase.
Different versions are F-term~\cite{Dvali:1994ms,Copeland:1994vg},
D-term~\cite{Binetruy:1996xj,Halyo:1996pp}
and P-term~\cite{Kallosh:2003ux} inflation, with supersymmetry during
the inflationary phase being broken by an F-term, a D-term or a mixture
of both, respectively.


Hybrid inflation is very attractive for a number of reasons.
It can be naturally embedded into grand unification, and the GUT scale
$M_{\rm GUT}$ yields the correct order of magnitude for the amplitude
of the primordial scalar fluctuations~\cite{Dvali:1994ms}.
Moreover, supergravity corrections are typically small, since during
inflation the value of the inflaton field is $\mathcal{O}(M_{\rm GUT})$,
i.e.\ much smaller than the  Planck scale.
Hybrid inflation ends by tachyonic preheating, a rapid `waterfall' phase transition
in the course of which a global or local symmetry is spontaneously broken~\cite{Felder:2000hj}.
Pre- and reheating  have recently been studied in detail for the case
where this symmetry is $B$$-$$L$, the difference between baryon and lepton number.
The decays of heavy $B$$-$$L$ Higgs bosons and heavy Majorana neutrinos can
naturally explain the primordial entropy, the observed baryon asymmetry 
and the dark matter abundance~\cite{Buchmuller:2010yy,Buchmuller:2011mw,Buchmuller:2012wn}.%
\footnote{For related earlier work, cf.\ Refs.~\cite{Asaka:1999yd,Senoguz:2005bc}.}
Finally, inflation, preheating and the formation of cosmic
strings are all accompanied by the generation of gravitational waves
that can be probed with forthcoming gravitational
wave detectors~\cite{Shafi:2010jr,GarciaBellido:2007dg,Dufaux:2008dn,
Hindmarsh:2011qj, Buchmuller:2013lra}.


The supersymmetric extension of the Standard Model with local
$B$$-$$L$ symmetry is described by the superpotential
\begin{align}
\label{eq_W}
 W = \lambda \Phi \left(\frac{v^2}{2} - S_1 S_2 \right) +
\frac{1}{\sqrt{2}} h_i^n n_i^c n_i^c S_1 + h^{\nu}_{ij} \textbf{5}^*_i
n_j^c H_u + W_{\text{MSSM}} \,.
\end{align}
The first term is precisely the superpotential of F-term hybrid inflation, with the
singlet superfield $\Phi$ containing the inflaton $\phi$ and the waterfall superfields
$S_1$ and $S_2$ containing the Higgs field $\chi$ responsible for
breaking $B$$-$$L$ at the scale $v$.
The next two terms involve the singlet superfields $n^c_i$ whose fermionic components
represent the charge conjugates of the three generations of right-handed neutrinos.
These two terms endow the singlet neutrinos with a Majorana mass term and a
Yukawa coupling to the MSSM Higgs and lepton doublets, denoted here
by $H_u$ and $\textbf{5}^*$ in $SU(5)$ notation.
$\lambda$ and $h$ are coupling constants.


In a universe with an (almost) vanishing cosmological constant,
F-term supersymmetry breaking leads to a constant term in the superpotential,
\begin{align}
W_0 = \alpha \, \mG\Mp^2 \,, \label{eq_W0}
\end{align}
where $\mG$ is the vacuum gravitino mass at low energies and
$\alpha$ a model-dependent $\mathcal{O}(1)$ parameter.
In the Polonyi model, one has $\alpha = \exp{(\sqrt{3}-2)}$~\cite{Polonyi:1977pj}.
For definiteness, we choose $\alpha \equiv 1$ in the following.
We assume that the supersymmetry breaking field is
located in its minimum and that its dynamics can be neglected during inflation. 
Together with the non-vanishing F-term of the inflaton field during inflation,
$F_\Phi = -\lambda \,v^2 / 2$, this constant term in the superpotential induces
a term linear in the real part of the inflaton field in the scalar
potential~\cite{Buchmuller:2000zm},
\begin{align}
V(\phi)
\supset -\left[3\,W(\phi) + F_\Phi^* \,\phi \right]\frac{W_0^*}{M_{\textrm{Pl}}^2} + \textrm{h.c.}
\supset -4 \, \alpha \, m_{3/2}\,\textrm{Re}\left\{-F_\Phi^* \, \phi\right\} \,, \quad
W(\Phi) = -F_\Phi^* \Phi + ... \,.
\label{eq:Vlinear}
\end{align}
The real and the imaginary part of the inflaton field are thus
governed by different equations of motion, requiring an analysis
of the inflationary dynamics in the complex inflaton plane.
As a consequence, all of the inflationary observables are sensitive to the
choice of the inflationary trajectory.
In this sense, the measured values of these quantities do not point to
a particular Lagrangian or specific values of the fundamental model parameters.
To large extent, they are the outcome of a random selection among
different initial conditions which has no deeper meaning within the model itself.
We emphasize that these conclusions apply in general to every
inflationary model in which inflation is driven by one or several large
F-terms.
In the presence of soft supersymmetry breaking, these F-terms will always
couple to the constant in the superpotential and thus induce linear terms
in the scalar potential of exactly the same form as in Eq.~\eqref{eq:Vlinear}.
The analysis in this paper can hence be easily generalized to other models
of inflation, in particular also to models of the large-field type.


Taking the two-field nature of hybrid inflation into account,
we find that the initial conditions problem of hybrid
inflation is significantly relaxed and we can
obtain successful inflation in accordance with the
PLANCK data~\cite{Ade:2013uln} without running into problems due to cosmic
strings~\cite{Ade:2013xla}.
First results of this two-field analysis were presented in Ref.~\cite{Buchmuller:2013dja}.
Non-supersymmetric multi-field hybrid inflation, commonly referred
to as `multi-brid' inflation, has been studied in Refs.~\cite{Sasaki:2008uc,Naruko:2008sq}.
The model investigated here
differs from multi-brid inflation in two regards: (i) we embed inflation into a realistic model
of particle physics and (ii) we study inflation in the context of softly broken supersymmetry.
Furthermore, we note that, during the final stages of preparing this paper,
evidence for a potentially primordial B-mode signal in the polarization of
the cosmic microwave background (CMB)
radiation was announced by the BICEP2 Collaboration~\cite{Ade:2014xna}.
In App.~\ref{app-bicep2}, we discuss the implications of this
very recent development on F-term hybrid inflation.


Our discussion is organized as follows.
In Sec.~\ref{sec_2}, we analyze the connection between $W_0$ and
the spectral index analytically for inflation along the real axis.
In Sec.~\ref{sec_complex_plane}, we then turn to the generic situation
of arbitrary inflationary trajectories in the complex plane.
We perform a full numerical scan of the parameter space, based on a customized
version of the $\delta N$ formalism, in order to determine the inflationary
observables and again reconstruct our results analytically.
Sec.~\ref{sec:inicon} demonstrates how these results relax
the initial conditions problem of F-term hybrid inflation
and Sec.~\ref{sec_gravitino_mass} is dedicated to an investigation
of the allowed range for the gravitino mass.
Finally, we conclude in Sec.~\ref{sec_conclusion}.
As a supplement, we derive in App.~\ref{app:index} simple analytical expressions
that allow to estimate the scalar amplitude as well as the scalar spectral tilt in
general multi-field models of inflation in the limit of negligible effects due
to isocurvature perturbations. 


\section{Hybrid inflation on the real axis}
\label{sec_2}


\subsection{Successes and shortcomings}
\label{subsec_real_overview}


The potential energy of the complex inflaton field
$\phi = \frac{1}{\sqrt{2}} \varphi e^{i \theta}$,
determined by the superpotential given in Eqs.~\eqref{eq_W} and \eqref{eq_W0},
receives contributions from the classical energy density of the false
vacuum~\cite{Copeland:1994vg}, from quantum corrections~\cite{Dvali:1994ms},
from supergravity corrections~\cite{Linde:1997sj}
and from soft supersymmetry breaking~\cite{Buchmuller:2000zm},%
\footnote{The full expression for the effective one-loop or
`Coleman-Weinberg' potential $V_{\textrm{CW}}$ is given in Eq.~\eqref{pot2}.
\smallskip}
\begin{align}
\label{eq:Vtot}
V (\phi) &= V_0 + V_{\rm CW}(\phi) + V_{\rm SUGRA}(\phi) + V_{3/2}(\phi) \,,\\
V_0 &= \frac{\lambda^2 v^4}{4} \,, \\ \label{eq:VCW}
V_{\rm CW}(\phi) &= \frac{\lambda^4 v^4}{\rule{0pt}{10pt} 32 \pi^2}
\ln \left( \frac{|\phi|}{\rule{0pt}{10pt} v/\sqrt{2}}\right) + \ldots \,, \\
V_{\rm SUGRA}(\phi) &= \frac{\lambda^2v^4}{8 \Mp^4} |\phi|^4 + \ldots \,,\label{eq:Vsg}\\
V_{3/2}(\phi) &= - \lambda v^2 \mG (\phi + \phi^*) + \ldots \,,
\label{eq_linear_term}
\end{align}
where $M_\text{Pl} \simeq 2.44 \times 10^{18}\,\textrm{GeV}$
denotes the reduced Planck mass.
During inflation, the energy density of the Universe is dominated
by the false vacuum contribution $V_0$, while the inflaton dynamics are
governed by the field-dependent terms $V_{\rm CW}, V_{\rm SUGRA}$ and/or $V_{3/2}$.
Inflation ends when the waterfall field $\chi$ becomes tachyonically unstable
at $\varphi =  v$.
The scalar potential determines the predictions for the amplitude $A_s$
and the spectral tilt $n_s$ of the scalar power spectrum as well as
the amplitude $f_\text{NL}^\text{local}$ of the local bispectrum.
These should be compared to the recent measurements by the PLANCK
satellite~\cite{Ade:2013xla,Ade:2013ydc},
\begin{equation}
A_s = (2.18^{+ 0.06}_{-0.05}) \times 10^{-9} \,, \quad n_s = 0.963 \pm 0.008 \,, \quad
f_\text{NL}^\text{local} = 2.7 \pm 5.8 \,.
\label{eq_planck_results}
\end{equation}
In the following, we shall consider Yukawa couplings $\lambda \gtrsim 10^{-5}$,
comparable to Standard Model Yukawa couplings, and $v \sim {\cal O} (M_\text{GUT}) $.
In this case, supergravity corrections are negligible, cf.\ Ref.~\cite{Buchmuller:2000zm}.%
\footnote{In our numerical analysis described in
Sec.~\ref{sec_numerics}, we however do incorporate the full supergravity expression.\smallskip}
Most analyses also neglect the linear term in Eq.~\eqref{eq_linear_term},
which arises due to soft supersymmetry breaking.
For small values of $\lambda$ and sufficiently large gravitino masses, this term
is however important and can even dominate the inflaton potential~\cite{Buchmuller:2000zm}.


Hybrid inflation with a linear term has been analyzed in detail in
Ref.~\cite{Nakayama:2010xf}. The authors focused on initial conditions
along the real axis with $\theta_i = \pi$,
to avoid fine-tuning of the initial conditions.%
\footnote{Note that our sign convention for the linear term differs
from the one in Ref.~\cite{Nakayama:2010xf}.
We also remark that the inflaton potential in Eq.~\eqref{eq:Vtot} is
invariant under reflection across the real axis, $\theta \rightarrow - \theta$.
This restricts the range of physically inequivalent values for, say, final
inflaton phases at the end of inflation, $\theta_f$, from $(-\pi,\pi]$ to
$[0,\pi]$, which is why we will not consider any further
negative $\theta_f$ values in the following.}
The linear term induces a local minimum
at large field values in the inflaton potential and for $\theta_i \neq \pi$ the
inflaton may get trapped in this false minimum, preventing successful inflation
if the initial conditions are chosen unfittingly.
For $\theta_i = \pi$, successful inflation is
difficult to achieve but possible for carefully chosen parameter values.
The observed spectral index can be obtained by resorting to a non-minimal
K\"ahler potential~\cite{BasteroGil:2006cm}.


\begin{figure}[t]
\centering
\includegraphics[width = 0.5\textwidth]{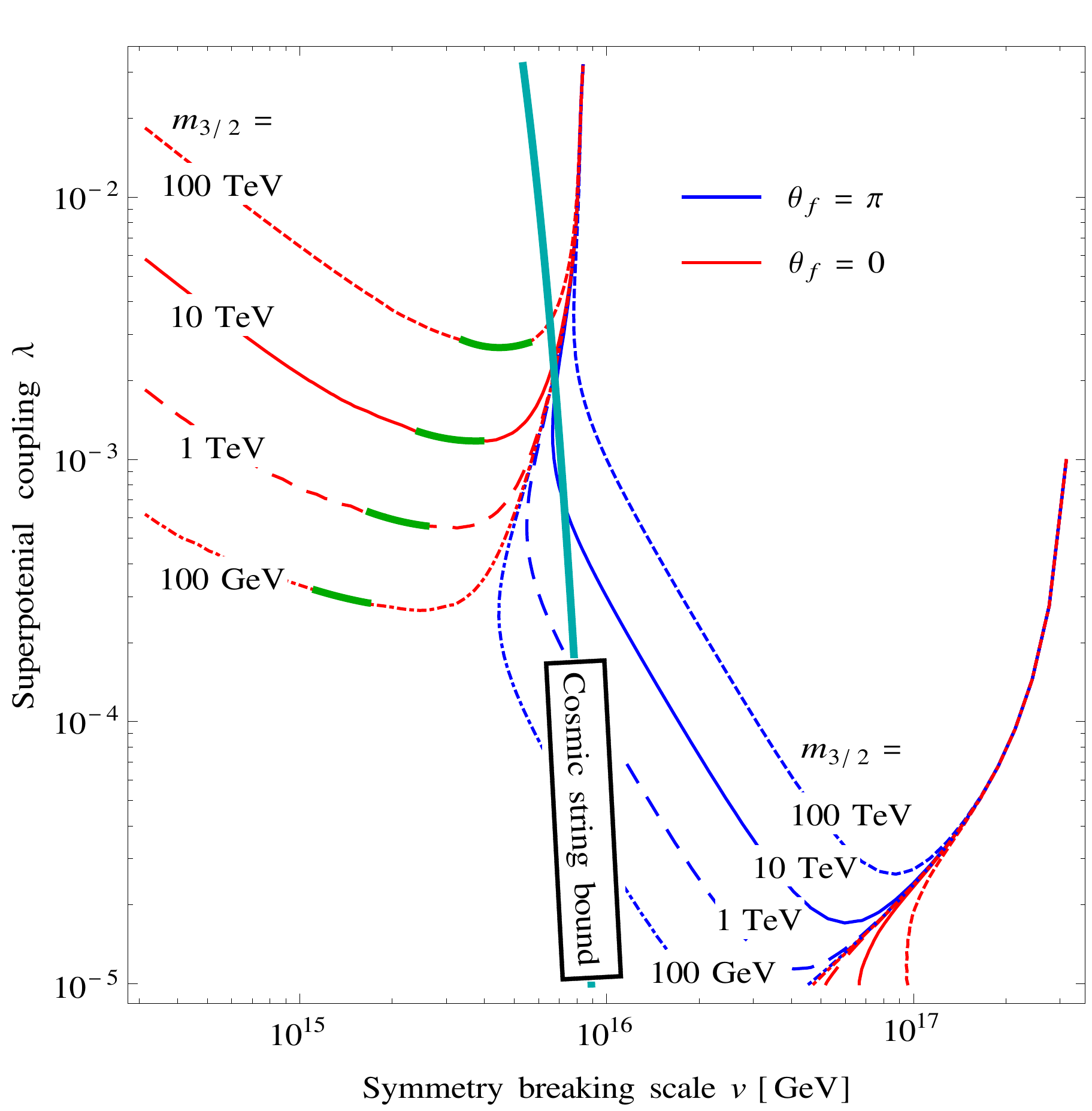}
\caption{Contour lines in the $(v,\lambda)$ plane along which inflation
on the real axis succeeds in reproducing the observed value of $A_s$;
the red and blue contours correspond to $\theta_f = 0$ and $\theta_f = \pi$,
respectively.
The gravitino mass is varied over four different values,
$\mG = 0.1$, $1$, $10$ and $100$~TeV, and consistency with the 
observed value of the scalar spectral index at $95\,\%$\,C.L.\
is indicated by the green contour segments.
The region to the right of the thick light-blue line is excluded due to
the non-observation of cosmic strings.}
\label{fig_vlpi2}
\end{figure}


Recently, it has been observed that for inflation along the real axis
with $\theta_i=0$ the observed spectral index can be obtained for a
canonical K\"ahler potential~\cite{Armillis:2012bs,Pallis:2013dxa} in
the hill-top regime of hybrid inflation~\cite{Boubekeur:2005zm},
if one allows for severe fine-tuning of the initial conditions.
Furthermore, the current bound on the tension of cosmic strings~\cite{Ade:2013xla}
is naturally satisfied in this case,%
\footnote{This bound derives from constraints on the string contribution to
the CMB power spectrum.
Meanwhile, even stronger bounds on $G\mu$ can be obtained from constraints
on the stochastic gravitational wave background induced by decaying string loops.
Using  data from the European Pulsar Timing Array
(EPTA)~\cite{vanHaasteren:2011ni,Sanidas:2012ee},
the authors of Ref.~\cite{Blanco-Pillado:2013qja} arrive, for instance, at
$G\mu < 2.8\times10^{-9}$, which translates into a roughly ten times stronger
constraint on the symmetry breaking scale $v$ than the bound in
Eq.~\eqref{eq_cosmic_string_bound}.
In our analysis, we will however stick to the bound in
Eq.~\eqref{eq_cosmic_string_bound} nonetheless.
While in principle this bound is based on simulations of the cosmic string network
in the Abelian-Higgs model, it turns out to be rather model-independent after all.
An analogous analysis based on string simulations in the Nambu-Goto model
arrives at a very similar result~\cite{Ade:2013xla}.
By contrast, the bound presented in Ref.~\cite{Blanco-Pillado:2013qja} strongly depends
on uncertain string physics such as the production scale of string loops and the nature
of string radiation.
In this sense, the bound in Eq.~\eqref{eq_cosmic_string_bound} is more 
conservative and hence also more reliable.}
\begin{align}
G\mu < 3.2 \times 10^{-7} \,,
\label{eq_cosmic_string_bound}
\end{align}
where $G = (8\pi\Mp^2)^{-1}$ is Newton's constant and $\mu \simeq 2\pi
v^2$ is the string tension~\cite{Hindmarsh:2011qj}.
These are  interesting results despite the fine-tuning problem of initial conditions.
In both cases, $\theta_i = 0$ and $\theta_i = \pi$,
the inflaton phase remains unchanged during inflation,
so that at the end of inflation the final phase $\theta_f$
either corresponds to $0$ or to $\pi$.
In Fig.~\ref{fig_vlpi2}, we compare the constraints on the parameters $v$ and $\lambda$
imposed by the normalization of the scalar power spectrum for these two situations.
In doing so, we also vary the gravitino mass and determine the parameter
combinations for which the scalar spectral index falls into the $2\,\sigma$ range
around the measured best-fit value.
The results shown in Fig.~\ref{fig_vlpi2} are based on the numerical analysis
described in Sec.~\ref{sec_complex_plane}.
We observe that, while the case $\theta_f = \pi$ (blue contours)
is almost excluded by the cosmic string constraint, this constraint is
automatically satisfied in most of the parameter space for
the case $\theta_f = 0$ (red contours).
To sum up, we find that hybrid inflation on the positive real axis
is able to reproduce the scalar spectral index for a canonical K\"ahler potential
and is in less severe tension with the non-observation of cosmic strings.
At the same time, hybrid inflation on the negative real axis has the virtue 
that it does not require the initial position of the inflaton to be finely tuned.


\subsection{Understanding the hill-top regime}
\label{sec_hilltop}


In this section, our goal is to analytically reconstruct our results for
$A_s$ and $n_s$ depicted in Fig.~\ref{fig_vlpi2} for the case of hybrid inflation on the real
axis in the the hill-top regime ($\theta_f = 0$) based on a canonical K\"ahler potential.
This analysis will prove to be a useful preparation for our general
investigation of hybrid inflation in the complex plane in Sec.~\ref{sec_complex_plane}.
As the amplitude of the local bispectrum $f_{\rm NL}^{\rm local}$
is slow-roll suppressed in the single-field case, we do not study it
in this section; for a discussion of $f_{\rm NL}^{\rm local}$ in
the general two-field scenario, cf. Sec.~\ref{subsec:parameterscan}.


The inflaton field is a complex scalar, $\phi = \frac{1}{\sqrt{2}}(\sigma + i \tau)$,
and the relevant variables are its real and imaginary parts normalized to the symmetry
breaking scale $v$, $x \equiv \sigma/v$ and $y \equiv \tau/v$.
During the inflationary phase, the inflaton potential is flat in
global supersymmetry at tree-level.
The one-loop quantum and tree-level supergravity corrections 
only depend on $|\phi|$, the absolute value of the inflaton field.
Supersymmetry breaking generates an additional term linear in $\sigma$,
such that one obtains for the scalar potential
\begin{align}\label{pot1}
V(x,y) \simeq V_0 + a f(z) - b x  \,, \quad z \equiv x^2 + y^2 \,, \quad
a \equiv \frac{\lambda^4 v^4}{128\pi^2} \,, \quad
b \equiv \sqrt{2}\lambda v^3 \mG \,.
\end{align}
where we have neglected the quartic supergravity term and with
the one-loop function
\begin{align}\label{pot2}
f(z) \equiv (z+1)^2 \ln (z+1) + (z-1)^2 \ln (z-1) - 2 z^2 \ln z - 1 \,. 
\end{align}
Here, $af$ is nothing but the Coleman-Weinberg potential,
$af \equiv V_{\textrm{CW}}$, cf.\ Eq.~\eqref{eq:VCW}.
We choose the sign convention such that $b > 0$.
For $z > 1$, i.e.\ $\sigma^2 + \tau^2 > v^2$, inflation can take place, ending
in a waterfall transition at $z=1$.%
\footnote{Typically, the slow-roll condition for the slow-roll parameter
$\eta$, cf.\ Eq.~\eqref{eta-0}, is violated slightly before $z=1$ is reached.
We will take this into account when solving the equations of motion
for the inflaton fields numerically.
For the purpose of the analytical estimates of this section, this effect
is negligible.}
In the slow-roll regime, the equations of motion for the two real inflation fields
$\sigma$ and $\tau$ as well as the Friedmann equation for the Hubble parameter $H$ read,
\begin{align}
3 H \dot \sigma = -\partial_{\sigma} V \,, \quad 3 H \dot \tau = -\partial_{\tau} V \,,
\quad H^2 = \frac{V}{3M_{\textrm{Pl}}^2}\,.
\end{align}
As $V_0$ vastly dominates the potential energy $V$ for all times during inflation,
we shall approximate $H^2$ by $H_0^2 = V_0 / (3M_{\textrm{Pl}}^2)$ in the following
for the purposes of our analytical calculations.


The number of $e$-folds between a critical point $\phi_c$, at which
inflation ends, and an arbitrary point $\phi$ in the complex plane
are given by a line integral along the inflationary trajectory,
\begin{align}\label{efolds}
N(\phi) = - \int_{t(\phi_c)}^{t(\phi)} H dt \,. 
\end{align}
As explained in App.~\ref{app:index}, in general multi-field models of inflation,
the scalar amplitude $A_s$ and the scalar spectral tilt $n_s$ are approximately
given by the simple single-field-like expressions
\begin{align}\label{ns-0}
A_s = \frac{H^2}{8\pi^2 \epsilon \, M_{\rm Pl}^2} \,, \quad
n_s = 1 - 6\,\epsilon + 2\,\eta \,,
\end{align}
if (and only if) isocurvature modes during inflation do not give a
significant contribution to the scalar power spectrum.
$\epsilon$ and $\eta$ are the slow-roll parameters along the inflationary trajectory,
\begin{align}
\epsilon &= \frac{1}{2} \Mp^2 \frac{\partial^a V\partial_a V}{V^2} \,,  \label{epsilon-0}\\
\eta &= \frac{\Mp^2}{V} \frac{1}{\partial^c V \partial_c V} \partial^a V
(\partial_a\partial_b V)\partial^b V \,, \label{eta-0}
\end{align}
with the inflaton `flavor' indices $a$, $b$ and $c$ all running over $\sigma$ and $\tau$.
In the following, we shall use these expression to obtain simple
analytical estimates for $A_s$ and $n_s$.
Hence, in order to make connection between our predictions and the measured values for
the inflationary observables, we need to evaluate $\epsilon$ and $\eta$
in Eqs.~\eqref{epsilon-0} and~\eqref{eta-0}
$N_* \simeq 50$ $e$-folds before the end of inflation, when the CMB pivot
scale $k_* = 0.05\,\textrm{Mpc}^{-1}$ exits the Hubble horizon.


In this section, we shall restrict ourselves to inflation along the real axis.
Since
\begin{align}
3 H \dot y = - \frac{1}{v^2} \partial_y V = - \frac{2a}{v^2} f'(z) \, y \,,
\end{align}
with $f'(z) = \partial_z f(z)$, the real axis with $y = 0$ is a
indeed a stable solution of the slow-roll equations.
In $x$~direction, one has
\begin{align}
3 H \dot x = - \frac{1}{v^2} \partial_x V = - \frac{1}{v^2}\left(2a f'(z)\,x - b\right) \,.
\end{align}
If the constant term $b$ can be neglected, one obtains the standard
form of hybrid inflation.
In this case, $N_* \simeq 50$ $e$-folds correspond in field
space to a point $\left|x_*\right| \gg 1$, where $f'(x_*^2) \simeq 2/x_*^2$,
which leads to the spectral index
\begin{align}
n_s \simeq 1 - \frac{1}{N_*} \simeq 0.98\,.
\end{align}
This value is disfavoured by the recent PLANCK data.
It deviates from the measured central value $n_s \simeq 0.96$ by about $3\,\sigma$.


\begin{figure}[t]
\centering
\includegraphics[width = 0.7\textwidth]{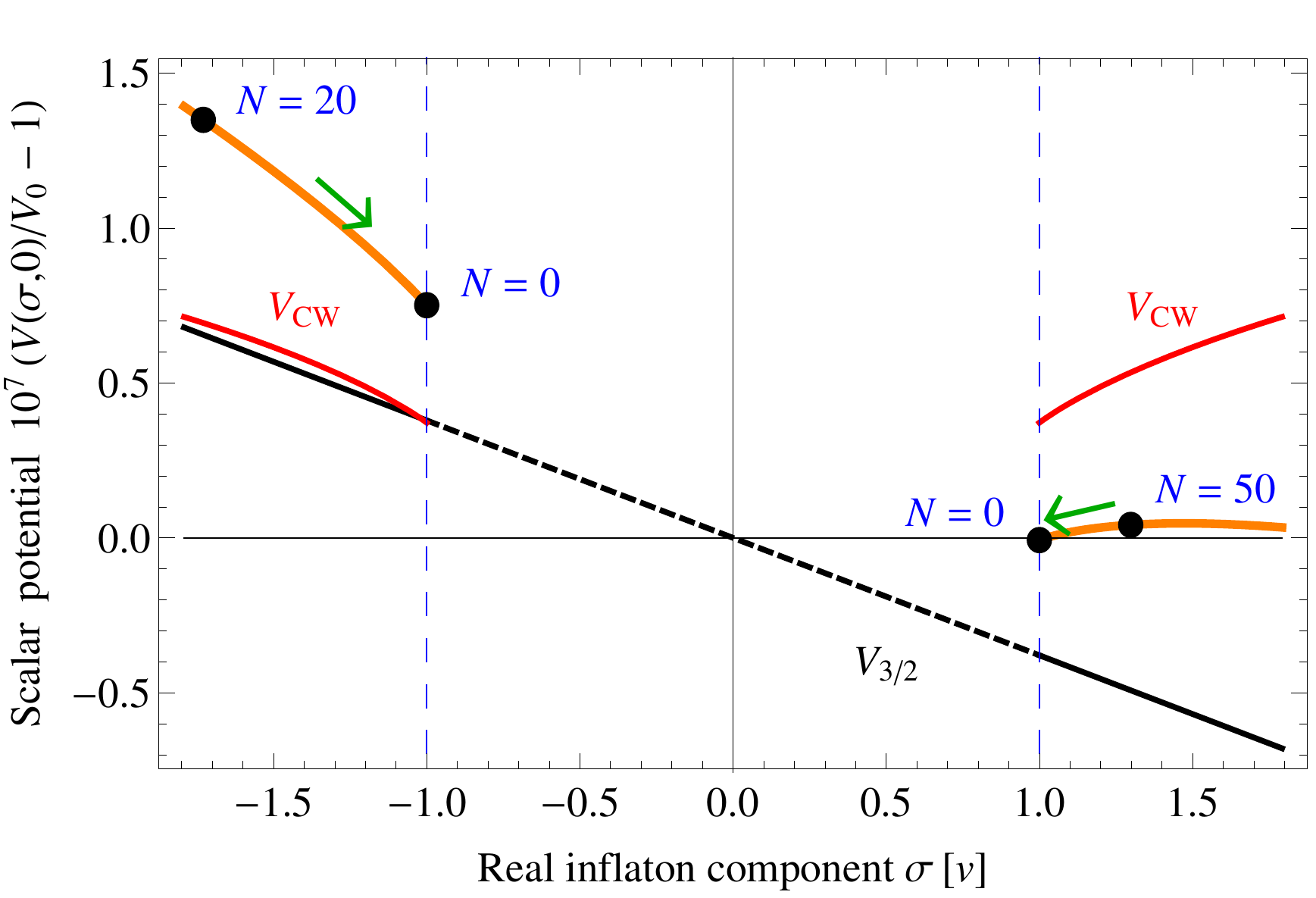}
\caption{Scalar potential for inflation along the real axis in the complex inflaton
field space after adding a constant term $W_0$ to the superpotential.
Slow-roll inflation is possible for both $\theta = 0$ and $\theta = \pi$.
Here, we have chosen parameter values $v = 3.6 \times 10^{15}$~GeV,
$\lambda = 2.1 \times 10^{-3}$ and $m_{3/2} = 50$~TeV.}
\label{fig_potential_1D}
\end{figure}


For sufficiently large values of $b$, an interesting new regime opens
up for field values very close to the critical point~\cite{Pallis:2013dxa}.
This is apparent from Fig.~\ref{fig_potential_1D}, where the
potential is displayed for representative values of $v$, $\lambda$ and $\mG$.
Note that the first derivative of the loop-induced potential is always positive,
\begin{align}
f'(z) = 2 (z+1) \ln {\left(1 + \frac{1}{z}\right)} 
+  2 (z-1) \ln {\left(1 - \frac{1}{z}\right)} > 0 \,.
\end{align}
As a consequence, for initial conditions $x_i > 1$, cancellations
between the gradients of the linear term and the one-loop potential
can lead to extreme slow roll.
The second derivative of the loop potential is always negative,
\begin{align}
f''(z) = 2 \ln \left(1 - \frac{1}{z^2}\right) \,< \,0 \,,
\end{align}
and diverges for $z \rightarrow 1$.
This allows small values of $n_s$,
if $x_*$ is sufficiently close to the critical point.
For the example shown in Fig.~\ref{fig_potential_1D},
the point of 50 $e$-folds is  $x_* \simeq 1.3$.
Note that successful inflation requires carefully
chosen initial conditions. The inflaton rolls in the direction of the
critical point only if $x_{i} \lesssim 1.5$.
We will come back to the problems related to the necessary tuning of the
initial conditions in more detail in Sec.~\ref{sec:inicon}.
Also for initial values $x_i < -1$, the linear term significantly
modifies the loop-induced potential, but qualitatively the picture does not change.


\newpage

Let us now consider the hill-top regime quantitatively.
Close to the critical point, i.e.\ for $x_* - 1 \ll 1$, one has
for the first and the second derivative of the one-loop function $f$
\begin{align}
\begin{split}\label{deriv1}
\left.\partial_x f(x^2)\right|_{x_*}
& = 4 x_* \left[\left(x_*^2-1\right) \ln\left(x_*^2-1\right) +
\left(x_*^2+1\right) \ln\left(x_*^2+1\right) - 2 x_*^2 \ln x_*^2\right] \\
& = 8 \ln 2 + \mathcal{O}(x_*-1) \,,
\end{split} \\
\begin{split}\label{deriv2}
\left.\partial^2_x f(x^2)\right|_{x_*}
& = 12 \, x_*^2 \ln\left(1 - \frac{1}{x_*^4}\right)
+ 4 \ln\left(\frac{x_*^2+1}{x_*^2-1}\right) \\
& = 8 \ln \left[8(x_* - 1)\right] + \mathcal{O}(x_*-1) \,.
\end{split}
\end{align}
The value of $x_*$ is determined by, cf. Eq.~\eqref{efolds},
\begin{align}
N_* = \frac{v^2}{\Mp^2} \int_1^{x_*} \frac{V}{\partial_x V}\,dx \,,
\end{align}
and using Eq.~\eqref{deriv1} one obtains
\begin{align}
N_* = \frac{v^2}{\Mp^2} \frac{4\pi^2}{\lambda^2  (1- \xi) \ln 2}\, (x_*-1)\,,
\label{eq:Nxrel}
\end{align}
where the parameter $\xi$ measures the relative importance of the
two contributions to the slope of the potential in Eq.~\eqref{pot1},
\begin{align}
\xi \equiv \frac{2^{9/2} \pi^2}{\lambda^3\,\ln 2}\,\frac{\mG}{v} \,.
\end{align}
Consistency (i.e.\ the inflaton rolling towards the critical line)
requires $\xi < 1$, which yields an upper bound on the
gravitino mass, cf.\ also the discussion in Sec.~\ref{sec_gravitino_mass},
\begin{align}
\mG\, < \, \frac{\lambda^3\,\ln 2}{2^{9/2} \pi^2} \,v \,.
\label{eq_mgbound1}
\end{align}

 
Clearly, tuning $\lambda$ and $\mG$, one can move $x_*$ very close to
the critical point. This enhances the amplitude of the scalar
fluctuations,
\begin{align}\label{As}
A_s = \left.\frac{H_0^2}{8\pi^2\epsilon\Mp^2}\right|_{x_* \approx 1}
= \frac{\pi^2}{3 (\ln 2)^2 \,\lambda^2\, (1 - \xi)^2} \
\left(\frac{v}{\Mp}\right)^6 \,.
\end{align}
From Eqs.~\eqref{ns-0}, \eqref{eta-0} and \eqref{deriv2}
one obtains for the spectral index
\begin{align}\label{ns}
n_s - 1 \simeq 2 \left.\eta\right|_{x_* \approx 1} \,
\simeq \frac{\lambda^2}{2\pi^2} \frac{\Mp^2}{v^2} \ln{\left(\frac{2 \ln2\
      \lambda^2}{\pi^2} \frac{\Mp^2}{v^2} N_* (1-\xi)\right)} \,.
\end{align}
Finally, eliminating $\xi$ by means of
Eq.~\eqref{As},  one obtains a relation
between the spectral index and the amplitude of scalar fluctuations,
which is independent of the gravitino mass,
\begin{align}\label{smallxns}
n_s - 1 \simeq - \frac{\lambda^2}{4\pi^2} \frac{\Mp^2}{v^2} \ln
\left(\frac{3\pi^2 A_s}{4\lambda^2 N_*^2} \frac{\Mp^2}{v^2}\right) \,.
\end{align}
Note that this relation is very different from standard hybrid
inflation, where $A_s$ and $n_s$ are determined by $v$ and $N_*$,
respectively, and where the dependence on $\lambda$ is very weak.


\begin{figure}[t]
\begin{center} 
\includegraphics[width = 0.48\textwidth]{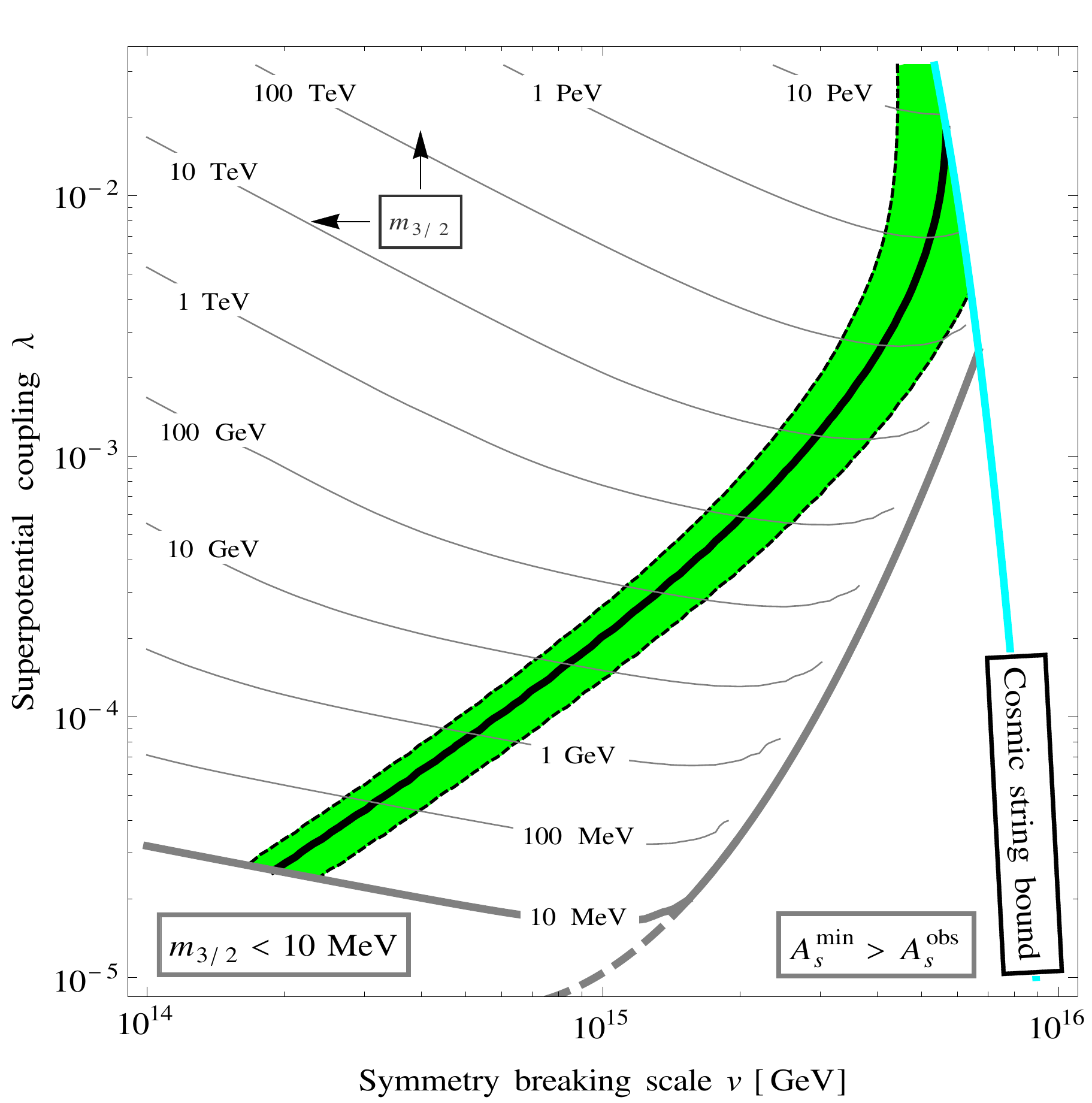}\hfil
\includegraphics[width = 0.48\textwidth]{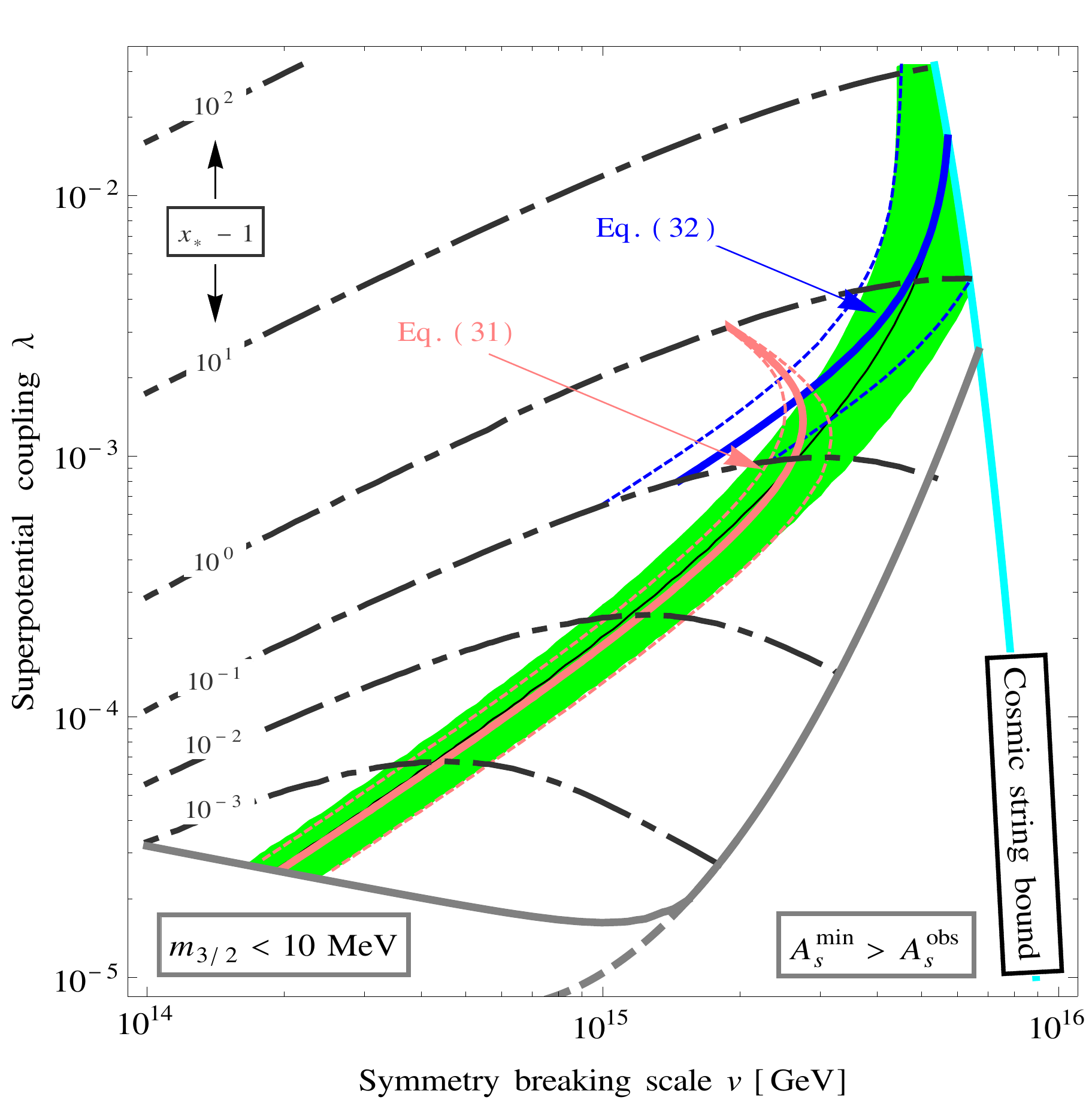}
\end{center}
\caption{Constraints on the model parameters of hybrid inflation, $v$, $\lambda$
and $m_{3/2}$, imposed by the measured values for the inflationary observables
and the cosmic string bound (light-blue curve) for $\theta_f = 0$.
For each ($v, \lambda$) pair, the gravitino mass is adjusted, as indicated
by the grey contour lines in the \textbf{left panel}, such that the
scalar amplitude $A_s$ comes out right.
In the region labeled $A_s^{\textrm{min}} > A_s^{\textrm{obs}}$, our prediction
for $A_s$ is always larger than the observed value $A_s^{\textrm{obs}}$, even if
$m_{3/2}$ is set to $0$.
Along the solid black lines, the best-fit value
for the scalar spectral index is reproduced, with the green band indicating the
corresponding $2\,\sigma$ confidence interval.
All of the black and grey contour curves in both panels are the result of our full numerical
calculation, cf.\ Sec.~\ref{sec_complex_plane}.
The red and blue curves in the \textbf{right panel} are by contrast based on our
(semi-)analytical results for $n_s$ in the small-$x_*$ and large-$x_*$ regime, respectively,
cf.\ Eqs.~(\ref{smallxns}) and \eqref{largexns}.
The initial field values $x_*$ are indicated by the grey dot-dashed contour lines.}
\label{fig_1field_results}
\end{figure}


For larger couplings $\lambda$, the gradient of the one-loop potential
increases and a longer path in field space is needed to obtain $N_*
\simeq 50$ $e$-folds.
To achieve this for GUT-scale field values, i.e.\
$x_*= \mathcal{O}(1)$, a larger gravitino mass is needed to reduce the
gradient of the total potential.
A rough estimate for the spectral index
can be obtained by using for the second derivative of the potential
the approximation for large field values, $\partial_x^2 f|_{x_*}
\simeq -4/x_*^2$, which yields
\begin{align}\label{largexns}
n_s - 1 \sim -\frac{\lambda^2}{4\pi^2} \frac{\Mp^2}{v^2}
\frac{1}{x_*^2} \,.
\end{align}  
This expression agrees with Eq.~\eqref{smallxns} up to an
$\mathcal{O}(1)$ factor.
Note that a numerical determination of
$x_*$ is needed in order to obtain quantitative result for $n_s$.


The domain of successful inflation in the $(v,\lambda)$ plane
reproducing the measured amplitude of the scalar fluctuations and the
spectral index is displayed in Fig.~\ref{fig_1field_results}.
The left panel shows the result of a numerical analysis.
Since the real axis is merely a special case of all possible trajectories
in the complex plane, these results were obtained using the two-field
method described in Sec.~\ref{sec_numerics}.
For each $(v, \lambda)$ pair, the measured amplitude of the primordial
fluctuations is used to fix the gravitino mass, cf.\ the grey contour lines.
In the green band, the spectral index lies in the range
$n_s = 0.963 \pm 0.016$, cf.\ Eq.~\eqref{eq_planck_results}.
In the right panel, the numerical results are compared with
the analytical estimates.
The small-$x_*$ approximation in Eq.~\eqref{smallxns} works approximately
up to $x_*-1 \sim 0.1,$ whereas Eq.~\eqref{largexns}, after inserting
numerical  values for $x_*$, provides a rough estimate
for $x_*-1 \gtrsim 0.1$.
The four parameter points discussed in Ref.~\cite{Pallis:2013dxa}
correspond to  $x_*-1 \sim 0.01$, i.e.\ they require a rather
strong fine-tuning of the initial position of the inflaton field.


\section{Hybrid inflation in the complex plane \label{sec_complex_plane}}


So far, we have considered inflation for $\theta = 0$.
Due to the linear term in the inflaton potential, a new interesting hilltop
region has emerged, which allows for a small spectral index consistent
with observation.
This improvement in $n_s$ is only achieved, however, at the price of a
considerable fine-tuning of the initial position of the inflaton field on the real axis.

\begin{figure}[t]
\includegraphics[width = 1.0\textwidth]{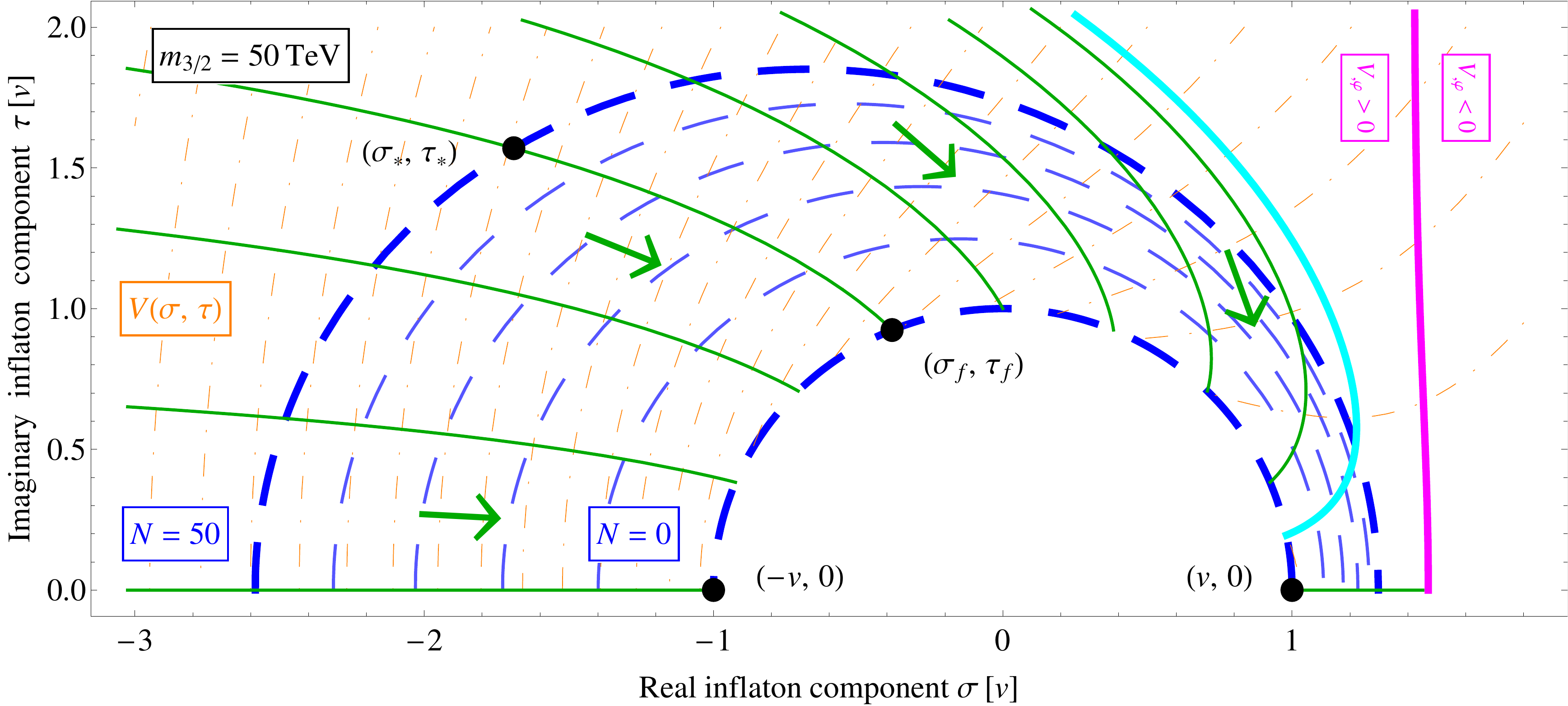}
\caption{Two-field dynamics of the complex inflaton in field space.
The solid green lines represent possible inflationary trajectories
in the scalar potential $V(\sigma, \tau)$ (dot-dashed orange contour lines.)
Lines of constant $N$ are marked by dashed blue contours, with the beginning
and end of inflation ($N = N_*$ and $N = 0$, respectively) marked by thicker contours.
Along the light-blue trajectory, the measured values of $A_s$ and $n_s$ are reproduced.
The model parameters are again set to
$v = 3.6 \times 10^{15}$~GeV, $\lambda = 2.1 \times 10^{-3}$ and $\mG = 50$~TeV,
cf.\ Fig.~\ref{fig_potential_1D}.}
\label{fig_trajectories_50}
\end{figure}
\begin{figure}
\begin{center} 
\includegraphics[width = 1.0\textwidth]{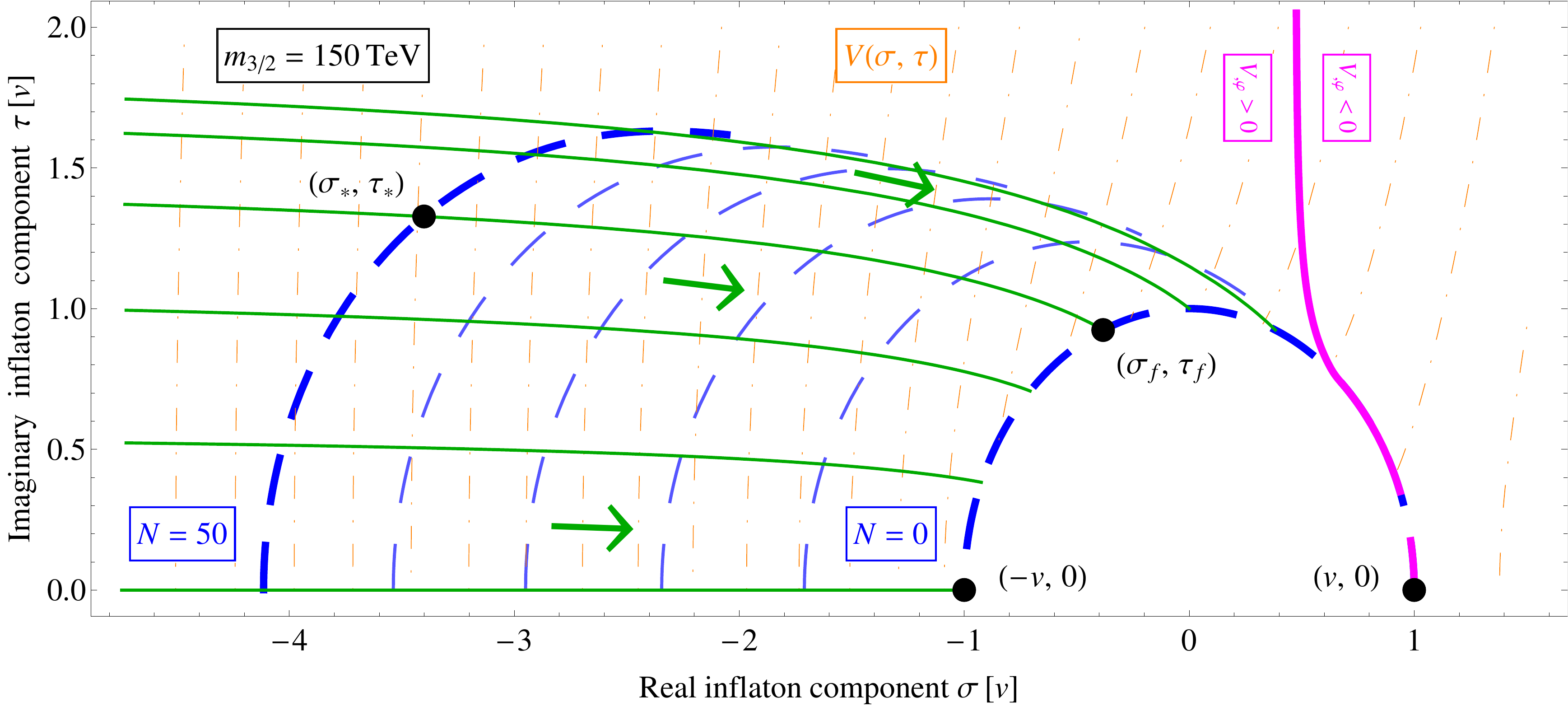}
\end{center}
\caption{Effect of a large constant term in the superpotential,
$W_0 = m_{3/2} M_{\textrm{Pl}}^2$, on the
inflationary trajectories in the complex plane. Colour code, labels
and parameter values as in Fig.~\ref{fig_trajectories_50}, but now with $\mG = 150$~TeV.}
\label{fig_trajectories_largem32}
\end{figure}

The situation changes dramatically once we take into account the fact that,
also due to the linear term in the inflaton potential, F-term hybrid
inflation is a two-field model of inflation:
As we have demonstrated in Sec.~\ref{sec_2}, the potential depends in fact differently on
the real and the imaginary part of the inflaton field $\phi$ and not only on
its absolute value $\varphi$.
The rotational invariance in the complex plane is thus broken, which is why,
depending on the initial value of the inflaton phase, $\theta_i$, the inflaton may actually
traverse the field space along complicated trajectories that
strongly deviate in shape from the simple trajectories on the real axis.%
\footnote{This is illustrated in Figs.~\ref{fig_trajectories_50} and
\ref{fig_trajectories_largem32}, which show a set of possible inflationary
trajectories in field space for typical parameter values.
We will come back to these plots in Sec.~\ref{subsec:parameterscan},
when presenting our numerical results.}
In order to obtain a complete picture of hybrid inflation, it is
therefore important to extend our analysis from the previous section to the
general case of inflation in the complex plane.
To do so, we will first introduce our formalism, by means of which we are able to
calculate predictions for the inflationary observables in the case of multi-field inflation.
Then, we will apply this formalism to hybrid inflation in the complex plane and present
our numerical results.
After that, we will finally demonstrate how our numerical findings can be roughly reconstructed
based on analytical expressions.


\subsection[Inflationary observables in the $\delta N$ formalism]
{Inflationary observables in the \boldmath$\delta N$ formalism \label{sec_numerics}}


The analytical estimates presented in the previous section were mostly based on an effective
single-field approximation.
However, in order to fully capture the two-field nature of hybrid inflation,
we have to go beyond this approximation and perform a numerical analysis of
the inflationary dynamics in the complex plane.
In doing so, we shall employ an extended version of the so-called `backward method'
developed by Yokoyama et al.~\cite{Yokoyama:2007uu,Yokoyama:2007dw}
in the context of the $\delta N$ formalism~\cite{Sasaki:1995aw,Starobinsky:1986fxa,
Salopek:1990jq,Sasaki:1998ug,Lyth:2004gb,Lyth:2005fi}.


The essence of the $\delta N$ formalism is that it identifies the curvature perturbation~$\zeta$
on uniform energy density hypersurfaces as the fluctuation $\delta N$ in the number of $e$-folds
which is induced by the fluctuation of the inflaton in field space, $\delta \phi$,
around its homogeneous background value,%
\footnote{More concretely, $\delta N$ is calculated as the fluctuation in the number of $e$-folds
between the initial flat hypersurface at $t = t_*$, i.e.\ at the time when the CMB pivot scale
$k_* = 0.05\,\textrm{Mpc}^{-1}$ exists the Hubble horizon, and some appropriately chosen
final uniform energy density hypersurface at $t = t_f$, on which all possible
inflationary trajectories have already converged.
This latter hypersurface is hence constructed such that, for all later times, the universe is
in the adiabatic regime and can be described by a single cosmic clock.
Consequently, the curvature perturbation $\zeta$ remains constant for all times $t \geq t_f$.}
\begin{align}
\zeta \approx \delta N \,.
\label{eq:zeta}
\end{align}
In calculating $\delta N$, one is free to either specify
a boundary condition $N = N^{(0)}$ at early
or at late times and then either evolve $N$ forward or backward in time.
Obviously, the backward method described by Yokoyama et
al.~\cite{Yokoyama:2007uu,Yokoyama:2007dw} pursues the latter approach,
cf.\ also the geometrical analysis presented in Ref.~\cite{Mazumdar:2012jj}.
The former approach is implemented in the `forward method' developed by
the authors of Refs.~\cite{Seery:2012vj,Anderson:2012em}.
Either way, it is important to notice that the $\delta N$ formalism in its
standard formulation, cf.\ Eq.~\eqref{eq:infobsdN}, comes
with intrinsic limitations.
For instance, the possible interference between different modes at the time of Hubble exit
is usually neglected and all perturbations are instead taken to be uncorrelated and Gaussian.
Likewise, the universe is assumed to eventually reach the adiabatic limit with
no isocurvature modes remaining at late times.
Finally, the decaying modes in the curvature perturbation spectrum
cannot be accounted for by the $\delta N$ formalism.
More advanced computational techniques to overcome this latter problem
have recently been proposed in the literature~\cite{Naruko:2012fe}.
But, as we do not have to deal with any, say, temporal violation of the
slow-roll conditions, these decaying modes are negligible in our case just
as in any other `standard scenario' of slow-roll inflation.
We therefore do not have to resort to a more
sophisticated method and can safely stick to the $\delta N$ formalism. 
Similarly, as the two slow-roll parameters $\epsilon$ and $\eta$ are always small
except during the last few $e$-folds of inflation, we will make use of the
slow-roll approximation for most of the inflationary period.
At the same time, the smallness of $\epsilon$ and $\eta$ also guarantees that
the `relaxed slow-roll conditions' stated in Ref.~\cite{Yokoyama:2007uu}
are satisfied for most times.
This justifies why Yokoyama et al.'s backward method
is applicable to our inflationary model in its slow-roll formulation.


In the $\delta N$ formalism, the inflationary observables $A_s$, $n_s$ and
$f_{\textrm{NL}}^{\textrm{local}}$ are all determined by the derivatives of
the function $N$ w.r.t.\ to the various directions in field space,
\begin{align}
A_s = \left(\frac{H}{2\pi}\right)^2 N^a N_a \,, \quad
n_s = 1 - 2 \left(\frac{H'}{H} + \frac{N^a N_a'}{N^b N_b}\right) \,, \quad
f_{\textrm{NL}}^{\textrm{local}} = \frac{5}{6}\frac{N^a N_{ab}N^b}{\left(N^c N_c\right)^2} \,,
\label{eq:infobsdN}
\end{align}
where $N_a$ and $N_{ab}$ are the first and second partial derivatives of $N$ in the sense
of a function on field space
and with a prime denoting differentiation w.r.t.\ to $N$ in the sense of a time coordinate.
For an arbitrary number of canonically normalized real inflaton fields $\phi^c$, we have%
\footnote{In the remainder of this paper, all `flavour' indices $a$, $b$, $c$, ...
always run over $\sigma$ and $\tau$, just as in Sec.~\ref{sec_hilltop}.\smallskip}
\begin{align}
N_a(N) = \frac{\partial N \left(\left\{\phi^c(N)\right\}\right)}{\partial\phi^a} \,, \quad
N_{ab}(N) = \frac{\partial^2 N
\left(\left\{\phi^c(N)\right\}\right)}{\partial\phi^a\partial\phi^b}\,, \quad
X'(N) = \frac{dX(N)}{dN} \,. 
\end{align}
As we shall not consider the possibility of a non-canonical K\"ahler potential,
we have assumed canonical kinetic terms for all scalar fields in writing down
Eq.~\eqref{eq:infobsdN}.%
\footnote{Typically, the most important consequence of a non-canonical K\"ahler potential
would  be the new Planck-suppressed terms it induces in the scalar potential.
Such terms could definitely still be included into our analysis without having to
modify Eq.~\eqref{eq:infobsdN}.}


In order to obtain predictions for $A_s$, $n_s$ and $f_{\textrm{NL}}^{\textrm{local}}$
which can be compared with observations, all quantities on the right-hand sides of the
relations in Eq.~\eqref{eq:infobsdN} need to be evaluated at $N = N_*$.
As for $N_a$ and $N_{ab}$,
the traditional way to do this, followed by many authors in the literature, is
to directly calculate $N$ as function on field space by solving the equations of
motion for the scalar fields $\phi^c$ and then to take the partial derivatives
of the such obtained expression for $N$.
This `brute force' approach is, however, prone to numerical imprecisions and in particular not
suited for comparing results from different authors.
Every author has to come up with his own
numerical procedure to compute $N$ and its derivatives, which impedes the
comparability of independent studies.
By contrast, the backward method by Yokoyama et al.\ is an
elegant and standardizable means of computing the derivatives $N_a$ and $N_{ab}$
directly as the solutions of simple first-order differential equations,
rendering the intermediate step of calculating the function $N$ first obsolete.
Let us now outline how we adapt this method to the scenario of hybrid inflation
in the complex plane.


It is convenient to divide the evolution of the
inflaton field in field space into three stages:
(i) the phase of slow-roll inflation at early times, (ii) the phase of fast-roll inflation
shortly before the instability in the scalar potential is reached, and
(iii) preheating in the course of the waterfall transition at the end of inflation.
In order to quantify the time at which the transition between the slow-roll and the fast-roll
stages takes place,
we generalize the slow-roll parameters $\epsilon$ and $\eta$
in Eqs.~\eqref{epsilon-0} and \eqref{eta-0} to the case of
multi-field inflation~\cite{Sasaki:1995aw},
\begin{align}
\epsilon_{\textrm{tot}} = \epsilon^a \epsilon_a \equiv \epsilon \,, \quad
\eta_{\textrm{tot}} = \left(\eta^{ab}\eta_{ab}\right)^{1/2} \sim \left|\eta\right| \,, \quad
\epsilon_a = \frac{M_{\textrm{Pl}}}{\sqrt{2}} \frac{V_a}{V}\,, \quad
\eta_{ab} = M_{\textrm{Pl}}^2 \frac{V_{ab}}{V} \,.
\label{eq:epsetatot}
\end{align}
The physical difference between these two sets of slow-roll parameters is the 
following:
While $\epsilon$ and $\eta$ parametrize the gradient and the curvature of
the scalar potential in the direction of the trajectory, cf.\ Eq.~\eqref{eq:epsetatraj}
in App.~\ref{app:index}, $\epsilon_{\textrm{tot}}$ and $\eta_{\textrm{tot}}$
quantify the total gradient and the total curvature of the scalar potential
at the momentary location of the inflaton.
Since in the slow-roll approximation the inflaton happens to roll in the direction
of the potential gradient, $\epsilon$ coincides with $\epsilon_{\textrm{tot}}$.
The parameter $\eta_{\textrm{tot}}$, however, is proportional to the
Frobenius norm of the Hessian matrix of the scalar potential,
$\eta_{\textrm{tot}} \propto \left\|V_{ab}\right\|$, and thus receives
contributions from the directions in field space perpendicular to
the trajectory, which are not contained in $\eta$.
In this sense, $\eta$ is only a good approximation to $\eta_{\textrm{tot}}$,
as long as the contributions from isocurvature perturbations to $\delta N$ are
negligible, i.e.\ as long as the inflationary dynamics in the complex plane
are effectively very similar to the dynamics of ordinary single-field inflation.
Slow-roll inflation is now characterized by both generalized slow-roll
parameters being at most of $\mathcal{O}(10^{-1})$.
As $\epsilon_{\textrm{tot}} \ll \eta_{\textrm{tot}}$ for all times during inflation,
the end of slow-roll inflation is marked by the time when
$\eta_{\textrm{tot}} = \eta_{\textrm{tot}}^0 \equiv 10^{-1/2}$.
The radial inflaton component at this time, $\varphi_\eta$, can be readily
estimated making use of the second derivative of the one-loop potential
in the limit of a large field excursion during inflation,
$\partial_x^2 f \simeq -4/x^2$.
To good approximation, we have%
\footnote{In principle, the linear term in the scalar potential induces
a slight dependence of $\eta_{\textrm{tot}}$ on the phase $\theta$.
For all relevant gravitino masses, this dependence is however completely
negligible.
In our numerical analysis, we employ the exact expression for $\varphi_\eta$
evaluated at $\theta = \pi/2$ for definiteness.\smallskip}
\begin{align}
\varphi_\eta = \varphi\left(\eta_{\textrm{tot}} = \eta_{\textrm{tot}}^0\right) \simeq
\begin{cases}
\left(\eta_{\textrm{tot}}^0\right)^{-1/2}\lambda/(2\sqrt{2}\pi)
M_{\textrm{Pl}} & ; \:\: \lambda \gg 2\sqrt{2}\,\pi \left(\eta_{\textrm{tot}}^0\right)^{1/2}
\, v / M_{\textrm{Pl}} \\
v & ; \:\: \lambda \ll 2\sqrt{2}\,\pi \left(\eta_{\textrm{tot}}^0\right)^{1/2}
\, v / M_{\textrm{Pl}}
\end{cases} \,.
\end{align}


As long as $\varphi \geq \varphi_\eta$, the slow-roll approximation is valid
and the evolution of $\varphi$ and $\theta$ is governed by the slow-roll equations,
\begin{align}
\varphi'(N) = M_{\textrm{Pl}}^2 \, \frac{V_{,\varphi}}{V} \,, \quad
\theta'(N) = \left(\frac{M_{\textrm{Pl}}}{\varphi}\right)^2 \, \frac{V_{,\theta}}{V} \,.
\label{eq:phithetaSR}
\end{align}
In order to solve these equations, we specify boundary conditions for them
at the end of slow-roll inflation, $\varphi = \varphi_\eta$ and $\theta = \theta_f$,
where $\theta_f$ is nothing but the free parameter labeling the different possible
trajectories in field space, which we introduced in Sec.~\ref{subsec_real_overview}.
At this point, it is worth emphasizing that technically $\theta_f$ is not defined
as the inflaton phase at the onset
of the waterfall transition, but as the phase at the end of slow roll.
If we were to define $\theta_f$ as the inflaton phase at the end of fast roll, it would
no longer suffice to parametrize the set of inflationary trajectories; in addition
to $\theta_f$, one would also have to know the final inflaton velocity $\dot{\phi}$ in order to
fully characterize a particular trajectory.
For small values of $\lambda$, this distinction between the different possibilities
to define $\theta_f$ is of course irrelevant, since
$\varphi_\eta \simeq v$.
In the large-$\lambda$ regime, the inflaton phase might however drastically change during
the stage of fast roll, in which case it is important to precisely define what is meant
by $\theta_f$.


In Eq.~\eqref{eq:phithetaSR}, we have omitted the interaction between the inflaton and 
the waterfall field.
This reflects the fact that we assume the waterfall field to be stabilized at its origin
throughout the entire inflationary phase.
Of course, unknown Planck-scale physics could result in the waterfall field
having a large initial field value and/or a large initial velocity.
But as long as we focus on the field dynamics around the GUT scale, 
it is natural to assume that the waterfall field has rolled down to its origin
before the onset of the last $N_*$ $e$-folds due to is inflaton-induced GUT-scale mass.
Guided by this expectation, we restrict ourselves to the study of
slow-roll inflation in the so-called `inflationary valley', in which the waterfall
field vanishes.
An extension of our analysis incorporating
arbitrary initial field values and velocities for the inflaton
as well as for the waterfall field is left for future work,%
\footnote{Neglecting the effect of spontaneous supersymmetry breaking on hybrid inflation,
i.e.\ working with $W_0 = 0$, arbitrary initial conditions for the inflaton-waterfall system
have been discussed in Refs.~\cite{Clesse:2008pf,Clesse:2009ur}, mainly in regard of the
question as to which initial conditions are capable of yielding
a sufficient number of $e$-folds during inflation.}
cf.\ also our discussion in Sec.~\ref{sec:inicon}.


For given values of $v$, $\lambda$, $m_{3/2}$ and $\theta_f$, the slow-roll
equations in Eq.~\eqref{eq:phithetaSR} have unique solutions, which describe
the time evolution of the homogeneous background fields $\varphi(N)$ and
$\theta(N)$.
At the same time, the slow-roll equations for the fluctuations $\delta \varphi(N)$
and $\delta\theta(N)$ together with the relation
$\delta N = N_a \delta \phi^a + \frac{1}{2!}N_{ab}\delta\phi^a\delta\phi^b +
\mathcal{O}\left(\delta\phi^3\right)$~\cite{Sasaki:1995aw,Sasaki:1998ug}
may be used to derive the following slow-roll transport equations for the
partial derivatives $N_a$~\cite{Yokoyama:2007uu} and $N_{ab}$~\cite{Seery:2012vj},
\begin{align}
\label{eq:NIJtransport}
N_a'(N) = - P_a^b N^b(N) \,, \quad
N_{ab}'(N) = - P_a^c N_{cb}(N) - P_b^c N_{ca}(N) - Q_{ab}^c N_c(N) \,.
\end{align}
Here, $P_a^b$ and $Q_{ab}^c$ are functions of $V$ and its
partial derivatives evaluated along the inflationary trajectory,
$P = P(N) = P(\varphi(N),\theta(N))$ and $Q = Q(N) = Q(\varphi(N),\theta(N))$,
\begin{align}
\begin{split}
P_a^b & \: = \eta_a^b - 2 \, \epsilon_a\, \epsilon^b \,, \\
Q_{ab}^c & \: = \frac{1}{M_{\textrm{Pl}}} \left[
M_{\textrm{Pl}}^3\frac{V_{ab}^c}{V} - \sqrt{2} \left(
\eta_a^c\epsilon_b + \eta_b^c\epsilon_a + \eta_{ab}\epsilon^c \right)
+ 4 \sqrt{2} \, \epsilon_a \, \epsilon_b \, \epsilon^c\right] \,.
\end{split}
\end{align}
According to Yokoyama et al.'s backward formalism, we specify the initial conditions
for the differential equations in Eq.~\eqref{eq:NIJtransport}
at the end of slow-roll inflation, when $\varphi=\varphi_\eta$.
In Cartesian coordinates, the hypersurface in field space on which
this condition is satisfied is given by
\begin{align}
\Sigma\left(\sigma,\tau\right) = 0 \,, \quad 
\Sigma\left(\sigma,\tau\right) = \varphi - \varphi_\eta = 
\left(\sigma^2 + \tau^2\right)^{1/2} - \varphi_\eta \,.
\end{align}
Often it is assumed that at the end of slow-roll inflation the universe has already
reached the adiabatic limit, which is equivalent to taking the energy density or
equivalently the Hubble rate on this hypersurface to be constant,
$\left.H\right|_{\Sigma = 0} = \textrm{const}$.
This renders Yokoyama et al.'s method insensitive to the further evolution of
the inflaton field at times after $\varphi = \varphi_\eta$.
As a consequence of this assumption, the conversion of isocurvature into
curvature perturbations during the final stages of inflation as well as
after inflation is neglected, which may however have important effects
in some cases such as, for instance, multi-brid
inflation~\cite{Sasaki:2008uc, Naruko:2008sq, Huang:2009vk}.
To remedy this shortcoming of the backward method in its original formulation, we explicitly
take into account the variation of the function $N$ on the $\Sigma = 0$ hypersurface.
Let us denote $\left.N\right|_{\Sigma = 0}$ by $N^{(0)}$, such that all
of the four following conditions are equivalent to each other,
\begin{align}
\eta_{\textrm{tot}} = \eta_{\textrm{tot}}^0 \,, \quad
\varphi = \varphi_\eta \,, \quad \Sigma\left(\sigma,\tau\right) = 0 \,, \quad
N\left(\sigma,\tau\right) = N^{(0)}\left(\sigma,\tau\right) \,.
\end{align}


After some algebra along the lines of Refs.~\cite{Yokoyama:2007uu,Yokoyama:2007dw},
we then find the initial values of $N_a$ and $N_{ab}$ at time $N = N^{(0)}$,
\begin{align}
\label{eq:NIJinicon}
N_a (N^{(0)}) = N^{(0)}_a + \frac{V}{\Sigma^b V_b} \frac{\Sigma_a}{M_{\textrm{Pl}}^2} \,, \quad
N_{ab} (N^{(0)}) = N^{(0)}_{ab} + \frac{V}{\Sigma^c V_c}
\frac{\Sigma_{ab} + \Xi_{ab}}{M_{\textrm{Pl}}^2} \,,
\end{align}
where all quantities on the right-hand sides of these two relations are to be evaluated at
$N = N^{(0)}$ and with $\Xi_{ab}$ being defined as
\begin{align}
\begin{split}
\Xi_{ab} = & \: \frac{1}{M_{\textrm{Pl}}} \Bigg[
\left(\frac{\Sigma^e\eta_{ef}\epsilon^f/\sqrt{2} +
M_{\textrm{Pl}}\,\epsilon^e\Sigma_{ef}\epsilon^f}{\Sigma^d \epsilon_d}
- \sqrt{2}\,\epsilon^d \epsilon_d \right)
\frac{\Sigma_a \Sigma_b}{2\,\Sigma^c \epsilon_c} \\
- & \: \left(\frac{1}{\sqrt{2}}\eta_b^d \Sigma_d + M_{\textrm{Pl}} \Sigma_b^d \epsilon_d
\right)\frac{\Sigma_a}{\Sigma^c \epsilon_c}
+ \sqrt{2}\, \Sigma_a \epsilon_b \Bigg] + \left(a\leftrightarrow b\right)\,.
 \end{split}
\end{align}
Our result for $N_a(N^{(0)})$ is identical to the one derived in Ref.~\cite{Mazumdar:2012jj},
which represents the first analysis properly taking care of the fact that $N^{(0)}$ is
in general actually not a constant.
By contrast, our expression for $N_{ab}(N^{(0)})$ has not been derived before.
It represents a straightforward generalization of the initial conditions for $N_{ab}$
stated in Refs.~\cite{Yokoyama:2007uu,Yokoyama:2007dw,Seery:2012vj,Anderson:2012em}
to the case of non-constant $N^{(0)}$.
As we will see shortly, the universe reaches the adiabatic limit in the course
of the preheating process.
This allows us to fix the origin of the $N$ time axis, $N = 0$, at some appropriate
time during preheating and distinguish between two contributions to the function $N^{(0)}$:
the number of $e$-folds elapsing during the final fast-roll stage of inflation, $N^{\textrm{FR}}$,
as well as the number of $e$-folds elapsing during preheating, $N^{\textrm{PH}}$,
\begin{align}
N^{(0)} = N^{\textrm{FR}} + N^{\textrm{PH}} \,. 
\end{align}
In this sense, our improved treatment of the initial conditions for $N_a$
and $N_{ab}$ now also includes the evolution of curvature and isocurvature
modes during fast-roll inflation as well as preheating. 


For a given slow-roll trajectory hitting the $\Sigma = 0$ hypersurface
for some inflaton phase $\theta_f$, we compute
$N^{\textrm{FR}}$ by solving the full equations of motion for the two inflaton
fields between the point $\phi = \varphi_\eta/\sqrt{2} \,e^{i\theta_f}$ and
the instability in the scalar potential.%
\footnote{In the case of critically large gravitino masses, not all trajectories hitting
the $\Sigma = 0$ hypersurface may also reach the instability.
Some trajectories may instead only approach a minimal $\varphi$ value,
$v < \varphi_{\textrm{min}} < \varphi_\eta$, and then `bend over'
in order to run towards a local minimum on the real axis located at
$\varphi \gg \varphi_\eta$, cf.\ Figs.~\ref{fig_thetas} and~\ref{fig_inicon}.
Such trajectories must then be discarded as they do not give rise to a possibility
for inflation to end.}
These equations are of second order and thus require us to specify the initial
velocities of the inflaton fields on the $\Sigma = 0$ hypersurface,
$\varphi'(N^{(0)})$ and $\theta'(N^{(0)})$.
The unique choice for these initial conditions ensuring consistency
with our treatment of the slow-roll regime obviously corresponds to
the expressions in Eq.~\eqref{eq:phithetaSR} evaluated at $N = N^{(0)}$
and it is precisely these velocities that we use in computing $N^{\textrm{FR}}$.
Nonetheless, we observe that our results for $N^{\textrm{FR}}$
are rather sensitive to the values we choose for $\varphi'(N^{(0)})$ and
$\theta'(N^{(0)})$.
This sensitivity becomes weaker once we lower $\eta_{\textrm{tot}}^0$,
the critical value of $\eta_{\textrm{tot}}$ dividing the fast-roll from the
slow-roll regime.
On the other hand, going to a smaller value of $\eta_{\textrm{tot}}^0$ also
reduces the portion of the inflationary evolution during which the transport
equations in Eq.~\eqref{eq:NIJtransport} are to be employed, the simplicity
of which motivated us to base our analysis on Yokoyama et al.'s backward
method in the first place.
It is therefore also under the impression of these observations that,
seeking a compromise between too large and too small $\eta_{\textrm{tot}}^0$,
we set $\eta_{\textrm{tot}}^0$ to an intermediate value such as
$10^{-1/2}$ rather than to $10^{-1}$ or $1$.


In order to compute $N^{\textrm{PH}}$, we solve the full second-order equations
of motion for the two inflaton fields $\sigma$ and $\tau$ as well as 
for the waterfall field $\chi$ from the onset of the phase transition up to the time
when the Hubble rate has dropped to some fraction $f$ of its initial value $H_0$
and the universe has reached the adiabatic limit.
Here, our numerical calculations indicate that a fraction of
$1-f \sim 1\,\% \cdots 10\,\%$ is enough, so as to obtain a sufficient
convergence of all inflationary trajectories.
Moreover, we note that, as $\chi \equiv 0$ is a classically stable solution, it is necessary to
introduce a small artificial shift $\Delta \chi$ of the field $\chi$ at the beginning
of preheating, so as to allow the waterfall field to reach the true vacuum.
The two parameters $f$ and $\Delta\chi$ are physically meaningless and just serve
as auxiliary quantities in our numerical analysis.
Their values must therefore be chosen such that our results for $N^{\textrm{PH}}$ 
remain invariant under small variation of these parameters.


Our procedure to determine  $N^{\textrm{PH}}$ captures of course only the classical
dynamics of the waterfall transition and misses potentially important
non-perturbative quantum effects.
A treatment of preheating at the quantum level however requires numerical
lattice simulations, which goes beyond the scope of this paper---and which is actually
also not necessary for our purposes.
As we are able to demonstrate numerically, $N^{\textrm{PH}}$ and its derivatives
never have any significant effect on our predictions for $A_s$, $n_s$, and
$f_{\textrm{NL}}^{\textrm{local}}$,
if solely computed based on classical dynamics.
Barring the unlikely possibility that quantum effects yield a substantial
enhancement of $N^{\textrm{PH}}$,
the evolution of the inflaton during the waterfall transition is thus completely
negligible from the viewpoint of inflationary physics.
Because of this, we will simply discard the contribution from preheating to the function
$N^{(0)}$ in the following and approximate it by its fast-roll contribution,
$N^{(0)} \approx N^{\textrm{FR}}$.
This also automatically entails that we do not need to consider the evolution of the
waterfall field $\chi$ any further.
As we focus on hybrid inflation in the inflationary valley, we can simply
set $\chi$ to $0$ at all times.


In conclusion, we summarize that, for given values of the parameters $v$, $\lambda$,
$m_{3/2}$ and $\theta_f$,
we have to perform four steps in order to compute our predictions for the observables
$A_s$, $n_s$ and $f_{\textrm{NL}}^{\textrm{local}}$.
(i)~First, we determine $N^{(0)} \approx N^{\textrm{FR}}$ by solving the
second-order equations of motion for $\sigma$ and $\tau$ from the
$\Sigma = 0$ hypersurface to the instability in the scalar potential.
Here, we specify the initial velocities of $\sigma$ and $\tau$ such that
they are consistent with Eq.~\eqref{eq:phithetaSR} evaluated on the $\Sigma = 0$ hypersurface.
(ii)~Subsequently, we solve the slow-roll equations for $\varphi$ and $\theta$
in Eq.~\eqref{eq:phithetaSR} starting on the $\Sigma = 0$ hypersurface and then going
backward in time up to the point when the CMB scales leave the Hubble horizon, i.e.,
in terms of the number of $e$-folds, from $N = N^{(0)}$ up to $N = N_*$.
(iii)~With the slow-roll solutions for $\varphi$ and $\theta$ at hand, we are able to solve the
transport equations for the partial derivatives $N_a$ and $N_{ab}$ in Eq.~\eqref{eq:NIJtransport}
in the interval $N^{(0)} \leq N \leq N_*$.
In doing so, we employ the initial conditions for $N_a$ and $N_{ab}$ at the time $N = N^{(0)}$ in
Eq.~\eqref{eq:NIJinicon}.
(iv)~The derivatives $N_a$ and $N_{ab}$ evaluated at time $N = N_*$ eventually allows us
to calculate the inflationary observables according to Eq.~\eqref{eq:infobsdN}.


\subsection{Phase dependence of the inflationary observables}
\label{subsec:parameterscan}


\subsubsection*{Inflationary trajectories in the complex plane}


As a first application of the above developed formalism, we are now able to
study the dynamics of the inflaton field in the complex plane.
In order to find all viable inflationary trajectories in the complex inflaton
field space, we impose two conditions:
(i) on the $\Sigma = 0$ hypersurface, the slope of the scalar potential in the
radial direction must be positive%
\footnote{This condition generalizes the requirement $\xi < 1$, which we imposed
in Sec.~\ref{sec_hilltop}, to the full two-field case.}
and (ii) the fast-roll motion during the last stages
of inflation must end on the instability in the scalar potential,
\begin{align}
V_{,\varphi}\left(\varphi_\eta,\theta_f\right) > 0 \,, \quad
\varphi_{\textrm{FR}}(N) \rightarrow v \,.
\label{eq:thetacond}
\end{align}
Together, these two requirements are sufficient to ensure that the inflaton does
not become trapped in the local minimum on the positive real axis.
For vanishing or small gravitino mass, they are always trivially fulfilled and
$\theta_f$ can take any value between $0$ and $\pi$.
However, once the slope of the linear term begins to exceed the slope of the one-loop potential,
the range of allowed $\theta_f$ values becomes more and more restricted, until eventually
only phases $\theta_f \sim \pi$ remain viable.
This effect is illustrated in Figs.~\ref{fig_trajectories_50} and
\ref{fig_trajectories_largem32}, which respectively show the set of possible
inflationary trajectories for an intermediate as well as for a large
value of the gravitino mass, while $v$ and $\lambda$ are set to
identical values in both plots.
Note that for Fig.~\ref{fig_trajectories_50} we have chosen the same
parameter values as for Fig.~\ref{fig_potential_1D}, which renders
this figure the continuation of Fig.~\ref{fig_potential_1D} from the real
axis to the complex plane.
Both Fig.~\ref{fig_trajectories_50} and Fig.~\ref{fig_trajectories_largem32}
demonstrate how the linear term distorts the rotational invariance of the scalar potential
by adding a constant slope in the direction of the real inflaton component $\sigma$.
As for Fig.~\ref{fig_trajectories_largem32}, the situation is however more extreme in consequence
of the enhanced gravitino mass compared to Fig.~\ref{fig_trajectories_50}.
Inflation on the positive real axis is, for instance, no longer possible for such a large
gravitino mass; instead, $\theta_f$ has at least to be slightly larger than $\pi/4$. 
Moreover, as an important consequence of our ability to determine all inflationary
trajectories, we are now in the position to identify the region in field space
which may provide viable initial conditions for inflation.
In fact, this region is nothing but the fraction of field space traversed by all inflationary
trajectories for $N > N_*$.
We will return to the issue of initial conditions for inflation in Sec.~\ref{sec:inicon}.


\subsubsection*{Inflationary observables for individual parameter points}


\begin{figure}
\centering
\includegraphics[width=.48\textwidth]{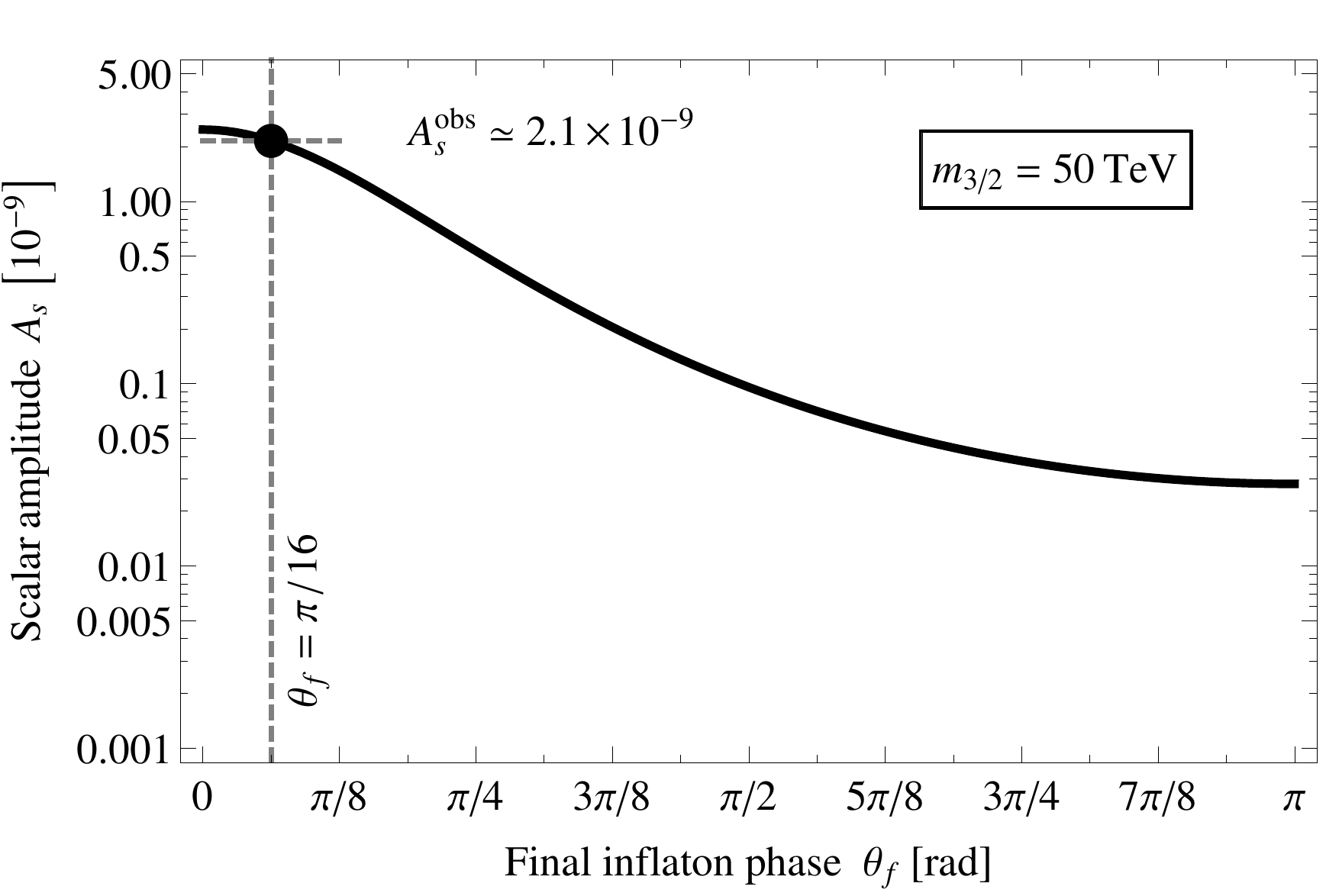}
\hspace{0.035\textwidth}
\includegraphics[width=.47\textwidth]{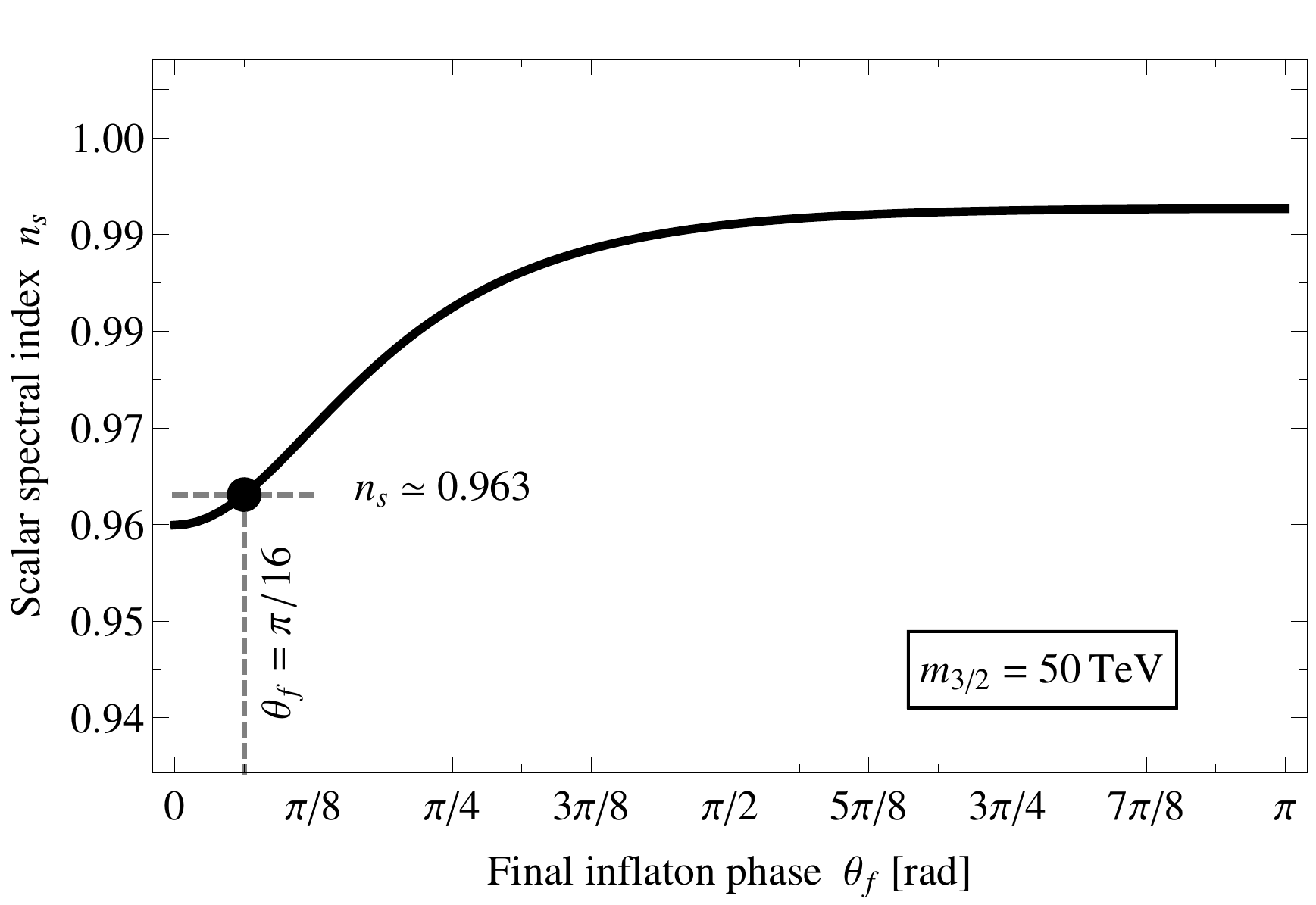}
\caption{Amplitude $A_s$ and spectral index $n_s$ of the primordial scalar 
power spectrum as functions of the inflaton phase at the end of slow-roll inflation,
$\theta_f$, for $v = 3.6 \times 10^{15}\,\textrm{GeV}$, $\lambda = 2.1 \times 10^{-3}$
and $m_{3/2} = 50\,\textrm{TeV}$.}
\label{fig_Asns}
\end{figure}

\begin{figure}
\centering
\includegraphics[width =.5\textwidth]{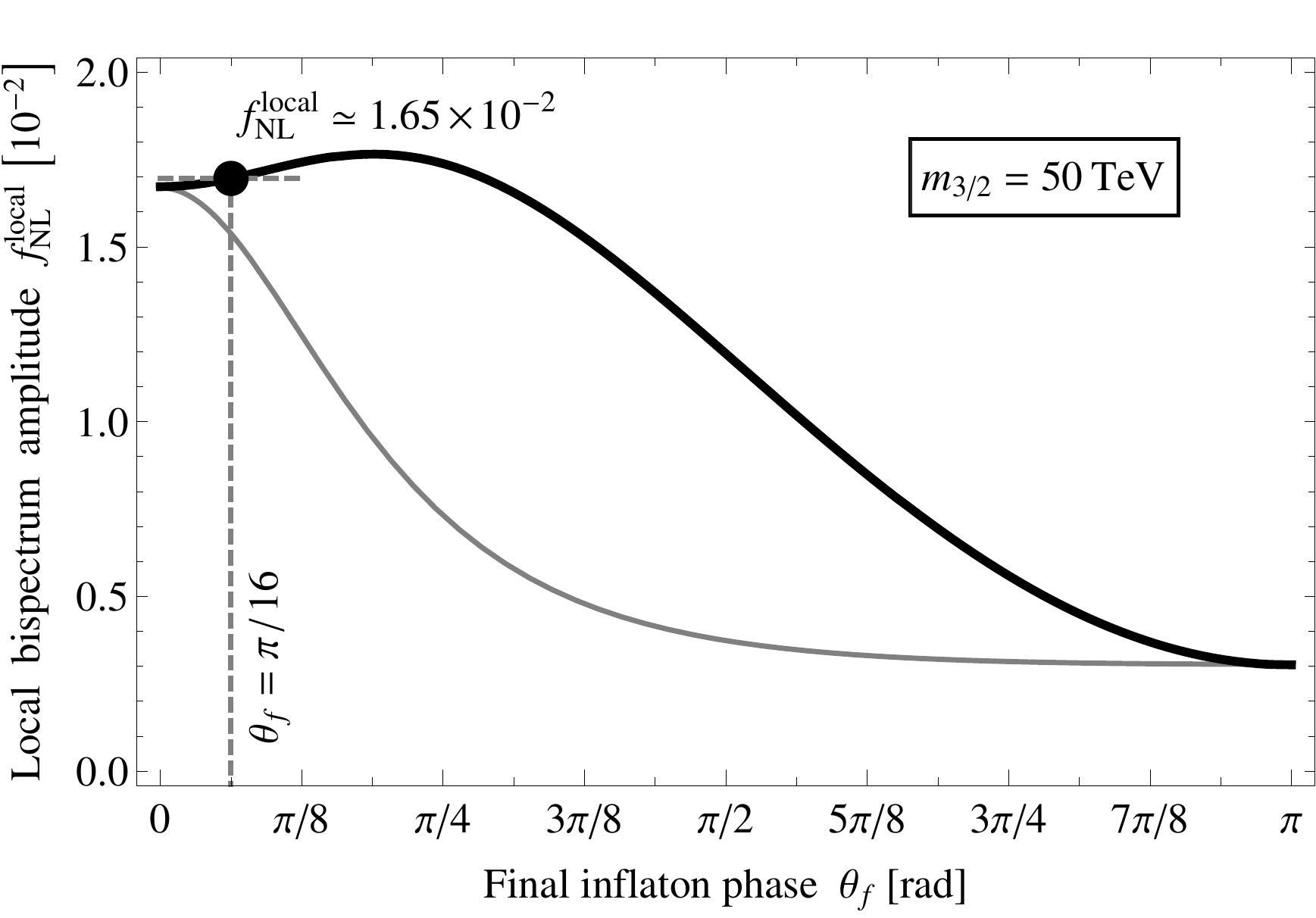}
\caption{Amplitude $f_{\textrm{NL}}^{\textrm{local}}$ of the local bispectrum (black curve)
and naive single-field slow-roll estimate for this quantity
(grey curve), cf.\ Eq.~\eqref{eq:fNLestimate},
as functions of the inflaton phase at the end of slow-roll inflation,
$\theta_f$, for $v = 3.6 \times 10^{15}\,\textrm{GeV}$, $\lambda = 2.1 \times 10^{-3}$
and $m_{3/2} = 50\,\textrm{TeV}$.}
\label{fig:fNL}
\end{figure}

In the next step, as we now know the trajectories along which the inflaton can
move across field space, we are able to compute the inflationary observables for
given values of $v$, $\lambda$ and $m_{3/2}$ and study their dependence on $\theta_f$.
In the limit of very small gravitino masses, when the slope of the inflaton potential
is dominated by the one-loop potential, this dependence becomes increasingly
negligible and $A_s$, $n_s$ and $f_{\textrm{NL}}^{\textrm{local}}$ as functions of
$\theta_f$ approach constant values.
On the other hand, for very large values of $m_{3/2}$, all viable trajectories start out
at a similar initial inflaton phase $\theta_i$ and run mostly in parallel to the real axis,
cf.\ Fig.~\ref{fig_trajectories_largem32}.
Due to this similarity between the different viable trajectories, the dependence of
the inflationary observables on $\theta_f$ is again rather weak for the most part.
There is however one crucial exception:
In the large-$m_{3/2}$ regime, $\theta_f$ is bounded from
below, $\theta_f \geq \theta_f^{\textrm{min}} > 0$ and once $\theta_f$ approaches
$\theta_f^{\textrm{min}}$, the scalar and the bispectrum amplitudes,
$A_s$ and $f_{\textrm{NL}}^{\textrm{local}}$, begin to rapidly increase.
This is due to the fact that for $\theta_f \gtrsim \theta_f^{\textrm{min}}$
the inflaton trajectory hits the instability in the scalar potential at a very 
shallow angle, so that initial isocurvature perturbations induce large shifts
$\delta N$ in $N$, and hence large curvature perturbations, at late times.
But at any rate, the most interesting case is the one of intermediate gravitino
masses, when the gradients of the one-loop potential and the linear term are of
comparable size and the inflationary observables strongly depend on $\theta_f$.
An example for such a situation is given in Figs.~\ref{fig_Asns} and
\ref{fig:fNL}, in which we show $A_s$, $n_s$ and $f_{\textrm{NL}}^{\textrm{local}}$
as functions of $\theta_f$ for the same parameter values that we also used
for Figs.~\ref{fig_potential_1D} and \ref{fig_trajectories_50}.
Now it becomes evident that for these parameter values and a final phase $\theta_f$
of $\pi/16$ the observed values for $A_s$ and $n_s$ can be nicely reproduced,
while $f_{\textrm{NL}}^{\textrm{local}}$ safely stays within the experimental bounds.


An important lesson which we learn from Figs.~\ref{fig_Asns} and \ref{fig:fNL}
is that the Lagrangian parameters, $v$, $\lambda$ and $m_{3/2}$, and hence the functional
form of the scalar potential do not fix the inflationary observables at all.
Under a variation of the inflationary trajectory, $A_s$, $n_s$ and
$f_{\textrm{NL}}^{\textrm{local}}$ vary over significant ranges, in which the observed
values are not singled out in any way.
We therefore conclude that the values for the inflationary observables realized
in our universe do not point to a particular Lagrangian, but rather seem to be 
a mere consequence of an arbitrary selection among different possible trajectories.
This is a very characteristic feature of hybrid inflation in the complex
plane, which distinguishes it from other popular inflation models.
In $R^2$~inflation~\cite{Starobinsky:1980te} or chaotic
inflation~\cite{Linde:1983gd}, for instance, the shape of the scalar potential
is the key player behind the predictions for the inflationary observables.
As we now see, the philosophical attitude in hybrid inflation is certainly a
different one:
Here, the main virtue of inflation are mainly its qualitative aspects---the fact that it
solves the initial conditions problems of big bang cosmology, explains the origin of
the primordial density perturbations and is consistent with a compelling model of
particle physics at very high energies.
Its quantitative outcome is the mere result of a selection process
that has no deeper meaning within the model itself.


\subsubsection*{Amplitude and spectral tilt of the scalar power spectrum}


In the third step of our numerical investigation, we perform a calculation
of the inflationary observables, as we just did it for one parameter point, for
all values of $v$, $\lambda$, $m_{3/2}$ and $\theta_f$ of interest.
In this scan of the parameter space, we shall cover the following parameter ranges,
\begin{align}
10^{14}\,\textrm{GeV} \leq v \leq 10^{16}\,\textrm{GeV} \,, \quad
10^{-5} \leq \lambda \leq 3 \times 10^{-2} \,, \quad
10\,\textrm{MeV} \leq m_{3/2} \leq 100\,\textrm{PeV} \,.
\end{align}
The ranges for $v$ and $\lambda$ are chosen such that on the one hand,
for values of $\lambda$ not much smaller than typical Standard Model Yukawa
couplings, the measured value of the scalar amplitude $A_s$ can be reproduced
and that on the other hand the bound on the cosmic string tension
in Eq.~\eqref{eq_cosmic_string_bound} is obeyed in most cases.
At the same time, the $m_{3/2}$ range covers all values of gravitino masses
which are commonly assumed in supersymmetric models of electroweak symmetry breaking.
As our results will confirm, the such defined parameter space contains all the
phenomenologically interesting parameter regimes for hybrid inflation.


\begin{figure}
\centering
\includegraphics[width = .45 \textwidth]{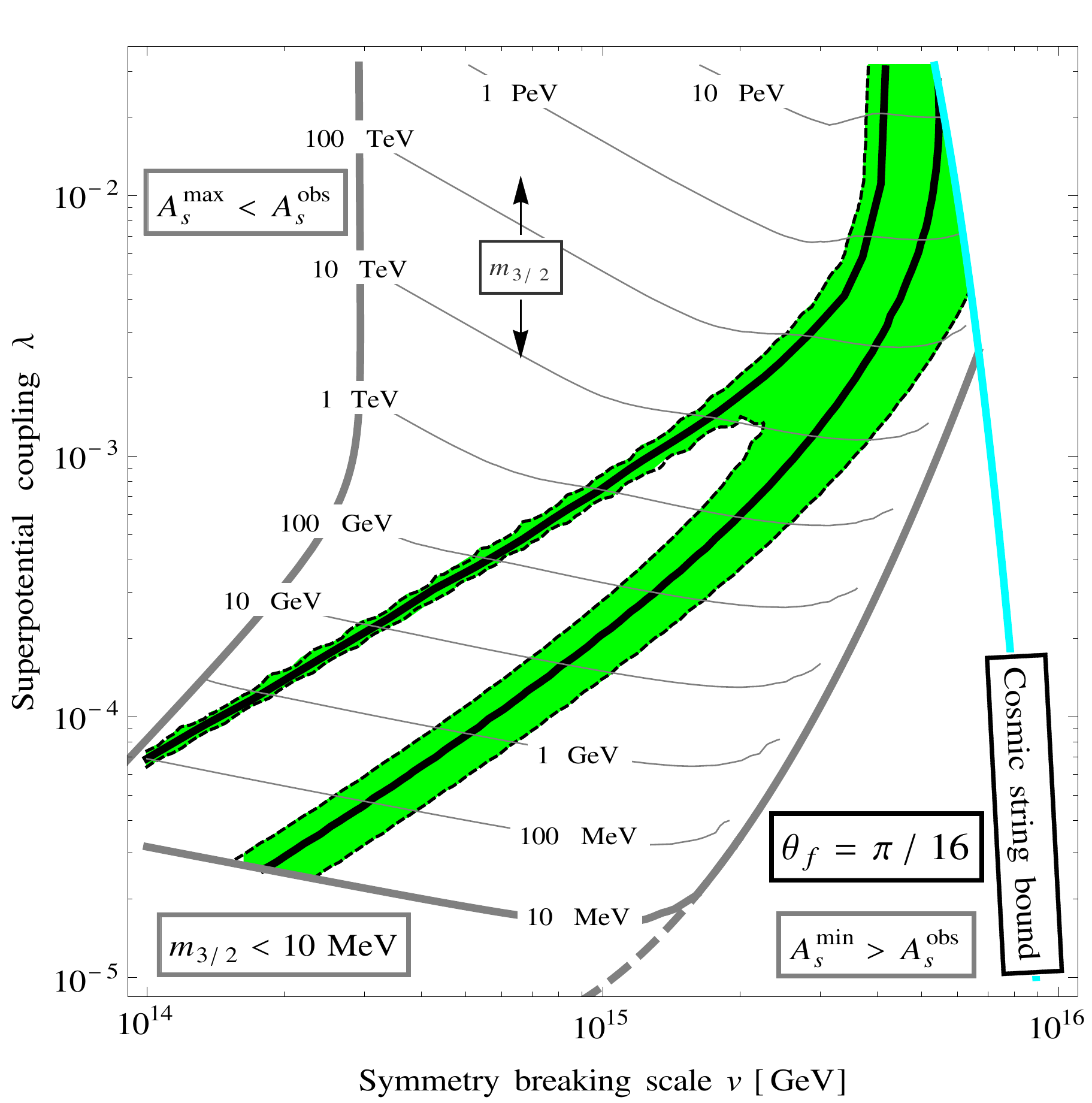}
\includegraphics[width = .45 \textwidth]{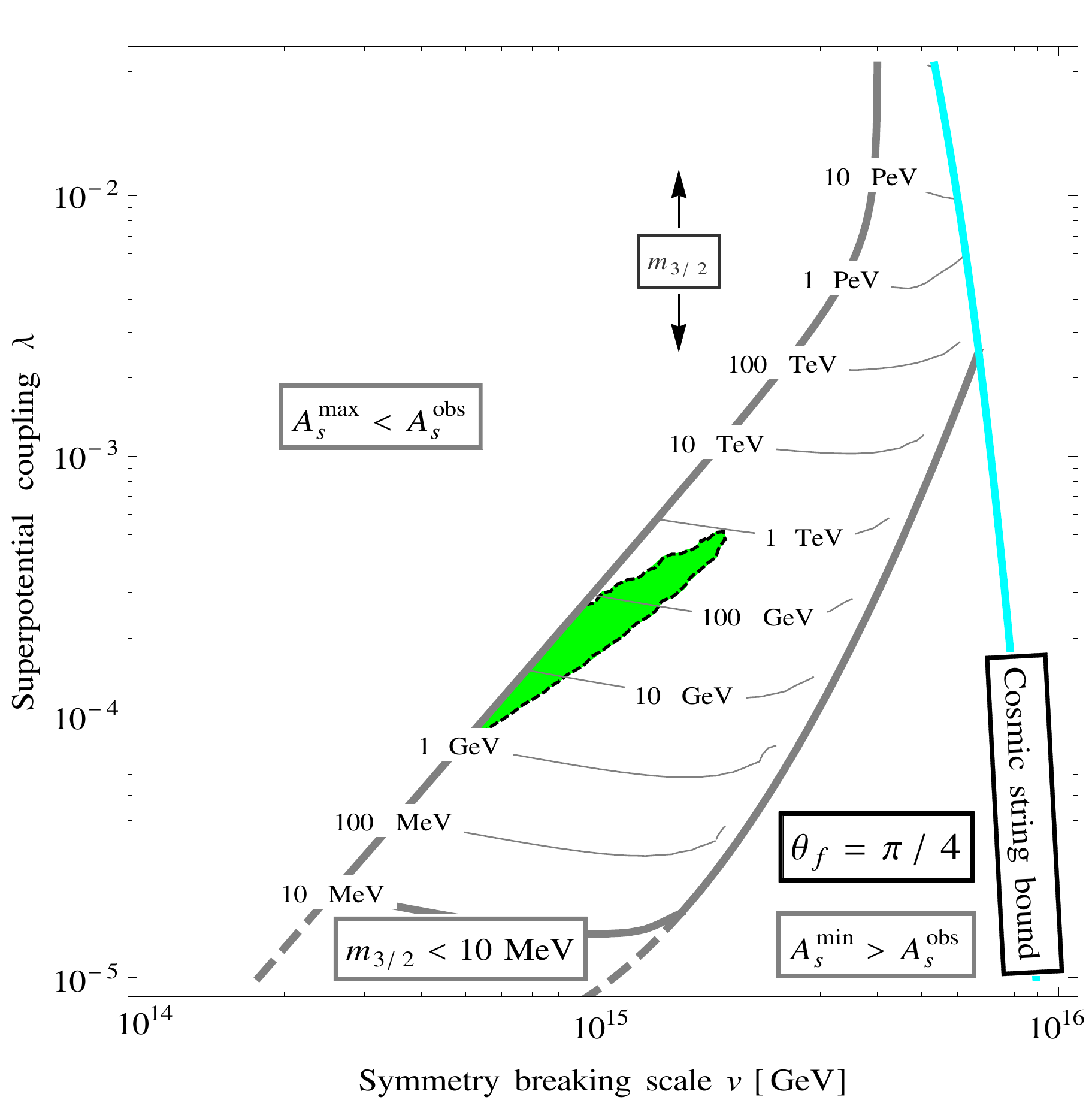}
\caption{Viable parameter space for hybrid inflation in the complex plane
for two different values of the final inflaton phase $\theta_f$.
Color code and labels as in the left panel of Fig.~\ref{fig_1field_results}.
In contrast to Fig.~\ref{fig_1field_results}, we now also find regions in parameter
space where our predictions for the scalar amplitude are always \textit{smaller}
than the observed value $A_s^{\textrm{obs}}$.
In these regions, it is impossible to increase $A_s$ by going to larger gravitino masses
as this would cause the inflaton to miss the instability in the scalar potential and reach
the local minimum on the positive real axis, cf.\ our discussion of Eq.~\eqref{eq:thetacond}.}
\label{fig_thetas}
\end{figure}


Let us first focus on $A_s$ and $n_s$, the two observables related to the scalar
power spectrum, before we then comment on $f_{\textrm{NL}}^{\textrm{local}}$, the
amplitude of the local bispectrum.
Both $A_s$ and $n_s$ depend on all three Lagrangian parameters $v$, $\lambda$ and $m_{3/2}$
as well as on the choice among the different inflationary trajectories, which we label by
$\theta_f$.
As $A_s$ has been measured very precisely by the various CMB satellite experiments,
cf.\ Eq.~\eqref{eq_planck_results}, we are able to eliminate one free parameter, say,
the gravitino mass, by requiring that our prediction for $A_s$ must always coincide
with the observed best-fit value for the scalar amplitude,
$A_s^{\textrm{obs}} = 2.18 \times 10^{-9}$,
\begin{align}
A_s \left(v,\lambda,m_{3/2},\theta_f\right) = A_s^{\textrm{obs}} \quad \Rightarrow \quad
m_{3/2} = m_{3/2} \left(v,\lambda,\theta_f\right) \,.
\label{eq:Asnorm}
\end{align}
This renders all remaining inflationary observables functions of $v$, $\lambda$ and $\theta_f$
only.
Next, we demand that our prediction for $n_s$ must fall into
the $2\,\sigma$ range around the measured best-fit value for the scalar spectral index,
$n_s^{\textrm{obs}} = 0.963$,
\begin{align}
n_s^{\textrm{obs}} - 2 \Delta n_s \leq n_s(v,\lambda,\theta_f)
\leq n_s^{\textrm{obs}} + 2 \Delta n_s \,, \quad \Delta n_s = 0.08 \,,
\label{eq:nscond}
\end{align}
which provides us with $95\,\%$\,C.L.\ exclusion contours in the $(v,\lambda)$
plane for every individual value of $\theta_f$.
As examples of such exclusion contours, we show the viable region in the
$(v,\lambda)$ plane for $\theta_f = \pi/16$ and $\pi/4$ in Fig.~\ref{fig_thetas}.
These two plots generalize the left panel of Fig.~\ref{fig_1field_results}
from hybrid inflation on the real axis to the full two-field scenario.


By comparing our parameter constraints in the two-field case with
the results obtained in Sec.~\ref{sec_hilltop}, we are able to identify the similarities
and differences between hybrid inflation on the real axis and hybrid inflation in
the complex plane.
These observations belong to the most important results of our analysis.
First of all, we note that for small but nonzero $\theta_f$ and fixed $v$,
we always find two pairs of $(\lambda,m_{3/2})$ values such that $A_s$ and $n_s$ are
successfully reproduced.
In our plots of the $(v,\lambda)$ plane, this is reflected by the appearance of
\textit{two} bands of viable parameter values stretching from small $v$ and small
$\lambda$ to large $v$ and large $\lambda$.
The lower one of these two bands directly derives from the band
in Fig.~\ref{fig_1field_results}.
The second band is however completely new, representing a genuine feature of hybrid
inflation in the complex plane.
We will qualitatively explain the origin of this second band in our semi-analytical discussion
in Sec.~\ref{analytic_reconst}.
For now, let us focus on its behaviour as we vary the inflaton phase $\theta_f$
and its physical implications.
In the limit $\theta_f \rightarrow 0$, the upper and the lower branch of the $95\,\%$\,C.L.\
region in the $(v,\lambda)$ plane move into opposite directions.
While the lower branch approaches the $95\,\%$\,C.L.\ region which we identified in
the single-field case, the upper branch moves to smaller values of $v$ and larger
values of $\lambda$.
In this process, it also becomes increasingly thinner.
On the other hand, as $\theta_f$ is further increased, the two bands move
closer together, until they fully merge and eventually shrink away to smaller
values of $v$ and $\lambda$, cf.\ right panel of Fig.~\ref{fig_thetas}.
Remarkably enough, for small $\theta_f$ and fixed $\lambda$, the upper branch of parameter
solutions makes smaller values of the symmetry breaking scale $v$ accessible.
These points in parameter space are hence further away from the cosmic string bound
and alleviate the tension between the predictions of hybrid inflation and 
the non-observation of cosmic strings.
In particular, if future observations should lead to an even more stringent bound on $G\mu$
that, for fixed value of $\lambda$, rules out symmetry breaking scales $v$ up to some
certain value, this $\lambda$ value might still be viable in combination with a smaller
value of $v$ and nonzero $\theta_f$.


As a second observation, we note that in the two-field case certain parts of the
$(v,\lambda)$ plane are excluded because they do not allow to reproduce the
spectral amplitude without violating the second condition in Eq.~\eqref{eq:thetacond}.
For small $v$ and large $\lambda$ values as well as gravitino masses as we would expect
them from the single-field case, the inflationary trajectories still hit the
$\Sigma = 0$ hypersurface. 
But during the fast-roll stage towards the end of inflation, they roll off the
hill-top in the scalar potential into the wrong direction, such that the inflaton
becomes eventually trapped in the false vacuum on the positive real axis.%
\footnote{We thank A.~Westphal for pointing out that, technically speaking, we have
to ensure that the inflaton never comes closer
to the ridge in the scalar potential than $H_0/(2\pi)$.
Otherwise, quantum fluctuations may let the inflaton tunnel to the other side
of the hill-top causing it to roll down towards the false vacuum.
The natural scale of the inflaton excursion, $v$, is however much larger
than the inflationary Hubble scale, $v \gg H_0$.
For all practical purposes, it is hence sufficient to make sure that the inflaton never
actually reaches the ridge.}
In the case of hybrid inflation on the real axis, such a behaviour of course never
occurs.
Here, once the inflaton starts out its journey on the correct side of the hill-top,
it will also always hit the instability.


Finally, we observe that for $\theta_f \gtrsim \pi/4$ the
scalar spectral index always comes out too large.
This is due to the fact that for such large values of $\theta_f$
the inflationary trajectories begin to look more and more similar
to the trajectory on the negative real axis.
Our inability to reproduce the scalar spectral index for
$\theta_f \gtrsim \pi/4$ is hence nothing but the original problem
of a too large value for $n_s$ in the case of standard hybrid inflation. 
As pointed out in Ref.~\cite{Nakayama:2010xf}, a viable possibility
to reduce the scalar spectral index on the negative real axis is
to resort to a non-canonical K\"ahler potential.
Therefore, it would be interesting to investigate by how much our upper bound
on $\theta_f$ might be relaxed in dependence of non-minimal couplings
in the K\"ahler potential.
Such a study is however beyond the scope of this paper and left for future work.
For the time being, we merely conclude that, while hybrid inflation in the complex
plane is not exclusively limited to the hill-top regime on the positive real axis,
it is still certainly necessary that the inflationary trajectories pass close to
this regime.


\begin{figure}
\centering
\includegraphics[width = 0.5 \textwidth]{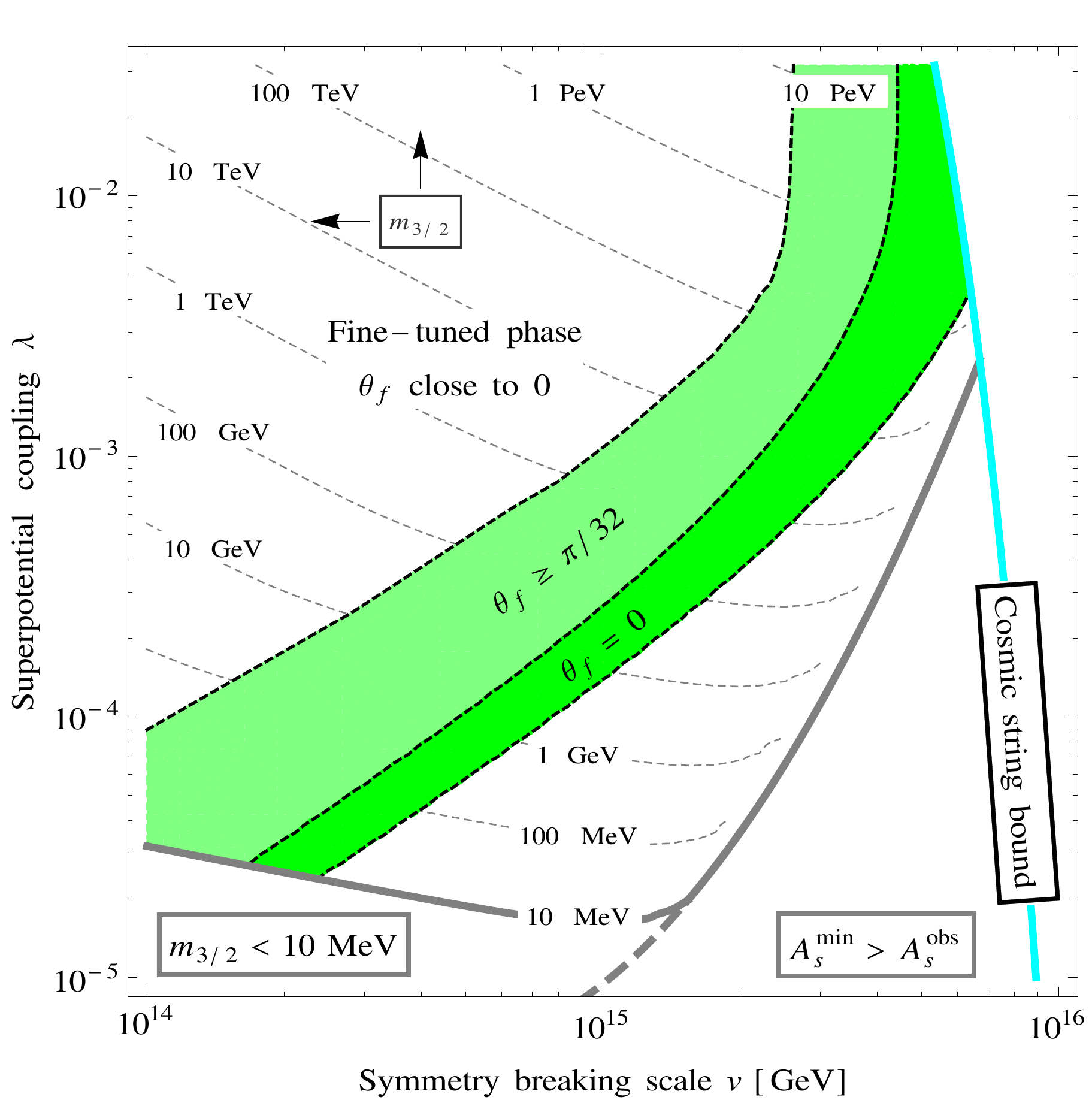}
\caption{Viable parameter space for hybrid inflation in the complex
plane for arbitrary values of the final inflaton phase $\theta_f$.
Color code and labels as in the left panel of Fig.~\ref{fig_1field_results}
as well as in Fig.~\ref{fig_thetas}.
The contour lines for the gravitino mass correspond to the case $\theta_f = 0$.
For larger  $\theta_f$, the gravitino mass at a fixed point in the $(v,\lambda)$ plane
slightly increases: at most by roughly half an order of magnitude,
but typically significantly less.}
\label{fig_overlay}
\end{figure}


In order to summarize our constraints on the model parameters of hybrid inflation imposed
by the inflationary observables as well as the cosmic string bound, we marginalize our results
over the inflaton phase $\theta_f$.
The result of this step is depicted in Fig.~\ref{fig_overlay}, in which we show
the union of all of our $95\,\%$\,C.L.\ regions.
The dark-green band marks the allowed parameter region in the case of
single-field hybrid inflation in the hill-top regime,
while the light-green region becomes available as soon as we allow for nonzero
$\theta_f$.
This increase in the totally accessible parameter region demonstrates
that the Lagrangian parameters of hybrid inflation are in fact not as tightly
constrained as has previously been thought.
Instead, it is possible to reproduce the inflationary observables
in a large fraction of parameter space, which certainly boosts the vitality
of the entire model.
Finally, we remark that, also in the white region on the
top left, it is in principle possible to obtain a viable value for the spectral index.
This merely requires a fine-tuning of $\theta_f$ very close to zero, so as to
push the upper branch of parameter solutions to ever smaller values of $\lambda$.
However, since one of the basic motivations for our study is to show
how the fine-tuning problem of single-field hybrid inflation in the hill-top
regime can be avoided or relaxed, we shall not discuss this possibility in more detail.


\subsubsection*{Primordial non-Gaussianities}


In the fourth and last step of our numerical analysis, we study our predictions for
the amplitude of the local bispectrum, $f_{\textrm{NL}}^{\textrm{local}}$.
As is well-known, $f_{\textrm{NL}}^{\textrm{local}}$ is suppressed by the slow-roll
parameters $\epsilon$ and $\eta$ in the case of single-field slow-roll
inflation~\cite{Maldacena:2002vr}, 
\begin{align}
f_{\textrm{NL}}^{\textrm{local}} = \frac{5}{12} \left(1 - n_s\right) = 
\frac{5}{6} \left(\eta - 3\epsilon\right)\,,
\label{eq:fNLestimate}
\end{align}
where $\epsilon$ and $\eta$ are to be evaluated at $N = N_*$.
Hence, as far as hybrid inflation on the real axis is concerned,
we expect $f_{\textrm{NL}}^{\textrm{local}}$ not
to be larger than of $\mathcal{O}(10^{-2})$, which is two orders of
magnitude below the sensitivity of the PLANCK satellite, cf.\ Eq.~\eqref{eq_planck_results}.
And indeed, requiring single-field hybrid inflation in the hill-top regime
to correctly reproduce the measured values of $A_s$ and $n_s$, we always find
an amplitude of the local bispectrum of $f_{\textrm{NL}}^{\textrm{local}} \sim 0.015$.


How does this situation now change in the full two-field case?
In answering this question, we shall restrict ourselves to values for $v$,
$\lambda$ and $m_{3/2}$, which already yield the correct values of $A_s$ and
$n_s$ for one specific final inflaton phase $\theta_f$.
This is to say that we will only investigate
our predictions for $f_{\textrm{NL}}^{\textrm{local}}$
in the respective $95\,\%$\,C.L.\ regions in the $(v,\lambda)$ plane.
In the lower branches of those $95\,\%$\,C.L.\ regions, our two-field model effectively behaves
like a single-field model, such that our predictions for $A_s$ and $n_s$ are well described
by the analytical expressions derived in Sec.~\ref{sec_hilltop}.
As expected, this is also reflected in our predictions for $f_{\textrm{NL}}^{\textrm{local}}$,
which are slow-roll suppressed to most extent in these regions of parameter space.%
\footnote{An exception are regions corresponding to very small gravitino
masses, $m_{3/2} \lesssim 1\,\textrm{GeV}$, in combination with
large final inflaton phases, $\theta_f \gtrsim 3\pi/32$.
Here, $f_{\textrm{NL}}^{\textrm{local}}$ can become roughly as large as $0.5$.
The origin of such large non-Gaussianities is the same as in the upper branches
of the $95\,\%$\,C.L.\ regions, cf.\ further below.}
In Fig.~\ref{fig:fNL}, we plot for instance $f_{\textrm{NL}}^{\textrm{local}}$ as a function
of $\theta_f$ for a parameter point in the lower
band of the  $95\,\%$\,C.L.\ region corresponding to
$\theta_f = \pi/16$ and it is clearly seen that $f_{\textrm{NL}}^{\textrm{local}}$ never
exceeds values of $\mathcal{O}(10^{-2})$.
Furthermore, Fig.~\ref{fig:fNL} illustrates that, in the limits $\theta_f\rightarrow 0$ and
$\theta_f \rightarrow \pi$, our numerical multi-field result nicely approaches
the single-field expectation according to Eq.~\eqref{eq:fNLestimate}.
For $\theta_f$ values in between $0$ and $\pi$, our multi-field prediction
is by contrast slightly larger than our naive single-field estimate; but the
deviation is always at most of $\mathcal{O}(1)$.
This slight enhancement of $f_{\textrm{NL}}^{\textrm{local}}$ is a direct
consequence of the inherent multi-field nature of hybrid inflation and
indicates that, for hybrid inflation off the real axis, effects such as the
inhomogeneous end of inflation or the late-time conversion of isocurvature modes
to curvature modes become important~\cite{Byrnes:2010em}.
Nonetheless, it is safe to conclude that in most of the lower bands of our
$95\,\%$\,C.L.\ regions also the generation of non-Gaussianities is,
up to  $\mathcal{O}(1)$ corrections,
well explained in an effective single-field picture.
Note that this is in contrast to the situation in multi-brid inflation,
where the simple single-field description breaks down and genuine multi-field
dynamics are responsible for a sizable value of
$f_{\textrm{NL}}^{\textrm{local}}$~\cite{Sasaki:2008uc, Naruko:2008sq, Huang:2009vk}.


The upper branches of our $95\,\%$\,C.L.\ regions are much closer to those
parts of parameter space in which $A_s$ cannot be reproduced
without violating the second condition in Eq.~\eqref{eq:thetacond}.
The trajectories corresponding to these parameter points are hence
much more strongly bent than the trajectories corresponding to the parameter
points in the lower branches of the $95\,\%$\,C.L.\ regions.
In these corners of parameter space, the multi-field character of hybrid inflation
hence comes much more into effect, resulting in the generation of quite sizable
non-Gaussianities up to values as large as $f_{\textrm{NL}}^{\textrm{local}} \sim 0.5$.
We are able to substantiate this qualitative understanding by studying the time
evolution of $f_{\textrm{NL}}^{\textrm{local}}$ in the course of inflation.
Generally speaking, if, in Yokoyama et al.'s backward method,
one does not fix the number of $e$-folds during inflation, $N_*$,
at $N_* = 50$, but allows it to freely vary, $N_* \rightarrow N \geq N^{(0)}$,
any given inflationary observable $\mathcal{O}$ turns into a time-dependent quantity
$\mathcal{O}(N)$.
Applying this procedure to $f_{\textrm{NL}}^{\textrm{local}}$ reveals that the final value
of the local bispectrum amplitude, $f_{\textrm{NL}}^{\textrm{local}}(N_*)$, is mostly
determined at late times when $N \gtrsim N^{(0)}$.
At earlier times, $N \lesssim N_*$, the variation of $f_{\textrm{NL}}^{\textrm{local}}(N)$
is by contrast rather weak.
This confirms our intuition that the large non-Gaussianities encountered in
the upper branches of our $95\,\%$\,C.L.\ regions mainly originate from the 
strong curvature of the inflationary trajectories towards the end of
inflation as well as from the conversion of isocurvature to
curvature modes associated with this curvature.
At the same, a similar analysis for $A_s$ indicates that
the final value of the scalar amplitude, $A_s(N_*)$ is in most cases already
fixed at early times, $N \lesssim N_*$.
In summary, we therefore conclude that, in hybrid inflation in the complex plane,
the scalar power spectrum is predominantly sourced by adiabatic perturbations around
the time when the CMB pivot scale exits the Hubble horizon, $N \sim N_*$,
while $f_{\textrm{NL}}^{\textrm{local}}$ is mainly generated by isocurvature
perturbations at the time when the inflationary trajectory bends around
at the end or after slow-roll inflation.
At the level of the observables related to the scalar power spectrum,
we are hence always free to work in an effective single-field approximation;
at the level of the local bispectrum, this approximation however breaks down
in certain parts of the parameter space.


\begin{figure}
\centering
\includegraphics[width =0.51\textwidth]{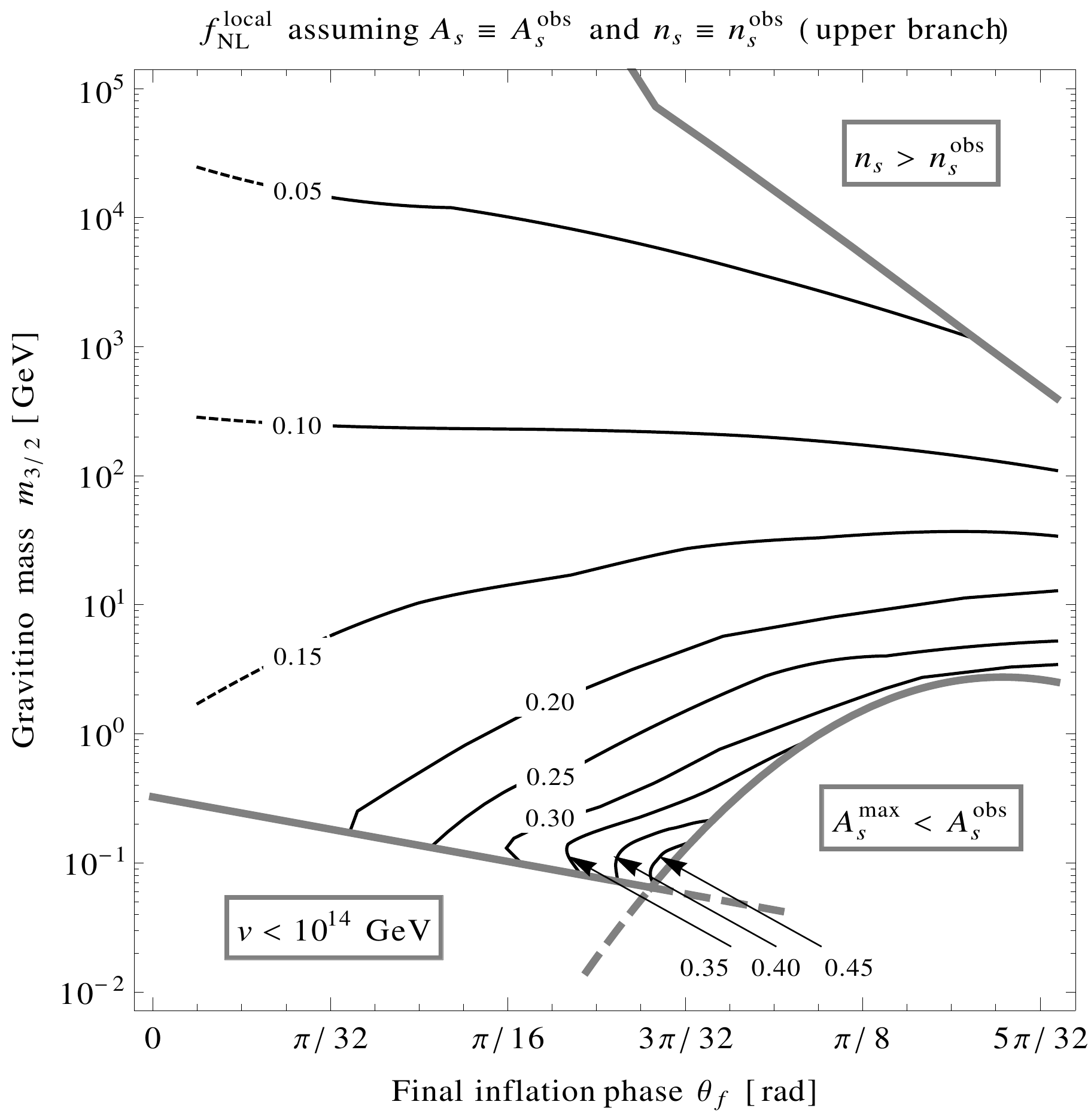}
\caption{
Prediction for the amplitude of the local bispectrum, $f_{\textrm{NL}}^{\textrm{local}}$,
in the upper branches of our $95\,\%$\,C.L.\ regions as a function of $\theta_f$ and $m_{3/2}$.
At any point in the $(\theta_f,m_{3/2})$ plane, the parameters $v$ and $\lambda$ are fixed such
that $A_s = A_s^{\textrm{obs}} \equiv 2.18 \times 10^{-9}$ and
$n_s = n_s^{\textrm{obs}} \equiv 0.963$.}
\label{fig_fNLscan}
\end{figure}


Finally, before concluding this section, we summarize our results for 
$f_{\textrm{NL}}^{\textrm{local}}$ in the upper branches of our $95\,\%$\,C.L.\ regions
in Fig.~\ref{fig_fNLscan}.
In this figure, we display our predictions for $f_{\textrm{NL}}^{\textrm{local}}$ as a
function of $\theta_f$ and $m_{3/2}$, with the parameters $v$ and $\lambda$ always
chosen such $A_s = A_s^{\textrm{obs}}$ and $n_s = n_s^{\textrm{obs}}$.
Remarkably enough, $f_{\textrm{NL}}^{\textrm{local}}$ can become as large as
roughly $0.5$.
At the same time, the $2\,\sigma$ uncertainty in the measured value of the scalar
spectral index results in an uncertainty in these predictions of at most a factor of $2$.
Together, these two observations imply a conservative upper bound on the
amplitude of the local bispectrum,
\begin{align}
f_{\textrm{NL}}^{\textrm{local}} \lesssim 1 \,.
\end{align}
This bound provides an interesting means to falsify hybrid inflation.
If future CMB experiments should reach a better sensitivity to primordial non-Gaussianities
and an $f_{\textrm{NL}}^{\textrm{local}}$ value larger than~$1$ should be measured,
hybrid inflation would be in serious trouble.


\subsection{Analytical reconstruction of the numerical results\label{analytic_reconst}}


Having presented the outcome of our numerical analysis in the previous section,
we now attempt to partly reconstruct our results by means of (semi-)analytical approximations.
Here, we will focus on the constraints on $v$ and $\lambda$ which we obtained
by requiring that inflation must yield the correct values for $A_s$ and $n_s$.
As for the non-Gaussianity parameter $f_{\textrm{NL}}^{\textrm{local}}$, we merely
remark that, as long as it is possible to work in an effective single-field picture,
$f_{\textrm{NL}}^{\textrm{local}}$ can be, up to $\mathcal{O}(1)$ corrections, well
approximated by the naive single-field expression in Eq.~\eqref{eq:fNLestimate}.
Once the multi-field dynamics of hybrid inflation come into effect, the only
possibility we see to determine $f_{\textrm{NL}}^{\textrm{local}}$ is a full-fledged
numerical analysis as we perform it in this paper.

\begin{figure}[t]
\centering
\includegraphics[width = 0.5\textwidth]{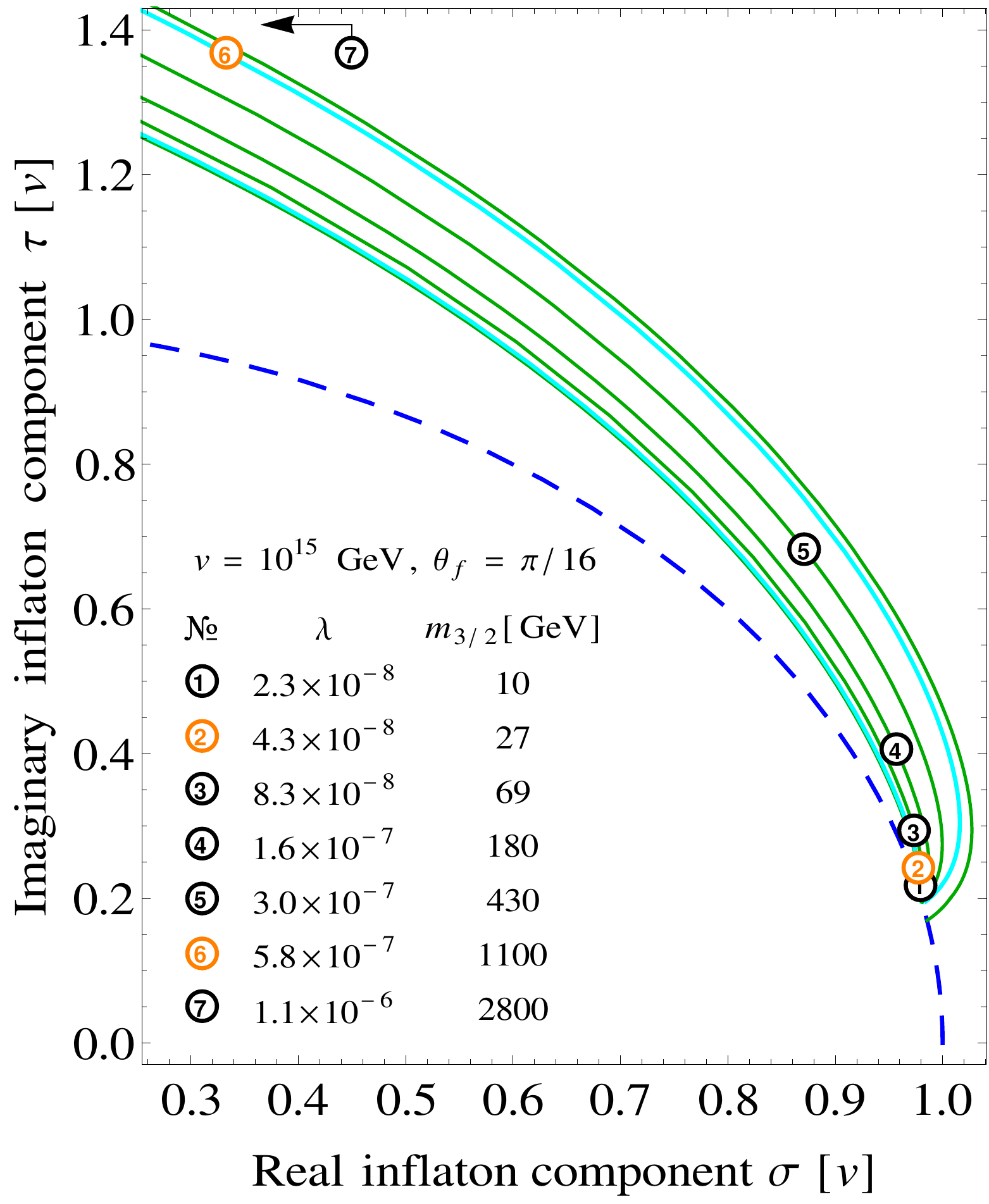}
\caption{Various possible inflationary trajectories for fixed symmetry breaking scale,
$v = 10^{15}\,\textrm{GeV}$, and final inflaton phase, $\theta_f = \pi/16$.
Recall that $\theta_f$ is defined on the $\Sigma = 0$ hypersurface,
where $\eta_{\textrm{tot}} = \eta_{\textrm{tot}}^0 \equiv 10^{-1/2}$,
not along the instability in scalar potential (blue dashed curve).
For each trajectory, $\lambda$ and $m_{3/2}$ are adjusted such that
$A_s = A_s^{\textrm{obs}} \equiv 2.18 \times 10^{-9}$.
The parameter points corresponding to trajectories \textnumero~2 and \textnumero~6
are located in the lower and the upper band of the $95\,\%$\,C.L.\
region for $\theta_f = \pi/16$, respectively, cf.\ the left panel of Fig.~\ref{fig_thetas}.
For these two trajectories, the scalar spectral index therefore comes out right,
$n_s = n_s^{\textrm{obs}} \equiv 0.963$.
At the same time, we have $n_s > n_s^{\textrm{obs}}$ for trajectories
\textnumero~1 and \textnumero~7 and $n_s < n_s^{\textrm{obs}}$ for trajectories
\textnumero~3, \textnumero~4 and \textnumero~5.
The black and orange circles mark the respective position of the inflaton
field at $N = N_*\equiv50$.}
\label{fig_trajectories_compare}
\end{figure}

In our (semi-)analytical discussion of the inflationary observables in Sec.~\ref{sec_hilltop},
we managed to reproduce the $95\,\%$\,C.L.\ region in the $(v,\lambda)$ for the special
case of single-field hybrid inflation in the hill-top regime,
i.e.\ for $\theta_f = 0$, cf.\ the right panel of Fig.~\ref{fig_1field_results}.
Now we attempt to extend this analysis to the full two-field case.
In a first step, it is important to understand the qualitative difference between
the inflationary trajectories respectively corresponding to points in the upper
and points in the lower branches of our $95\,\%$\,C.L.\ regions.
To do so, note that, for fixed symmetry breaking scale $v$, larger gravitino
masses are required in the upper branches than in the lower branches so as to
keep the potential flat enough by compensating for the comparatively larger
values of $\lambda$, cf.\ Sec.~\ref{sec_hilltop}.
Therefore, the individual contributions to the slope of the inflaton
potential all have a larger magnitude in the upper branches, which effectively
results in a larger field excursion during inflation in these parts of parameter space.
This is illustrated in Fig.~\ref{fig_trajectories_compare}, in which we display
several inflationary trajectories corresponding to parameter points along
a vertical cross section through the $(v,\lambda)$ plane for $\theta_f = \pi/16$,
cf.\ the left panel of Fig.~\ref{fig_thetas}.
While the last $50$ $e$-folds of inflation along trajectory \textnumero~2,
which belongs to a point in the lower band of the $95\,\%$\,C.L.\ region
for $\theta_f = \pi/16$, easily fit into a very small field range,
$z_* - 1 \simeq 1.6 \times 10^{-2}$, the same number of $e$-folds along
trajectory \textnumero~6, which belongs by contrast to a point in the
associated upper band, require a much larger field excursion, $z_* - 1 \simeq 0.98$.


This difference in the field excursion during inflation also explains
the absence of the second branch of parameter solutions in the case of
single-field hybrid inflation in the hill-top regime.
Such large $z_*$ values as they are required in the new band of parameter solutions
simply clash with the position of the local maximum on the positive real axis.
In other words, on the real axis, only one successful inflationary
trajectory fits in between the instability and the ridge in the scalar potential
for a fixed value of $v$.
On the other hand, allowing the inflaton to freely move in the full complex plane,
the possibility of reproducing the inflationary observables along an alternative,
much longer trajectory opens up, cf.\ Fig.~\ref{fig_trajectories_compare}.


\begin{figure}[t]
\includegraphics[width = 1.0\textwidth]{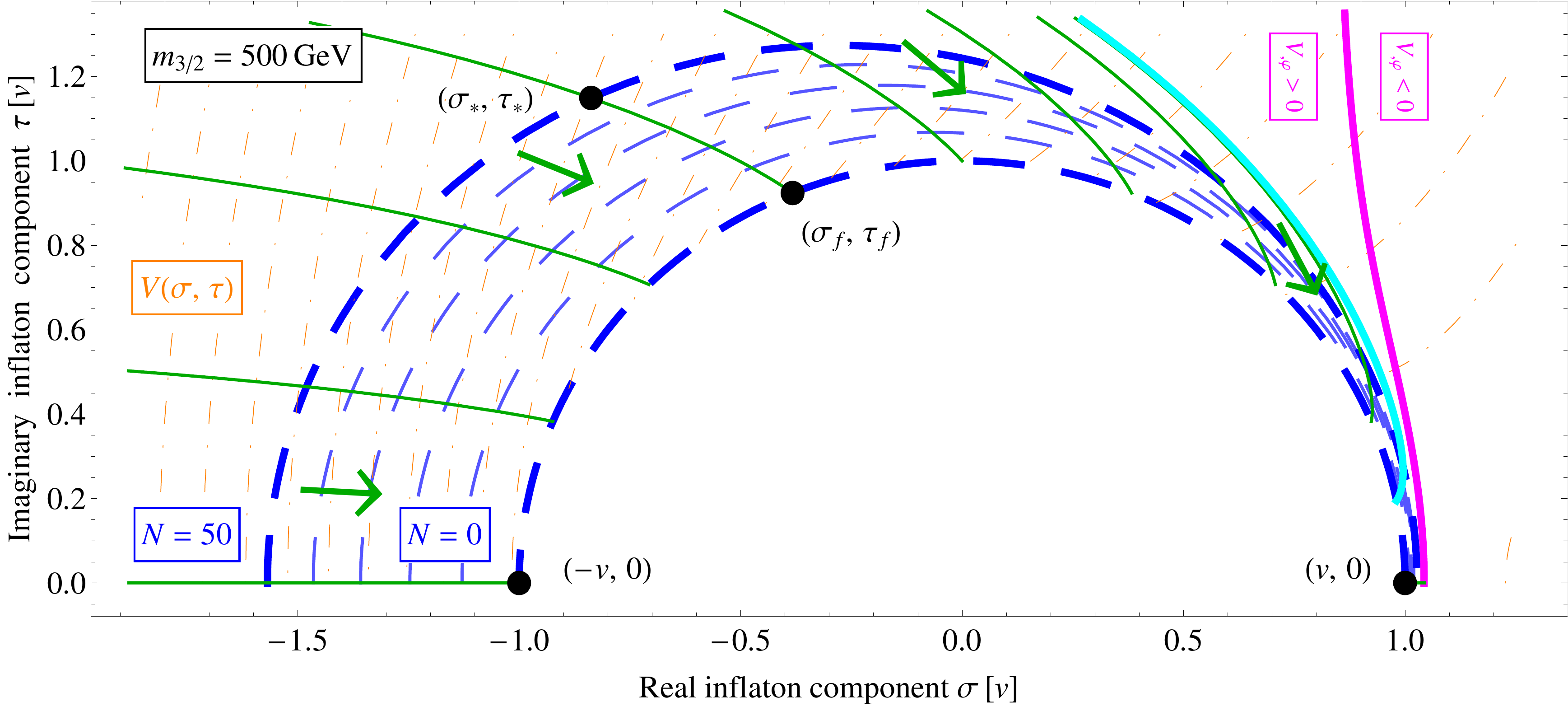}
\caption{Two-field dynamics of the complex inflaton in field space for
$v = 1.8 \times 10^{15}\,\textrm{GeV}$, $\lambda = 4.8 \times 10^{-4}$
and $m_{3/2} = 500 \,\textrm{GeV}$, to be compared to Fig.~\ref{fig_trajectories_50}.
Both the parameter point used for this plot as well as the one in Fig.~\ref{fig_trajectories_50}
are located in the lower band of the $95\,\%$\,C.L.\ region for $\theta_f = \pi/16$,
cf.\ the left panel of Fig.~\ref{fig_thetas}.
Note how the decrease in the parameter values results in a much smaller initial value
of the inflaton field at $N = N_* \equiv 50$.
For $\theta_f=\pi/16$, we have $z_* - 1 \simeq 6.7 \times 10^{-2}$ in this figure and
$z_* -1 \simeq 7.5 \times 10^{-1}$ in Fig.~\ref{fig_trajectories_50}.}
\label{fig_trajectories_05}
\end{figure}


In summary, the lower branches of our $95\,\%$\,C.L.\ regions come in
general with smaller values of $z_*$ than the corresponding upper branches.
In addition to that, $z_*$ also decreases as we move along the lower branches to
smaller and smaller values of $v$ and $\lambda$.
This behaviour is analogous and has the same origin as the behaviour of $x_*$
in the single-field case, cf.\ the right panel of Fig.~\ref{fig_1field_results}.
Moving to smaller values of $\lambda$, one has to simultaneously
reduce the gravitino mass to maintain the balance between the logarithmic
and the linear contribution to the slope of the scalar potential. 
Both contributions then become smaller in magnitude, which results in a smaller
field excursion.
This can be seen by comparing Fig.~\ref{fig_trajectories_50} and
Fig.~\ref{fig_trajectories_05}, which display the two-field dynamics of hybrid inflation
in the complex plane for two different points in the lower band of the
$95\,\%$\,C.L.\ region for $\theta_f = \pi/16$, cf.\ the left panel of Fig.~\ref{fig_thetas}.

\begin{figure}
\centering
\includegraphics[width = 0.5\textwidth]{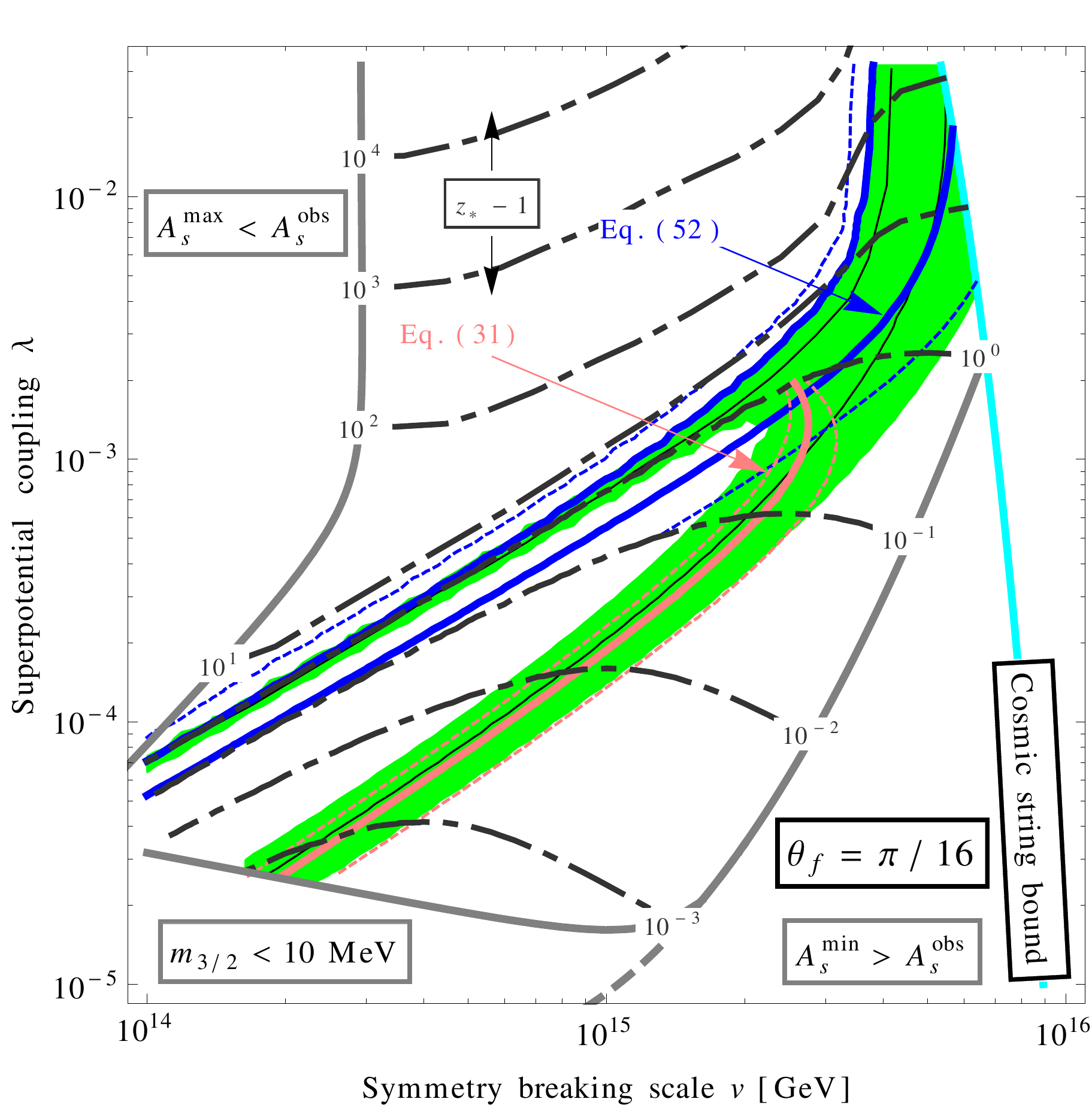}
\caption{Comparison of our analytical and numerical results 
for the $95\,\%$\,C.L.\ region in the $(v,\lambda)$ plane for a
final inflaton phase of $\theta_f = \pi/16$.
Color code and labels as in the right panel of Fig.~\ref{fig_1field_results}
as well as in Fig.~\ref{fig_thetas}.
The black and grey contour curves are the result of our full numerical
calculation.
The red and blue curves are by contrast based on our
(semi-)analytical results for $n_s$ in the small-$z_*$ and large-$z_*$ regime, respectively,
cf.\ Eqs.~(\ref{smallxns}) and \eqref{eq:nszstar}.
The initial field values $z_*$ are indicated by the grey dot-dashed contour lines.}
\label{fig_comparison_2}
\end{figure}

The smallness of $z_*$ in the lower branches of our $95\,\%$\,C.L.\ regions suggests
that, in these regions of parameter space, an effective single-field description might apply.
The inflationary trajectories do not significantly deviate in shape from those in
the single-field case and hence it appears feasible to describe the lower bands
to first approximation by the small-field expression in Eq.~\eqref{smallxns}.
And indeed, Eq.~\eqref{smallxns}, although it has been derived in the context of
single-field inflation, provides a fair description of the location of the lower bands
in the $(v,\lambda)$ plane, especially for $\theta_f$ values close to zero.
As an example, we show in Fig.~\ref{fig_comparison_2} how well we are able to reproduce
the lower band of the $95\,\%$\,C.L.\ region for $\theta_f = \pi / 16$, assuming that
$n_s$ can be still calculated according to Eq.~\eqref{smallxns}.
This result is of course no surprise.
Already in Sec.~\ref{subsec:parameterscan}, we noted that the lower branches 
of the $95\,\%$\,C.L.\ regions asymptotically approach the band of parameter
solutions for $\theta_f = 0$, as soon as $\theta_f$ is lowered to ever smaller values. 


How do we now go about reproducing the upper branches of our $95\,\%$\,C.L.\ regions?
In this case, the situation is unfortunately much more complicated.
The relevant inflationary trajectories often times run very close by
the ridge in the scalar potential before reaching the instability and are
therefore usually strongly curved.
This renders it difficult to integrate the slow-roll trajectories analytically
in order to obtain a two-field analogue of the relation between the inflaton
field value and the number of $e$-folds which we managed to derive in the
single-field case, cf.\ Eq.~\eqref{eq:Nxrel}.
Besides that, any appropriate generalization of Eq.~\eqref{eq:Nxrel}
would presumably look rather convoluted and not lead to further insights.
We therefore decide to pursue a different, already well-tested
semi-analytical approach and intend to make use of the fact that $z_*$ is always very large
in the upper bands of the $95\,\%$\,C.L.\ regions.


Let us assume for a moment that the effective single-field description also holds
in the upper branches of the $95\,\%$\,C.L.\ regions.
Given the curved shape of the trajectories, it is \textit{a priori} not
obvious that this simplified picture indeed applies; but the comparison
with the full numerical results will shortly justify our assumption.
In the effective single-field picture, we can then determine the scalar
spectral index based on the effective slow-roll parameters $\epsilon$
and $\eta$, cf.\ Eq.~\eqref{epsilon-0} and \eqref{eta-0} as well as
App.~\ref{app:index}.
In the large-$z_*$ regime and for small phases, we have
\begin{align}
n_s - 1 \simeq 2 \,\eta =
\frac{2M_\text{Pl}^2}{V}\frac{V^a V_{ab} V^b}{V^c V_c}
\overset{z_* \gg 1}{\longrightarrow}
-\frac{\lambda^2}{4\pi^2} \frac{\Mp^2}{v^2}
\frac{1}{z_*} 
+ \mathcal{O}\left(z_*^{-3/2},\theta_*^2\right)\,, \label{eq:nszstar}
\end{align}
which we immediately recognize as the straightforward generalization of
Eq.~\eqref{largexns}.
Similarly as in Sec.~\ref{sec_hilltop}, also this large-field
approximation of $n_s$ neglects the isocurvature contributions
to the scalar power spectrum.
Owing to our numerical analysis in Sec.~\ref{subsec:parameterscan},
we also know the values of $z_*$ for every point in the various $(v,\lambda)$ planes.
Inserting these numerical results for $z_*$ into Eq.~\eqref{eq:nszstar},
we always find two curves in the $(v,\lambda)$ plane,
at least as long as $\theta_f$ is not too large,
along which the observed value for $n_s$ is reproduced,
cf.\ the blue contour lines in Fig.~\ref{fig_comparison_2}.
In the vicinity of the upper branches of the $95\,\%$\,C.L.\ regions, 
$z_*$ as a function of $v$ and $\lambda$ depends to good approximation
only on the ratio $\beta = \lambda\,M_{\textrm{Pl}}/v$,
\begin{align}
z_* \approx f(\beta) \,.
\end{align}
Here, the function $f$ is such that Eq.~\eqref{eq:nszstar}
yields the correct value of $n_s$ not only for one $\beta$ value,
but actually for two distinct values of this ratio.
For instance, in the case of $\theta_f = \pi/16$, we have
the two solutions $\beta_1 \simeq 1.3$ and $\beta_2 \simeq 1.7$.
In between these two $\beta$ values, the spectral index is smaller,
outside the interval $[\beta_1,\beta_2]$, it is larger than the observed value.
The two cases $\beta = \beta_1$ and $\beta = \beta_2$ just represent
the two curves along which the correct $n_s$ value can be reproduced.
Here, the larger $\beta$ value always results in a very good description of
the upper band in the $(v,\lambda)$ plane.
At the same time, the smaller of the two $\beta$ solutions induces a contour
line in the $(v,\lambda)$ plane, which generalizes the blue contour in
the right panel of Fig.~\ref{fig_1field_results} and which therefore describes
the lower band very well in the large-$z_*$ regime.


As anticipated, our semi-analytical approach based on Eq.~\eqref{eq:nszstar}
succeeds in reproducing the numerical result for $n_s$.
This indicates that, at the level of the power spectrum, the curvature of
the trajectory as well as the conversion of isocurvature to curvature
modes are negligible in most parts of the parameter space and
we are allowed to work in a simplified effective single-field picture.
In general, this effective single-field description however breaks
down at the level of the bispectrum as
discussed at the end of Sec.~\ref{subsec:parameterscan}.


\section{Initial conditions}
\label{sec:inicon}


The hill-top regime of single-field
hybrid inflation is plagued by two problems related to the initial position
of the inflaton field on the real axis.
First, the initial value of the inflaton field, $\varphi_i$, must be carefully tuned.
Using the approximation $x_* -1 \ll 1$, we find that the CMB pivot scale exits the
Hubble horizon roughly at a field value, cf.\ Eq.~\eqref{eq:Nxrel},
\begin{equation}
\frac{\varphi_*}{v} \simeq 1 + \frac{\lambda^2}{4\pi^2}
\ln 2 \, (1-\xi) N_* \left(\frac{M_{\rm Pl}}{v}\right)^2 \,.
\end{equation}
At the same time, the scalar potential exhibits a local maximum and a local minimum at
\begin{equation}
\frac{\varphi_{\rm max}}{v} \simeq \frac{1}{2\ln 2} \frac{1}{\xi} \,, \quad
\frac{\varphi_{\rm min}}{M_{\textrm{Pl}}} \simeq
\left(\frac{\lambda^2}{2\pi^2} \ln 2 \, \xi \, \frac{M_{\textrm{Pl}}}{v}\right)^{1/3} \,,
\end{equation}
where the position of the local minimum is determined by the interplay between the
linear term $V_{3/2}$ and the supergravity correction $V_{\textrm{SUGRA}}$ to the scalar potential,
cf.\ Eqs.~\eqref{eq:Vsg} and \eqref{eq_linear_term}.
Hence, successful inflation can only be achieved for an initial field value
$\varphi_i$ satisfying
\begin{align}
\varphi_* < \varphi_i < \varphi_\text{max} \,.
\end{align}
The same conclusion also holds when the small-$x_*$ approximation
is no longer applicable, as can for instance be seen
from Fig.~\ref{fig_potential_1D}.
An initial field value slightly larger than $\varphi_{\textrm{max}}$
would result in a trajectory leading into the false vacuum on the other
side of the hill-top at $\varphi = \varphi_{\textrm{min}}$, whereas an
initial value smaller than $\varphi_*$ does not allow for a sufficiently
long period of inflation.
Note that this problem cannot be simply solved by setting $\xi$
to a small value, as the scalar amplitude could no longer be
reproduced in such a case, cf.\ Eq.~\eqref{As}.


Second, at the onset of inflation, a sufficiently homogeneous
region with volume $H^{-3}$ is generally required~\cite{Goldwirth:1991rj}.
While chaotic inflation~\cite{Linde:1983gd} can naturally accommodate
the existence of such a homogeneous region, even if the universe starts
out from chaotic initial conditions,
$V(\varphi)\sim {\dot \varphi_i}^2/2 \sim (\vec{\nabla} \varphi_i)^2/2 \sim M_{\rm Pl}^4$,
F-term hybrid inflation fails to do so
because it is associated with relatively low values
of the inflationary Hubble parameter, $H_0 \sim 10^4 \cdots 10^{11}~\text{GeV}$.
This is to say that hybrid inflation taking place at the GUT scale
comes with an `initial' horizon problem~\cite{Ijjas:2013vea}.
Assuming that at pre-inflationary times the energy
density of the universe decreases from some Planckian
value down to the GUT scale in consequence of an ordinary
radiation-dominated expansion, one finds that around
$N \sim N_*$ one Hubble patch (for instance, the one which will
inflate to become the observable universe) consists of roughly
$10^7 \,(10^{16}{\rm GeV}/M_{\rm GUT})$
nonequilibrated Planck domains~\cite{Ijjas:2013vea}.
It is thus hard to explain why the same fine-tuned initial
value should independently occur in each separate Planck domain.%
\footnote{If the real fundamental scale $M_*$ actually lies somewhere
between the Planck scale and the GUT scale, $M_{\textrm{GUT}} \lesssim M_* \ll M_{\textrm{Pl}}$,
as for instance in extra-dimensional theories, the required fine-tuning can be relaxed.
Also, we point out that very recently, the authors of Ref.~\cite{Easther:2014zga} have
shown for the example of nonsupersymmetric hybrid inflation that sub-horizon
inhomogeneities do not \textit{necessarily} spoil the success of inflation.
In certain cases, initial field configurations which do not inflate in the homogeneous
limit may in fact yield successful inflation after all, once inhomogeneous perturbations
are added.
However, this does not change the fact that too large inhomogeneities certainly
do pose a threat to field configurations which actually inflate in the homogeneous limit.
A small amplitude of sub-horizon inhomogeneities before the onset of inflation
is therefore still desirable.}


An attractive way out of these two initial conditions problems
is to have some kind of pre-inflation before the
onset of hybrid inflation, during which the energy density of
the universe falls from the Planck scale to the GUT scale.
In this case, the initial homogeneous region required for hybrid
inflation can be generated during the earlier phase of pre-inflation.
In particular, if pre-inflation corresponds to eternal inflation realized
in a local minimum of the potential, probably even the one located at
$\varphi = \varphi_{\textrm{min}}$, the inflaton can reach the
hybrid inflation regime, $\varphi\sim M_{\rm GUT}$, through a
tunneling process.
Similar ideas have for instance been developed in the context of
locked inflation~\cite{Easther:2004qs}, chain inflation~\cite{Freese:2004vs}
and multiple inflation~\cite{Burgess:2005sb}.

\begin{figure}
\begin{center} 
\includegraphics[width = 0.48\textwidth]{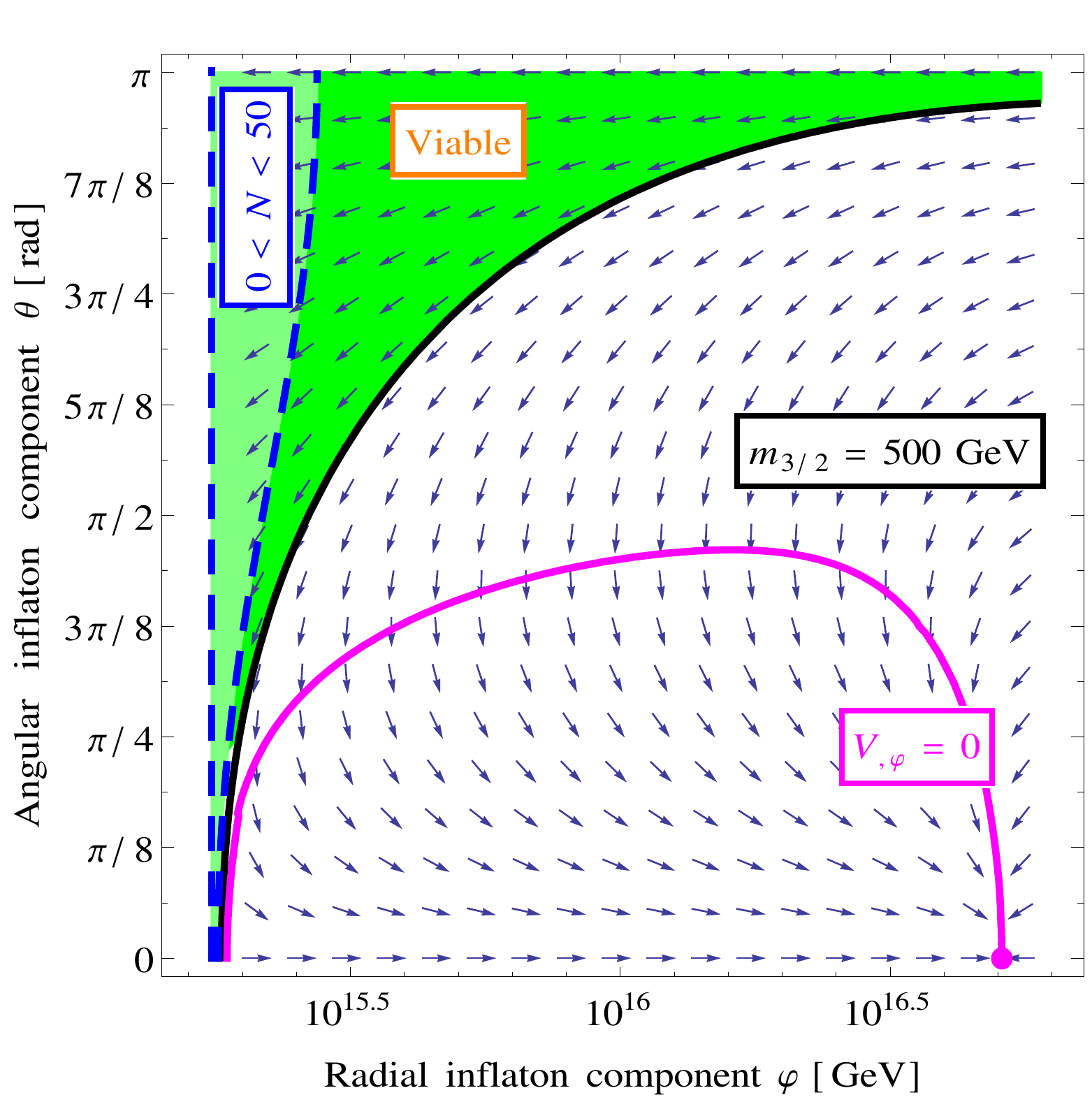} \hfil
\includegraphics[width = 0.48\textwidth]{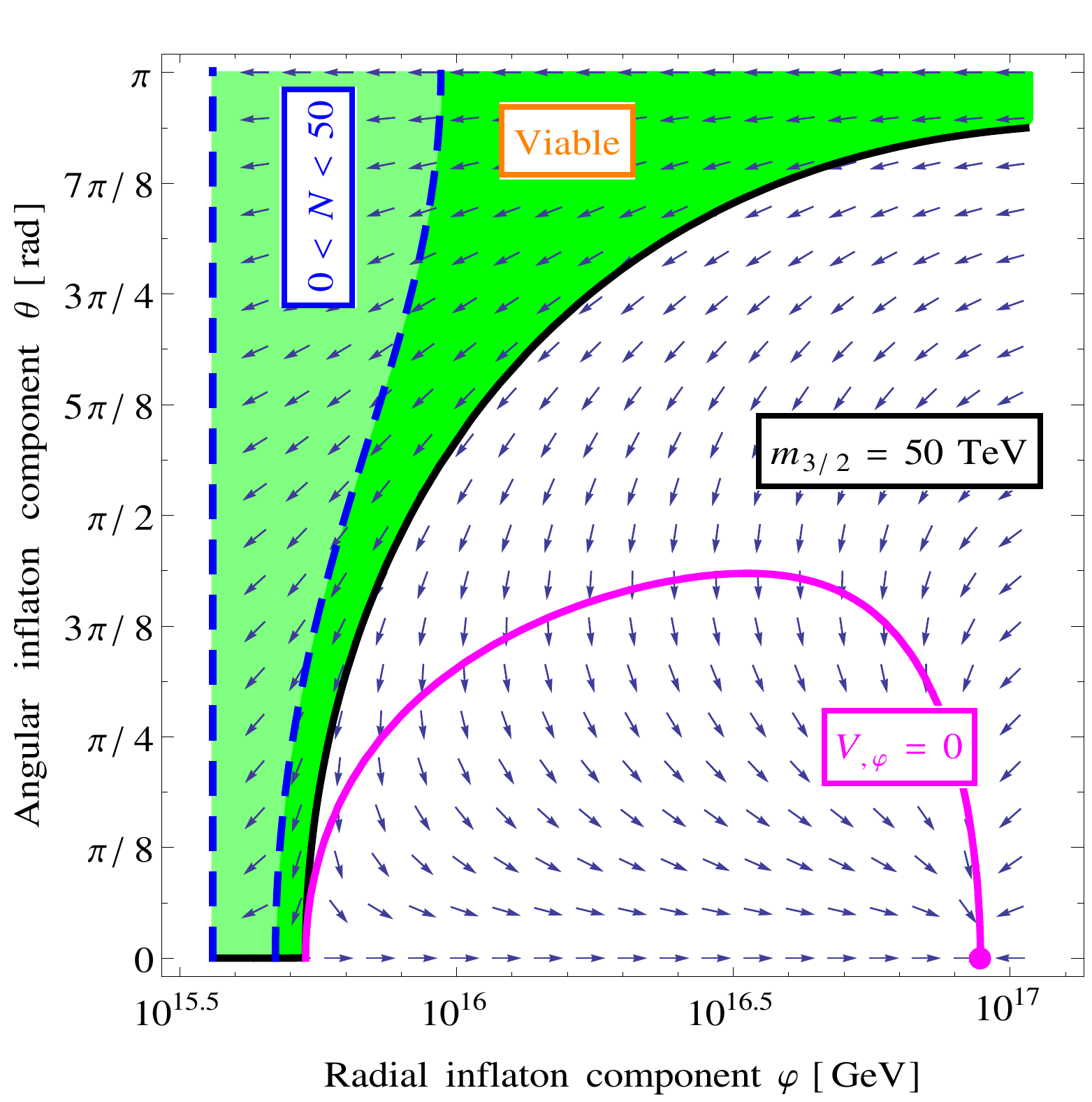}
\end{center}
\caption{Initial conditions in the complex plane.
The blue arrows indicate the direction of the slow-roll trajectories at
any given point in field space.
The dark-green regions mark all possible
initial conditions which lead to inflation ending on the instability.
The light-green regions, encircled by the dashed blue lines, show the last
50 $e$-folds of inflation.
If the inflaton starts out its journey in
the white regions, it becomes eventually trapped in the false vacuum.
The values for $v$, $\lambda$ and $\mG$ are chosen as
in Figs~\ref{fig_trajectories_50} and \ref{fig_trajectories_05}, respectively.}
\label{fig_inicon}
\end{figure}

What is now the situation in the two-field case?
We can best answer this question by studying the course of the
inflationary trajectories; cf.\ Fig.~\ref{fig_inicon}, which shows
the inflaton field space in polar coordinates for a large and a
rather small gravitino mass, respectively.
The small blue arrows represent the gradient field of the scalar potential,
i.e.\ the direction of the slow-roll trajectories at any given point in field space.
The pink lines denote a vanishing slope in the radial direction, which, as can
be seen from the arrows, can be either a maximum, a minimum or a saddle point.
The familiar hill-top and the false vacuum appear along the positive real
axis, i.e.\ for $\theta = 0$.
Correspondingly, the green-shaded regions indicate all initial field values
whose trajectories lead to the critical line, whereas in the white-shaded
regions, the trajectories lead into the false vacuum at $\varphi = \varphi_{\textrm{min}}$.
The dashed blue lines denote the instability in the scalar potential
and the initial flat hypersurface at $t = t_*$, respectively.


In this paper, our goal is not to precisely quantify the amount of
fine-tuning required in the initial conditions of hybrid inflation.
Before we could do that, we would need to define a suitable measure
in field space, taking also possible displacements of the waterfall fields as
well as variations in the initial velocities into account.
For analyses along these lines, cf.\ for instance Refs.~\cite{Tetradis:1997kp,
Mendes:2000sq,Clesse:2008pf,Clesse:2009ur,Easther:2013bga}.%
\footnote{Besides that, it has recently been claimed~\cite{Easther:2013rva} that,
in generic models of multi-field inflation, the concrete choice of
a prior in specifying initial conditions may, after all, only have
little effect on the eventual predictions for the observables.
If that should also hold true for our inflationary scenario, our
choice of a `slow-roll' prior~\cite{Easther:2013rva} may actually not go along
with a loss of generality.
The fact that we fix the initial velocities by the slow-roll equations
of motion, cf.\ Eq.~\eqref{eq:phithetaSR}, would then merely correspond
to a technical specification necessary for definiteness.\smallskip}
Instead, we here merely intend to make the point that recognizing hybrid inflation
as a two-field model in the complex plane significantly relaxes
the two problems related to the initial conditions for the inflaton field,
in particular the fine-tuning problem, cf.\ Fig.~\ref{fig_inicon}.
Now, a significant part of the field space yields initial conditions
leading to a sufficient amount of inflation ending in the right vacuum
(darker green region).
The initial position of the inflaton field no longer needs to be fine-tuned.
This is to be compared with the situation for $\theta_f = 0$, where suitable
initial conditions lie between the right dashed blue and the pink line.
In the left panel of Fig.~\ref{fig_inicon}, this segment of the
real axis is hardly visible.

 
As far as the horizon problem is concerned, we note
that now, where the inflaton is allowed to freely move in the
complex plane, inflation can also start out at the Planck scale.%
\footnote{Typically, this requires an initial phase close to $\theta_i \sim \pi$,
which re-introduces the necessity of a mild fine-tuning.
However, the further the inflaton moves down from the Planck scale, the
more do the trajectories spread in polar field space.
The amount of fine-tuning can therefore always be controlled and
reduced by specifying the initial conditions for the inflaton field
at lower and lower energy scales.}
One could therefore imagine that hybrid inflation
begins with $\varphi_i\sim M_{\rm Pl}$ within
a sufficiently homogeneous initial region, whereby the horizon problem
would be solved.
But, this picture rests upon the assumption that the initial velocity
of the inflaton is suppressed for some reason.
In the case of hybrid inflation starting out at the Planck scale,
we would expect that $|{\dot \varphi}| \sim M_{\rm Pl}^2$, such that
the inflaton would easily overshoot the inflationary region.
In order to fully solve the horizon problem in hybrid inflation,
we must therefore explain why the initial velocity of the
inflaton at the Planck scale is suppressed.
Compared to the challenge of fine-tuning the inflaton field
value in $10^7$ nonequilibrated regions at the GUT scale,
the task of suppressing $|{\dot \varphi}|$ in one Planck domain
seems however much more manageable.
Therefore, also without invoking any extension of the model,
the initial horizon problem appears to be relaxed as well.
As an alternative to a suppressed initial velocity, one
could also attempt to come up with an explicit scenario of pre-inflation.
Both options seem promising and interesting; a more detailed
investigation is left for future work.


\section{Bounds on the gravitino mass \label{sec_gravitino_mass}}


Up to this point, we have considered the gravitino mass,
$\mG \geq 10$~MeV, as a free input parameter.
We consequently found parameter solutions over a wide range of
gravitino mass scales, $\mG \sim 10\,\textrm{MeV}\cdots 10\,\textrm{PeV}$.
Interestingly, this range of gravitino masses includes all relevant values
commonly employed in supersymmetric models of electroweak symmetry breaking.
In this section, we now specify in more detail the allowed range of
gravitino masses, discussing in particular the consequences
of the production of gravitinos in the early universe.


\subsection{Supersymmetry breaking and slow-roll inflation}


Throughout our analysis, we assume that supersymmetry becomes softly
broken in a hidden sector already before the onset of inflation.
During inflation, supersymmetry is in addition broken by the tadpole term
for the inflaton field $\Phi$ in the superpotential,
$W_{\rm inf} = \lambda\,v^2/2\,\Phi \sim \lambda v^3$.
Our decision to ignore the dynamics of vacuum supersymmetry breaking is justified
as long as $|W_{\rm inf}| > |W_0|$, which translated into an upper bound
on the gravitino mass,
\begin{align}
\mG \lesssim  \frac{\lambda v^3}{\Mp^2} \sim 100~\mathrm{TeV} 
\left(\frac{\lambda}{10^{-3}}\right)
  \left(\frac{v}{10^{15}~\mathrm{GeV}}\right)^3 \,.
\end{align}


Two further bounds on the gravitino mass can be derived from requiring 
successful slow-roll inflation to occur.
In Sec.~\ref{sec_hilltop}, we derived a first upper bound
on the gravitino mass in the hill-top regime on the real axis,
cf.\ Eq.~\eqref{eq_mgbound1}, from the requirement that at least 50 $e$-folds
of inflation must fit in between the instability in the scalar potential
and the hill-top,
\begin{equation}
\mG < \frac{\lambda^3\,\ln 2}{2^{9/2}\,\pi^2} v
\simeq 3~\text{TeV} \left(\frac{\lambda}{10^{-3}}\right)^3
\left(\frac{v}{10^{15}~\mathrm{GeV}}\right) \,.
\label{eq_mgbound11}
\end{equation}
Also outside the hill-top regime, the requirement of consistent
slow-roll inflation imposes an upper bound on the gravitino mass.
For large values of $\mG$, all trajectories leading to successful
inflation run initially in parallel to the $\theta_f = \pi$ trajectory,
cf.\ Fig.~\ref{fig_trajectories_largem32}, so that we can restrict
our discussion to this case.
Increasing $\mG$ steepens the scalar potential, thereby
pushing the initial field value $z_*$ to ever larger values.
As $z_*$ approaches the Planck scale, the supergravity
contributions to the scalar potential become important, until at
$z_* \sim M_\text{Pl}^2$ the slow-roll condition for $\eta$ is violated.
The requirement of achieving $50$ $e$-folds of inflation at sub-Planckian field
values without violating the slow-roll conditions therefore yields
an upper bound on $m_{3/2}$.
An analytical analysis of the inflaton slow-roll equation
on the negative real axis leads us to
\begin{equation}
\mG \lesssim 3 \times 10^{-3} \, H_0 \sim 350~\text{TeV}
\left( \frac{\lambda}{10^{-3}} \right) \left( \frac{v}{10^{15}~\text{GeV}} \right)^2\,,
\label{eq_mgbound2}
\end{equation}
where $H_0$ denotes the inflationary Hubble scale.
Note that both Eq.~\eqref{eq_mgbound11} and Eq.~\eqref{eq_mgbound2}
are obtained solely from requiring at least 50 $e$-folds of
slow-roll inflation.
Demanding in addition that the correct values for $A_s$ and $n_s$
be successfully reproduced yields even tighter upper bounds on $m_{3/2}$,
which vary as functions of the final inflaton
phase $\theta_f$, cf.\ Fig.~\ref{fig_thetas}.


\subsection{Nonthermal gravitino production}


After the end of inflation, gravitinos are generated thermally and nonthermally.
A too large abundance of these gravitinos leads to the infamous cosmic gravitino
problem~\cite{Weinberg:1982zq, Khlopov:1984pf}, with the precise
bounds depending on the mass hierarchies of the theory.
In the following, we briefly review
nonthermal~\cite{Kawasaki:2006gs, Kawasaki:2006hm, Endo:2007sz}
and thermal~\cite{Bolz:2000fu} gravitino production and derive the
resulting constraints on the parameter space of hybrid inflation.
Then we comment on possibilities to relax or avoid these constraints.


After the end of inflation (and preheating), the energy density of the
universe is dominated by the contributions from the
non-relativistic scalar particles of the waterfall-inflaton sector.
As a consequence of the constant term in the superpotential, the resulting
mass eigenstates $\varphi_{1,2}$ are a maximal mixture of the inflaton and 
waterfall gauge eigenstates $\varphi$ and $\chi$.
They hence both obtain a large vacuum expectation value (vev),
$\langle \varphi_{1,2} \rangle = v/\sqrt{2}$, and their masses
are given by $m_\varphi \simeq \lambda v$, with a small mass 
splitting proportional to the gravitino mass $m_{3/2}$.
These particles can decay into gravitinos, thereby yielding a
population of nonthermal gravitinos. 
The decay rate into a pair of gravitinos is
given as~\cite{Endo:2006zj, Nakamura:2006uc, Endo:2007sz}
\begin{equation}
\Gamma_{3/2} = \frac{c}{96 \pi} \left(\frac{\langle \varphi \rangle}
{M_\text{Pl}}\right)^2 \frac{m_\varphi^3}{M_\text{Pl}^2}\,,
\label{eq_gamma32}
\end{equation}
where we have assumed that the fields $\varphi_{1,2}$ are lighter
than the sgoldstino $z$, i.e.\ the complex scalar in the hidden-sector
chiral multiplet responsible for soft supersymmetry breaking.%
\footnote{If on the contrary the fields $\varphi_{1,2}$ were heavier than
the sgoldstino, i.e.\ if $m_\varphi > m_z$, cf.\ Ref.~\cite{Endo:2007ih},
the gravitino abundance could be significantly suppressed~\cite{Nakayama:2012hy},
depending on the details of the supersymmetry breaking sector.}
Note that this decay channel only opens up as soon as $H < m_{3/2}$,
since it requires a helicity flip which is not possible for effectively
massless gravitinos~\cite{Kawasaki:2006gs}.
Exploiting the conservation of the comoving entropy density
after the end of reheating, the resulting abundance of nonthermally
produced gravitinos normalized to the entropy density $s$ is given by
\begin{equation}
Y_{3/2}^{\text{nt}} = \frac{n_{3/2}}{s} = 2 \,
\frac{\Gamma_{3/2}}{\Gamma_\varphi} \frac{n_\varphi(t_f)}{s} \,, 
\end{equation}
with $\Gamma_\varphi$ denoting the total decay rate of $\varphi$ bosons,
$n_\varphi(t_f)$ their comoving number density at the end of reheating,
i.e.\ at matter-radiation equality,
$\rho_\text{rad}(t_f) = \rho_\text{m}(t_f) = n_\varphi(t_f) \, m_\varphi$.
Expressing the energy and entropy density of the thermal bath in terms of
the reheating temperature, $T_R(t_f) = (45/(\pi^2 g_*))^{1/4}
\sqrt{\Gamma_\varphi M_\text{Pl}} $,
we find the resulting gravitino abundance to be inversely proportional
to the reheating temperature,
\begin{equation}
Y_{3/2}^{\text{nt}, 0} = \frac{3}{2} \left( \frac{90}{\pi^2 g_*} \right)^{1/2}
\frac{\Gamma_{3/2}}{m_\varphi} \frac{M_\text{Pl}}{T_R} \,.
\label{eq_Ylit}
\end{equation}


The thermal contribution on the other hand, stemming mainly from supersymmetric
QCD 2-to-2 scatterings in the thermal bath, can be expressed as,
cf.\ App.\ D of Ref.~\cite{Buchmuller:2011mw}, 
\begin{equation}
Y_{3/2}^{\text{th}} = \frac{\rho_c}{m_{3/2} s_0} \, \epsilon \, C_1
\left(\frac{T_{RH}}{10^{10} \text{ GeV}} \right) \left[ C_2
\left(\frac{m_{3/2}}{100 \text{ GeV}} \right) +  \left(\frac{100
\text{ GeV}}{m_{3/2}} \right) \left(\frac{m_{\tilde g}}{1
\text{ TeV}} \right)^2\right] \,,
\label{eq_thermal}
\end{equation}
with $s_0 = 2.9 \times 10^3~\text{cm}^{-3} $ and
$\rho_c/h^2 = 1.052 \times 10^{-5}~\text{GeV/cm$^3$}$
denoting the entropy and critical energy densities today.
The coefficient functions $C_1$ and $C_2$ can be calculated analytically
and feature a weak dependence on the reheating temperature, the
parameter $\epsilon$ accounts for details of the reheating process
which cannot be taken into account analytically.
For the analysis here, it is sufficient to choose representative,
constant values for these parameters,
$C_1 =0.26 $, $C_2 = 0.13 $ and $\epsilon = 1$~\cite{Buchmuller:2012wn}.
Moreover, we will set the gluino mass to $m_{\tilde g} = 1$~TeV.


\begin{figure}[t]
\center
\includegraphics[width = 0.64\textwidth]{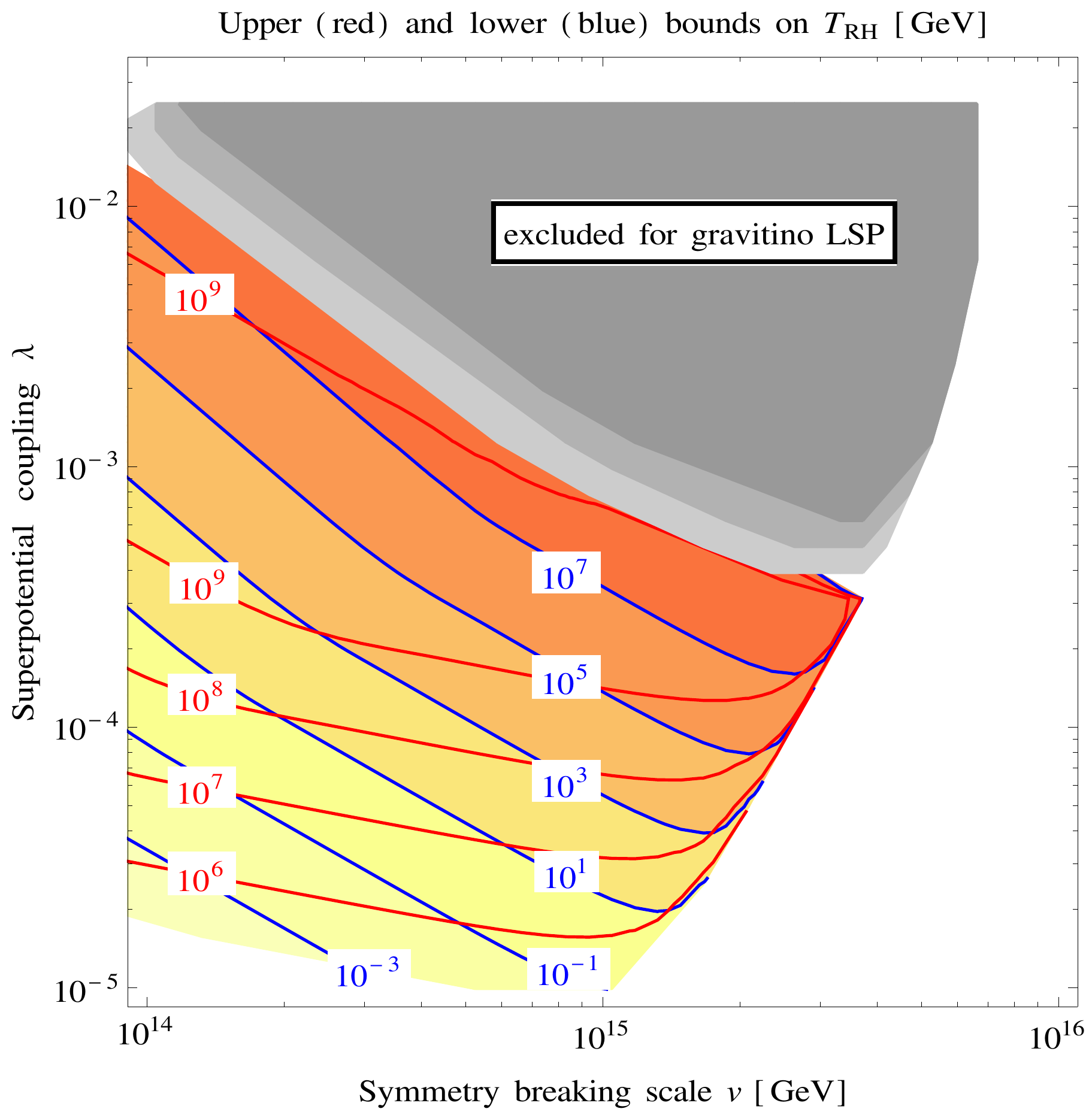}
\caption{Bounds on the reheating temperature in the gravitino LSP scenario.
The yellow/red-shaded region marks the viable parameter space, with the blue
(red) contour lines referring to the corresponding lower (upper) bounds on the
reheating temperature.
The grey region is excluded if the gravitino is the LSP.
The darker grey regions show how this constraint can be relaxed for
$\Gamma_\varphi/\Gamma_X = 10$ (medium grey) or
$\Gamma_\varphi/m_{3/2} = 10$ (darkest grey).}
\label{fig_gravitinos}
\end{figure}


Stringent  bounds on the gravitino mass are obtained when assuming that 
the gravitino is the lightest supersymmetric particle (LSP).
In this case, the total gravitino abundance is restricted by the measured
relic abundance of dark matter,
\begin{equation}
\left(Y_{3/2}^{\text{nt}} + Y_{3/2}^{\text{th}}\right) \, m_{3/2} <
\frac{\rho_c}{s_0}\, \Omega_{\text{DM}} = 4.3 \times
10^{-10}~\text{GeV} \,,
\label{eq_bound_gravitinolsp}
\end{equation}
where we have used $\Omega_\text{DM} h^2 = 0.12$~\cite{Ade:2013zuv}.
Making use of the relation between $m_{3/2}$, $v$ and $\lambda$
imposed by the correct normalization of the scalar power spectrum,
cf.\ Eq.~\eqref{eq:Asnorm}, and treating the reheating temperature
as a free parameter, this bound can be translated into constraints on
$v$ and $\lambda$, cf.\ Fig.~\ref{fig_gravitinos}.
As it turns out, it is the region of small $v$ and $\lambda$ values, corresponding
to gravitino masses $m_{3/2} \lesssim 1$~TeV, that is in agreement with the bound
in Eq.~\eqref{eq_bound_gravitinolsp}.
The resulting lower and upper bounds on the reheating temperature
(driven by the nonthermal and thermal contributions, respectively)
are depicted by the blue and red contour lines.
Comparing Fig.~\ref{fig_gravitinos} with Fig.~\ref{fig_overlay},
we see that for very light gravitinos stringent bounds on the
reheating temperature apply, e.g.\ $m_{3/2} \lesssim 1$~GeV requires
$T_\text{RH} \lesssim 10^8$~GeV.
Note that the decrease of the upper bound on $T_\text{RH}$ for
large values of $\lambda$, responsible for extending the excluded region,
is due to the first term in the squared brackets on the right-hand side
of Eq.~\eqref{eq_thermal}, which takes over at $m_{3/2} \gtrsim 100$~GeV.
For this figure, we chose $\theta_f = \pi/8$ for concreteness.
As $m_{3/2}$ however changes only very mildly for fixed $v$ and $\lambda$
as well as $\theta_f$ varying in the interval $0\lesssim \theta_f \lesssim \pi/4$,
our conclusions are independent of this particular choice.


From Figs.~\ref{fig_overlay} and \ref{fig_gravitinos}, we see that
the region corresponding to $m_{3/2} \gtrsim 1$~TeV is excluded in
the case of a gravitino LSP.
However, for such large gravitino masses, we would anyway expect
that the gravitino is not the LSP in the supersymmetric mass spectrum.
Unstable gravitinos in the mass range $\mG \sim 1\ldots 10$~TeV are
subject to severe constraints  from primordial
nucleosynthesis~\cite{Kawasaki:2004qu,Jedamzik:2006xz}, which are difficult
to circumvent.
Gravitinos heavier than $\mG \gtrsim 10$~TeV however
decay before the onset of big bang nucleosynthesis.
Then the requirement that the abundance of the LSP produced in gravitino
decays does not exceed the measured dark matter abundance imposes a bound
similar to Eq.~\eqref{eq_bound_gravitinolsp},
\begin{equation}
m_{3/2} Y_{3/2} \lesssim 4 \times 10^{-10}~\text{GeV}
\, \left(\frac{\mG}{m_\text{LSP}} \right) \,,
\end{equation}
which can however be loosened for large LSP annihilation cross sections and/or
high gravitino decay temperatures, cf.\ e.g.\ Ref.~\cite{Nakayama:2010xf}.
Fully worked examples of the gravitino
LSP and the heavy gravitino setup can be
found in Refs.~\cite{Buchmuller:2012wn, Buchmuller:2012bt}, respectively.


There are two further loopholes in the derivation of the bounds from nonthermal
gravitino production which we want to point out here.
The derivation leading to Eq.~\eqref{eq_Ylit} assumes that the $\varphi_{1,2}$
mass eigenstates decay directly into light particles forming the thermal bath.
However, in a generic scenario, the $\varphi$ particles may first decay
into another heavy particle species $X$ which becomes the dominant contribution
to the energy density of the universe before decaying into the thermal bath itself.
For example, in Ref.~\cite{Buchmuller:2012wn} the $\varphi$ particles decay
into heavy Majorana neutrinos, thereby setting the stage for nonthermal leptogenesis.
Assuming that the vev of the $X$ particles remains at $0$, a
sufficiently long intermediate period governed by these particles
($\Gamma_X \ll \Gamma_\varphi$) can significantly reduce the resulting
gravitino abundance,
\begin{equation}
Y_{3/2}^{\text{nt}} = \begin{cases} \frac{\Gamma_X}{\Gamma_\varphi}
\, Y_{3/2}^{\text{nt}, 0}  \hspace{8mm} \quad \quad \text{for $X$ non-relativistic}  \\
\rule{0pt}{24pt}\left(\frac{\Gamma_X}{\Gamma_\varphi}\right)^{1/2}
Y_{3/2}^{\text{nt}, 0} \hspace{2.5mm} \quad \text{for $X$ relativistic} \end{cases} \,.
\end{equation}
The reason for this suppression is that during the $X$-dominated phase
no gravitinos are produced according to Eq.~\eqref{eq_gamma32}, while
the onset of the radiation dominated era, crucial to linking the
produced gravitino abundance to the reheating temperature, is delayed.


A further suppression of the final gravitino abundance arises
if $m_{3/2} < \Gamma_\varphi$.
In this  case, part of the initial abundance of $\varphi$ particles will
have decayed before the nonthermal gravitino production sets in, leading to
\begin{equation}
Y_{3/2}^{\text{nt}}  = \begin{cases} \exp
\left[ - \frac{2 \Gamma_\varphi}{3 m_{3/2}}\right] Y_{3/2}^{\text{nt}, 0}
\quad \text{for matter domination after } H = \Gamma_\varphi \\
\rule{0pt}{24pt}  \exp \left[ - \frac{ \Gamma_\varphi}
{2 m_{3/2}}\right] Y_{3/2}^{\text{nt}, 0} \quad
\text{for radiation domination after } H = \Gamma_\varphi \end{cases} \,.
\end{equation}
Fig.~\ref{fig_gravitinos} illustrates the resulting relaxation of the
constraints in the $(v,\lambda)$ plane.
The light-grey contour marks the excluded region for single-stage reheating,
the darker shaded regions show how this bound relaxes taking into
account the two above mentioned effects, with
$\Gamma_\varphi/\Gamma_X = 10$ and $\Gamma_\varphi/m_{3/2} = 10$,
respectively.
Of course, this also enlarges the allowed range for
the reheating temperature.


In summary, while at first sight the gravitino problem seems to exclude a
significant part of the parameter space in the case of gravitino LSP,
cf.\ Fig.~\ref{fig_gravitinos}, there are several ways to avoid these
constraints, e.g.\ assuming $m_\varphi \gg m_z$ with some assumptions
on the supersymmetry breaking sector or particular mass hierarchies
in the reheating process.
However, when embedding hybrid inflation into a more complete model
of particle physics and the early universe, these options may not
all be available and in particular, the reheating temperature may
not be a free parameter.
The above mentioned bounds and possible loopholes
must then carefully be taken into account. 
In any case, the gravitino mass range suitable for rendering hybrid
inflation in accordance with the PLANCK data is interesting both from
a particle physics as well as from a cosmological point of view,
as it contains the mass range relevant for supersymmetric electroweak
symmetry breaking and at the same time mass scales which can be
restricted by early universe cosmology.


\section{Conclusions and outlook \label{sec_conclusion}}


Supersymmetric hybrid inflation models typically feature a true vacuum in
which supersymmetry is fully restored.
A simple and straightforward way to accommodate soft low-energy supersymmetry
breaking in this Minkowski vacuum is to assume that supersymmetry is
spontaneously broken by non-vanishing F-terms in a hidden sector,
whose dynamics are already completely fixed during inflation.
This effectively results in a constant term in the superpotential proportional
to the vacuum gravitino mass.
Since the mass scale of the gravitino is typically
expected to be much smaller than the energy scale of inflation, the effect of
this term on the inflationary dynamics has been widely neglected.
However, since the inclusion of this term breaks the rotational invariance
of the scalar potential in the complex inflaton plane, its effects can
be very important even for small gravitino masses.
F-term hybrid inflation is consequently a two-field model of inflation,
such that its predictions for the observables related to the primordial
fluctuations depend not only on the parameters of the scalar 
potential, but in particular also on the choice of the inflationary trajectory.
This puts the measured values of the amplitude of the scalar power spectrum,
the scalar spectral index and the amplitude of the local bispectrum into new light:
their precise values are no longer dictated by the fundamental model parameters, but are
rather strongly influenced by a selection process at very early times that
appears to be random within the model itself.
As these insights only rely on the presence of a large F-term driving inflation
and the assumption of soft symmetry breaking in a hidden sector at very high
scales, similar conclusions should apply in comparable inflationary scenarios.
We expect that our study and in particular our analysis of the linear
term in the scalar potential can be easily generalized to other models
of inflation, including large-field models, in which supersymmetry breaking
turns an originally single-field model into a multi-field model.


In this paper, we analyzed the inflationary dynamics of F-term
hybrid inflation in the complex plane based on the $\delta N$ formalism.
After extending the method presented in Refs.~\cite{Yokoyama:2007uu,Yokoyama:2007dw}
so as to explicitly take into account the contributions to the
curvature perturbation spectrum produced after the end of slow-roll inflation,
we calculated the inflationary observables related to the scalar power
spectrum and the local bispectrum as functions of the symmetry breaking scale $v$,
the superpotential coupling $\lambda$, the gravitino mass $\mG$ and
the choice of the inflationary trajectory, labeled by the final inflaton
phase $\theta_f$.
We found that the predictions for the scalar power spectrum are well described
in an effective single-field approximation, whereas the bispectrum can obtain
large contributions from the inherently multi-field dynamics.
In ordinary single-field slow-roll inflation, we would expect the primordial
non-Gaussianities to be suppressed by the slow-roll
parameters.
By contrast, in hybrid inflation in the complex plane,
we partly obtained $f_{\textrm{NL}}^{\textrm{local}}$ values roughly
as large as $0.5$.
We cross-checked the results of our numerical analysis by means of
analytical calculations, which provided us with accurate analytical
formulas for the hill-top regime on the real axis as well as with
semi-analytical formulas for the two-field case.


The results of our analysis demonstrate that F-term hybrid inflation is in
much better shape than widely believed in two important points.
First, the fine-tuning in the initial conditions necessary to obtain successful
inflation is greatly reduced.
Second, the measured scalar spectral index can be reproduced in a significant part
of the parameter space without resorting to a non-canonical K\"ahler potential.
Roughly speaking, a correct spectral index is obtained when the contributions
to the slope of the scalar potential from one-loop corrections and
from supersymmetry breaking have opposite sign, but are of comparable size.
This is typically accomplished along trajectories
in the complex plane corresponding to $\theta_f \lesssim \pi/4$, i.e.\
trajectories which pass through the vicinity of the hill-top region
on the real axis.
Taking into account the effect of supersymmetry breaking hence links
the CMB observables to the mass scale of soft supersymmetry breaking.
The resulting mass range for the gravitino mass turns out to lie in a region which
is very interesting, including the mass range relevant for supersymmetric
electroweak symmetry breaking, for gravitino LSP dark matter as well as for nonthermal
dark matter production through the decay of heavy gravitinos.
A crucial further development which will have an important impact on
F-term hybrid inflation is the ongoing search for primordial
B-mode polarization of the CMB radiation.
If the recent results of the BICEP2 experiment are confirmed, an explanation
within the framework of small-field inflation will be challenging.


\subsubsection*{Acknowledgements}

The authors thank A.~Hebecker, V.~Mukhanov, T.~Suyama, T.~Takahashi, A.~Westphal,
M.~Yamaguchi, T.~Yanagida and S.~Yokoyama for helpful discussions and comments.
This work has been supported in part by the German Science Foundation (DFG)
within the Collaborative Research Center 676 ``Particles, Strings and the Early Universe''
(W.B.),  by the European Union FP7-ITN INVISIBLES
(Marie Curie Action PITAN-GA-2011-289442-INVISIBLES) (V.D.), by the
JSPS Postdoctoral Fellowships for Research Abroad (K.K.) and by the World
Premier International Research Center Initiative (WPI Initiative) of the Ministry
of Education, Culture, Sports, Science and Technology (MEXT) of Japan (K.S.).


\begin{appendix}


\section{Simple estimate for the scalar spectral index}
\label{app:index}


Consider a set of scalar fields $\phi^a$ with canonical kinetic terms,
\begin{align}
S[\phi] = \int d^4x \sqrt{g} \left(\frac{1}{2}\,g^{\mu\nu}\partial_\mu
\phi^a \partial_\nu \phi_a - V(\phi) \right) \,.
\end{align}
In the slow-roll regime of an inflationary phase, the trajectories in
field space are determined by the equations of motion
\begin{align}
3 H \dot \phi_a = -\partial_a V \,,
\end{align}
where $H$ is the Hubble parameter obeying the Friedmann
equation and $\partial_a = \partial/\partial \phi^a$.
To obtain the number of $e$-folds of expansion as a function of an initial
point $\phi$ in field space, one has to evaluate a line integral along
the inflationary trajectory, cf.\ Eq.~\eqref{efolds}, as well as
throughout the preheating process until the point in time when the universe reaches
the adiabatic limit.


The amplitude as well the the spectral index for the primordial scalar fluctuations
are then given by the compact expressions~\cite{Sasaki:1995aw}:
\begin{align}\label{SaSt}
A_s = \left(\frac{H}{2\pi}\right)^2 \partial^a N \partial_a N \,, \quad
n_s-1 = \frac{\Mp^2}{\partial^c N \partial_c N}\left(2 \,\partial^a\partial^b \ln V -
\delta^{ab} \frac{\partial^d V \partial_d V}{V^2}\right) \partial_a
N\partial_b N \,.
\end{align}
In general, the calculation of $N(\phi)$ is difficult since it requires
knowledge of the entire trajectory including the transition from
inflation to preheating.
However, in effective single-field cases%
\footnote{In general, the predictions of multi-field models of
inflation can strongly deviate from the single-field
estimate, cf.\ for instance Refs.~\cite{Sasaki:2008uc,Naruko:2008sq,Huang:2009vk}.
However, as our numerical analysis, in which we take into account all
potentially important effects, shows, this is typically
not the case for the power spectrum in F-term hybrid inflation.}
where fluctuations orthogonal to the
trajectory yield a negligible contribution to $\delta N$,
cf.\ Eq.~\eqref{eq:zeta}, one can use as an approximation
\begin{align}
\partial_a N \propto \partial_a V \,,
\end{align}
evaluated at $N = N_*$.
The expressions in Eq.~\eqref{SaSt} can then be written in a form familiar
from single-field inflation,
\begin{align}\label{ns-simple}
A_s = \left(\frac{H}{2\pi}\right)^2 \frac{1}{2\,\epsilon\,M_{\textrm{Pl}}^2} \,, \quad
n_s = 1 - 6\, \epsilon + 2\,\eta \,,
\end{align}
where $\epsilon$ and $\eta$ are the slow-roll parameters along
the inflationary trajectory given in Eqs.~\eqref{epsilon-0} and \eqref{eta-0}.
In fact, performing a field redefinition from the fields $\phi^a$
to a new basis $\varphi^a$, such that $\varphi^0$ points along the inflationary
trajectory and all other fields $\varphi^a$ with $a\neq0$ are orthogonal
to the trajectory, $\epsilon$ and $\eta$ can be simply written as
\begin{align}
\label{eq:epsetatraj}
\epsilon = \frac{1}{2} \frac{M_{\textrm{Pl}}^2}{V^2}
\left(\frac{\partial V}{\partial\varphi^0}\right)^2 \,, \quad
\eta = \frac{M_{\textrm{Pl}}^2}{V} \frac{\partial^2V}{(\partial\varphi^0)^2} \,.
\end{align}
Using Eq.~\eqref{ns-simple} as an approximation, one obtains for the
spectral index in hybrid inflation, cf.~Eqs.~\eqref{pot1} and \eqref{pot2},
\begin{align}\label{ns-explicit}
n_s-1 \simeq 2\eta 
= \frac{2}{V} \frac{M_\text{Pl}^2}{v^2} \frac{\partial^2_x V(\partial_x V)^2
+ 2 \,\partial_x V (\partial_x\partial_y V)\partial_y V
+ \partial^2_y V(\partial_y V)^2}{(\partial_x
V)^2 + (\partial_y V)^2} \,,
\end{align}
where
\begin{align}
\partial_x V &= 2a f'x - b \,, \quad 
\partial_y V = 2a f'y \,, \quad \partial_x\partial_y V = 4a f'' xy \,,\\
\partial_x^2 V &= 4a f'' x^2 + 2a f' \,, \quad \partial_y^2 V = 4a f''
y^2 + 2a f'\,.
\end{align}


\section{Comment on the recent evidence for CMB B-modes  \label{app-bicep2}}


During the final stages of preparing this paper, the BICEP2 collaboration
reported on a measurement of the CMB B-mode power spectrum with
unprecedented sensitivity~\cite{Ade:2014xna}.
The observed power spectrum is well fit by the
$\Lambda$-Cold-Dark-Matter model (which already features B-modes simply
because of gravitational lensing) including an additional contribution
from primordial tensor perturbations due to inflation with a
tensor-to-scalar ratio of $r=0.20^{+0.07}_{-0.05}$.
According to this result and a conservative estimate of the foreground
dust polarization, the null hypothesis $r=0$ is ruled out at a
confidence level of $5.9\,\sigma$.


Before arriving at a final conclusion, we will have to wait if
these ground-breaking results are confirmed by other upcoming experiments.
Nevertheless, it is indispensable to comment here on the implications
of this measurement on the inflation model discussed in this paper.
In F-term hybrid inflation, the tensor-to-scalar ratio is given as
\begin{align}
r & =\frac{A_T}{A_s}=\frac{2H^2}{\pi^2 A_s M_{\rm Pl}^2} \simeq 2.2 \times 10^{-6} 
\left(\frac{2.18 \times 10^{-9}}{A_s}\right)
\left(\frac{\lambda}{10^{-2}}\right)^2 \left(\frac{v}{10^{16}{\rm GeV}}\right)^4,
\end{align}
which, by itself, is obviously much too small to explain the BICEP2 result.
There are several attempts for model building that can produce
larger tensor perturbations in the context of hybrid inflation,
for example, by introducing a non-minimal K\"ahler
potential~\cite{Shafi:2010jr,Rehman:2010wm,Brummer:2014wxa}
or switching to smooth~\cite{Rehman:2012gd} or
shifted hybrid inflation~\cite{Civiletti:2011qg}.
However, almost all of these modifications can only enhance the tensor-to-scalar
ratio to at most $r\simeq 0.03$ and thus not explain the 
signal measured by the BICEP2 experiment in the framework 
of F-term hybrid inflation.
The only viable possibility to reach $r$ values as large as $r\sim0.1$ appears to
be the inclusion of non-minimal $R$ symmetry-breaking terms in the
K\"ahler potential, as recently demonstrated in Ref.~\cite{Brummer:2014wxa}.


Another source for CMB B-modes, which is inherent to F-term hybrid
inflation ending in a phase transition associated with the spontaneous
breaking of a $U(1)$ symmetry, is the cosmic string
network formed at the end of inflation.
It leads to a signal in the B-mode spectrum which is peaked at larger
multipoles than the signal expected from primordial gravitational waves,
cf.\ Ref.~\cite{Kuroyanagi:2012jf} for a recent analysis.
Generically, cosmic strings with a tension close to the current experimental upper bound,
cf.\ Eq.~\eqref{eq_cosmic_string_bound}, can have a significant effect on
the B-mode power spectrum at multipoles $\ell \sim 100$, which is currently under
investigation~\cite{Lizarraga:2014eaa,Moss:2014cra}.


Finally, we note that the BICEP2 result is in tension
with the upper bound on $r$ deduced from the PLANCK data,
$r<0.11$~\cite{Ade:2013uln}.
This tension can be relaxed by going to less minimal theoretical models,
for instance by allowing for a large running of the scalar spectral index.
But in particular upcoming experimental data from B-mode observation
experiments such as PLANCK~\cite{Planck:2006aa}, the Keck Array~\cite{Sheehy:2011yf},
ABS~\cite{EssingerHileman:2010hh}, SPTpol~\cite{SPTpol:2012aa} or
POLARBEAR~\cite{Errard:2010bn} will be crucial for any final conclusion.    


\end{appendix}



\end{document}